\documentclass[a4paper,11pt]{article}
\usepackage{fullpage}
\usepackage{color}
\usepackage{latexsym}
\usepackage{mathrsfs}
\usepackage{amsmath,amssymb,amsthm}
\usepackage{bbm} 

\usepackage{tikz-cd}                   
\usepackage{xspace,bbold}              
\usepackage[margin=1in]{geometry}      
\usepackage[utf8]{inputenc}  

\numberwithin{equation}{section}       
\allowdisplaybreaks                    

\usepackage{jheppub}                   

\makeatletter
\gdef\@fpheader{\ }                    
\makeatother

\newcommand{\dd}{\mathrm{d}}

\newcommand{\ii}{\mathrm{i}}

\newcommand{\der}{\partial}

\newcommand{\bbZ}{\mathbb{Z}}
\newcommand{\bbR}{\mathbb{R}}

\newcommand{\U}{\mathit{U}}
\newcommand{\Uni}[1]{\mathit{U}(#1)}
\newcommand{\SU}{\mathit{SU}}

\newcommand{\SO}{\mathit{SO}}
\newcommand{\USp}{\mathit{USp}}
\newcommand{\SL}{\mathit{SL}}
\newcommand{\GL}{\mathit{GL}}

\newcommand{\Spin}{\mathit{Spin}}
\newcommand{\uni}[1]{\mathfrak{u}_#1}
\newcommand{\so}{\mathfrak{so}}
\newcommand{\su}{\mathfrak{su}}
\newcommand{\usp}{\mathfrak{usp}}

\newcommand{\Cliff}{\mathrm{Cliff}}

\newcommand{\rep}[1]{\boldsymbol{#1}}
\newcommand{\repp}[2]{(\rep{#1}, \rep{#2})}
\newcommand{\id}{\mathbb{1}}

\DeclareMathOperator{\tr}{tr}

\DeclareMathOperator{\im}{Im}

\DeclareMathOperator{\vol}{vol}

\DeclareMathOperator{\Lie}{Lie}
\DeclareMathOperator{\End}{End}

\newcommand{\Gs}[1]{\Gamma(#1)}

\DeclareMathOperator{\Comm}{C}

\newcommand{\Com}[2]{\Comm_{#2}(#1)}

\newcommand{\Lgen}{{L}}

\DeclareMathOperator{\adj}{ad}

\newcommand{\GM}[2]{\big<#1,#2\big>}

\newcommand{\Edd}{\mathit{E_{d(d)}}}

\newcommand{\Ex}[1]{\mathit{E}_{#1(#1)}}
\newcommand{\ex}[1]{\mathfrak{e}_{#1(#1)}}

\newcommand{\Gst}{G_S}

\newcommand{\Ggauge}{G_{\text{gauge}}}

\newcommand{\Tint}{{T_{\text{int}}}}

\newcommand{\SUH}{\SU(2)_H}

\newcommand{\be}{\begin{equation}}
\newcommand{\ee}{\end{equation}}
\newcommand{\bea}{\begin{eqnarray}}
\newcommand{\eea}{\end{eqnarray}}
\newcommand{\nn}{\nonumber}

\newcommand{\bpm}{\begin{pmatrix}}
\newcommand{\epm}{\end{pmatrix}}

\newcommand{\ti}{\widetilde}
\newcommand{\diff}{\mathrm{d}}

\newcommand{\rme}{\mathrm{e}} 

\def\tOmega{\widetilde{\Omega}}
\def\kk{{\boldsymbol\kappa}}

\newcommand{\DW}[1]{\textcolor{green} {{\bf #1}}}
\usepackage{booktabs}

\title{$\boldsymbol{\mathcal{N}\!=\!2}$ consistent truncations from wrapped M5-branes}

\author[a]{Davide Cassani,}
\emailAdd{davide.cassani@pd.infn.it}
\author[b]{Gr\'egoire Josse,}
\emailAdd{josse@lpthe.jussieu.fr}
\author[b]{Michela Petrini,}
\emailAdd{petrini@lpthe.jussieu.fr}
\author[c]{and Daniel Waldram}
\emailAdd{d.waldram@imperial.ac.uk}

\affiliation[a]{INFN, Sezione di Padova, Via Marzolo 8, 35131 Padova, Italy}
\affiliation[b]{Sorbonne Universit\'e, UPMC Paris 05, UMR 7589, LPTHE, 75005 Paris, France}
\affiliation[c]{Department of Physics, Imperial College London,
London, SW7 2AZ, UK}

\subheader{\hfill\textrm{Imperial/TP/20/DW/1}}

\abstract{We discuss consistent truncations of eleven-dimensional supergravity on a six-dimensional manifold $M$, preserving minimal $\mathcal{N}=2$ supersymmetry in five dimensions. These are based on $G_S \subseteq\USp(6)$ structures for the generalised $E_{6(6)}$ tangent bundle on $M$, such that the intrinsic torsion is a constant $G_S$ singlet. We spell out the algorithm defining the full bosonic truncation ansatz and then apply this formalism to consistent truncations that contain warped AdS$_5 \times_{\rm w}M$ solutions arising from M5-branes wrapped on a Riemann surface. The generalised $\U(1)$ structure associated with the $\mathcal{N}=2$  solution of Maldacena--Nu\~nez leads to five-dimensional supergravity with four vector multiplets, one hypermultiplet and $\SO(3)\times \U(1)\times \mathbb{R}$ gauge group. The generalised structure associated with ``BBBW'' solutions yields two vector multiplets, one hypermultiplet and an abelian gauging. We argue that these are the most general consistent truncations on such backgrounds.
}

\begin{document}

\maketitle

\section{Introduction}

Consistent Kaluza-Klein truncations are a precious tool for constructing compactifying solutions to ten or eleven-dimensional supergravity using a simpler lower-dimensional theory. Given a splitting of the higher-dimensional spacetime into an internal manifold $M$ and an external spacetime $X$, a consistent truncation selects a finite subset of the KK modes of the higher-dimensional theory on $M$ and provides an effective theory on $X$ describing their non-linear dynamics. The selected KK modes must form a protected sector, in the sense that they do not mix in the equations of motion with the modes that have been truncated out. In this way all solutions of the lower-dimensional theory on $X$ are guaranteed to also be solutions of the original higher-dimensional theory.

For such non-trivial reduction to be possible, the internal manifold $M$ should have a special geometric structure. The simplest case is when $M$ admits a homogeneous action of a group $\mathscr{G}$, that is $M=\mathscr{G}/\mathscr{H}$ for some subgroup $\mathscr{H}\subset \mathscr{G}$. Then one can decompose all higher-dimensional fields into representations of $\mathscr{G}$ and truncate to the $\mathscr{G}$-singlets. This $\mathscr{G}$-invariant truncation is consistent, since the singlet fields can never source  the non-singlet fields. When in particular $M$ is a group manifold, $M=\mathscr{G}$, one has a conventional Scherk--Schwarz reduction \cite{Scherk:1979zr}.
 Examples of such consistent truncations in the context of M-theory---which is our principal interest in this paper---can be found in~\cite{DallAgata:2005fjb,DallAgata:2005zlf,Hull:2006tp,Cassani:2012pj}.

As a step towards more general classes of truncations, it is convenient to think about reductions on homogeneous manifolds using the language of $G$-structures.  Let us consider Scherk--Schwarz reductions for definiteness. A group manifold $M=\mathscr{G}$ admits a basis of globally  defined left-invariant one-forms,  $\{ e^a \}$,  $a=1, \ldots , \dim M$,  that reduces the structure group to the identity (i.e.~$M$ is parallelisable). Furthermore, the group action implies that $\dd e^a = \frac12 f_{bc}{}^a e^b \wedge e^c$, where $f_{bc}{}^a$ are the structure constants of the Lie algebra $\Lie\mathscr{G}$. This means that the left-invariant identity structure has singlet, constant intrinsic torsion  (singlet because $\dd e^a$ is expressed in terms of the invariant $\{ e^a \}$ basis, and constant because the coefficients of the expansion are constant).
The truncation ansatz is defined by expanding all higher-dimensional fields in the basis of invariant tensors of the structure. When this is plugged into the equations of motion, we can again invoke the argument that only singlet tensors are generated and conclude that the truncation is consistent.
Since the spin bundle is also trivialised, Scherk--Schwarz reductions preserve the full supersymmetry of the higher-dimensional theory. More generally, $\mathscr{G}$-invariant consistent truncations on coset manifolds $M =\mathscr{G}/\mathscr{H}$ are based on the existence of an $\mathscr{H}$ structure with constant, singlet intrinsic torsion, and preserve only a fraction of supersymmetry or none at all.

Interestingly, the  argument based on $G$-structure applies also to internal manifolds $M$ that are not homogeneous. It is sufficient that $M$ has a structure group $G_S$ with only constant, singlet intrinsic torsion; then the truncation to the $G_S$-singlets is guaranteed to be consistent.
 This can preserve different fractions of supersymmetry, depending on how many $G_S$-invariant spinors exist on $M$. In fact the $G_S$ structure data determine the full field content and gauge interactions of the truncated theory. Examples of this type in M-theory are the truncations based on Sasaki--Einstein and weak-$G_2$ holonomy manifolds of \cite{Gauntlett:2009zw}, and the tri-Sasakian reduction of \cite{Cassani:2011fu}.

However there are consistent truncations that are not captured by conventional $G$-structures. Classic examples are the maximally supersymmetric consistent truncations on spheres, such as eleven-dimensional supergravity on $S^7$ \cite{deWit:1986oxb} and $S^4$ \cite{Nastase:1999kf}. M-theory truncations preserving less supersymmetry  and containing warped AdS solutions can be found in~\cite{Gauntlett:2006ai,Gauntlett:2007ma,Gauntlett:2007sm,OColgain:2011ng}.\footnote{Note that whenever there is non-trivial warping the truncation falls out of the conventional $G$-structure framework.}
 Building on the evidence emerging from these examples, a conjecture was formulated in \cite{Gauntlett:2007ma} stating that for any warped, supersymmetric AdS$_D\times_{\rm w}M$ solution to higher-dimensional supergravity, there is a consistent truncation on $M$ down to $D$-dimensional pure gauged supergravity with the same amount of supersymmetry.

Exceptional Generalised Geometry and Exceptional Field Theory offer an understanding of these more complicated examples that unifies them with the conventional ones. Exceptional Generalised Geometry uses an extension of the ordinary tangent bundle $TM$ to a larger bundle $E$ on $M$, whose fibres transform in a representation of the exceptional group $\Ex{d}$. In this way the diffeomorphism and gauge symmetries of higher-dimensional supergravity are unified as generalised diffeomorphisms on $E$. The notion of generalised $G_S$ structure, that is a $G_S$ structure of $E$, rather than of $TM$, leads to a new systematic approach to consistent truncations with different amounts of supersymmetry: 
one can argue that there is a consistent truncation any time a supergravity theory is reduced on a manifold $M$ admitting a generalised  $G_S$  structure with constant singlet intrinsic torsion \cite{Cassani:2019vcl}.
In particular, all maximally supersymmetric truncations, both conventional Scherk--Schwarz reductions  and sphere truncations, can be seen as generalised Scherk--Schwarz reductions on generalised parallelisable manifolds~\cite{Lee:2014mla,Baron:2014yua,Hohm:2014qga,Baron:2014bya,Baguet:2015sma,Lee:2015xga,Ciceri:2016dmd,Cassani:2016ncu,Inverso:2017lrz}. This also provides a connection to Poisson--Lie T-duality as described in~\cite{Demulder:2019vvh} (see also \cite{Severa:2018pag}). Truncations preserving less supersymmetry are based on generalised structures larger than the identity, the half-maximal case having been explored rather extensively by now \cite{Malek:2016bpu,Malek:2017njj,Ciceri:2016hup,Malek:2018zcz,Malek:2019ucd,Cassani:2019vcl}.
 Moreover, a proof of the conjecture of \cite{Gauntlett:2007ma} was given in this framework \cite{Malek:2017njj,Cassani:2019vcl}, based on the fact that the conditions for a supersymmetric AdS$_D\times_{\rm w}M$ vacuum can be rephrased as the requirement that $M$ admits a generalised $G_S$ structure with vanishing non-singlet intrinsic torsion  \cite{Coimbra:2014uxa,Coimbra:2015nha,Ashmore:2016qvs}. 

Although the general ideas were illustrated in \cite{Cassani:2019vcl} for any amount of supersymmetry, the Exceptional Generalised Geometry approach to consistent truncations has been developed just for maximal and half-maximal supersymmetry so far.
In this paper we enlarge this framework and discuss in detail truncations of eleven-dimensional supergravity preserving minimal $\mathcal{N}=2$ supersymmetry in five dimensions. 

While a strict $\USp(6)\subset \Ex{6}$ generalised structure leads to a truncation to minimal $\mathcal{N}=2$ gauged supergravity in five dimensions, smaller $G_S\subset \USp(6)$ structures lead to matter coupled supergravity. We show how the $G_S\subset \USp(6)$ structure defines a continuous family of $\USp(6)$ structures, and identify the moduli space of this family with the vector multiplet and hypermultiplet scalar manifolds in the truncated five-dimensional theory. We also show how the generalised Lie derivative acting on the generalised tensors defining the $G_S$ structure specifies the isometries of the scalar manifold that are being gauged. This fully determines the truncated $\mathcal{N}=2$ supergravity theory. 

We then derive general expressions that encode the uplift formulae for how the eleven-dimensional bosonic fields are encoded in terms of the moduli and the generalised tensors defining the $G_S$ structure. In order to make this truncation ansatz explicit we need to solve a number of technical issues. One is that, in contrast to the maximal and half-maximal case, the structure is not entirely characterised by the generalised vectors $K_I$ (i.e.~sections of $E$, transforming in the fundamental of $\Ex{6}$) which control the vector multiplet sector of the truncated theory. We also need to  consider generalised tensors $J_A$ belonging to the $\Ex{6}$ adjoint bundle, which eventually control the hypermultiplet sector. A related point, that is crucial to derive the scalar truncation ansatz, is the construction of the generalised metric on $E$, which receives contributions both from the $K_I$ and the $J_A$. A significant advantage of the formalism however, is that the expressions are universal. The ansatz can be applied to any $\mathcal{N}=2$ background once one identifies the $K_I$ and $J_A$ singlets.   

As application, we discuss M-theory truncations on geometries associated with M5-branes wrapping a Riemann surface $\mathbb{\Sigma}$. The near-horizon geometry of this brane configuration is given by a warped AdS$_5\times_{\rm w} M$ solution to eleven-dimensional supergravity, where $M$ is a fibration of a deformed $S^4$ over $\mathbb{\Sigma}$ \cite{Maldacena:2000mw,Bah:2012dg}. The fibration corresponds to a topological twist in the dual superconformal field theory on the M5-branes, where the holonomy of the Riemann surface is cancelled by a $\U(1)$ in the $\SO(5)$ R-symmetry, which in the supergravity background is realised geometrically as the isometries of $S^4$. Depending on which $\U(1)$ is chosen, one obtains different AdS$_5\times_{\rm w} M$ solutions, and correspondingly different $\U(1)_S$ generalised structures.

We start with the $\mathcal{N}=2$ background of Maldacena--Nu\~nez \cite{Maldacena:2000mw}: specifying its $\U(1)_S$ generalised structure and discussing its singlet intrinsic torsion, we obtain a consistent truncation to five-dimensional \hbox{$\mathcal{N}=2$} supergravity including four vector multiplets, one hypermultiplet, and a non-abelian $\SO(3)\times\U(1)\times\mathbb{R}$ gauging.  This extends the abelian truncation of \cite{Faedo:2019cvr} (see also \cite{Gauntlett:2006ai,Szepietowski:2012tb} for previous subtruncations) by adding $\SO(3)$ vector multiplets, which in the dual superconformal field theory source $\SO(3)$ flavour current multiplets. We also spell out the full bosonic truncation ansatz.  The same construction also applies to the ``BBBW'' solutions \cite{Bah:2011vv,Bah:2012dg}, as 
the corresponding  generalised structure is a simple deformation of the  Maldacena--Nu\~nez one, controlled by a (discrete) parameter describing the choice of $\U(1)_S$ in $\SO(5)$. The corresponding truncation features only two $\mathcal{N}=2$ vector multiplets, one hypermultiplet and an abelian gauging. We show that the Maldacena--Nu\~nez truncation admits a new non-supersymmetric AdS$_5$ solution when the Riemann surface is a sphere, which turns out to be perturbatively unstable. We also find new non-supersymmetric vacua in the BBBW truncations. Together with the consistent truncation including the $\mathcal{N}=4$ solution of \cite{Maldacena:2000mw}, whose $\U(1)_S$ generalised structure embeds in $\USp(4)\subset\USp(6)$ and leads to half-maximal supergravity \cite{Cheung:2019pge,Cassani:2019vcl}, the present study completes the landscape of what we believe are the most general consistent truncations that can be derived from eleven-dimensional supergravity on known smooth solutions associated with M5-branes wrapped over Riemann surfaces.\DW{\footnote{It may be possible to find other consistent truncations, that are not subsectors of the ones given here by using large structure groups, in analogy with the consistent truncation on $S^7$ viewed as a Sasaki--Einstein manifold~\cite{Gauntlett:2009zw} rather than a generalised parallelised sphere. However such truncations will have fewer fields.}}

The rest of the paper is organised as follows. In Section~\ref{sec:genN2} we characterise the generalised structure relevant for M-theory truncations on a six-dimensional manifold preserving $\mathcal{N}=2$ supersymmetry. In Section~\ref{sec:gen_str_trunc_ansatz} we specify the truncation ansatz and discuss how the gauging is determined from the generalised structure. In Sections~\ref{sec:MN1section} and~\ref{sec:BBBWsection} we apply our formalism to consistent truncations associated with M5-branes wrapping a Riemann surface, first for Maldacena--Nu\~nez backgrounds and then for BBBW ones. We conclude in Section~\ref{sec:Conclusions}. The appendices contains a brief account of $\Ex{6}$ generalised geometry, a summary of five-dimensional $\mathcal{N}=2$ gauged supergravity and some \hbox{technical details of our computations.}


\section{M-theory generalised structures and $\mathcal{N}=2$ supersymmetry}
\label{sec:genN2}

In this section we first recall some basic notions of Exceptional Generalised Geometry for the case of interest here, namely eleven-dimensional supergravity on a six-dimensional manifold, and then we illustrate how the general procedure described in \cite{Cassani:2019vcl} applies to consistent truncations to five-dimensional $\mathcal{N}=2$ gauged supergravity. A more extended review of the relevant generalised geometry can be found in Appendix~\ref{PreliminariesE66_Mth}.

\subsection{The HV structure}\label{sec:HVstr}

Consistent  truncations of  eleven-dimensional supergravity  on a six-dimensional manifold  $M$ are based on 
 $\Ex{6} \times \mathbb{R}^+$ generalised geometry. 
This extends the tangent bundle $TM$ to the generalised tangent bundle $E$ on $M$, and the corresponding structure group $\GL(6)$ to $\Ex{6}$. 
The group $\Ex{6}$ contains $\GL(6)$ as its geometrical subgroup, and we can use the latter to decompose the generalised tangent bundle as 
 \begin{equation}
 \label{gentan} 
   E \,\simeq \, TM \oplus \Lambda^2T^*M \oplus \Lambda^5T^*M \, .
\end{equation}
Therefore the sections of $E$ consist, locally, of the sum of a vector, a two-form and a five-form on $M$,
\be
V = v + \omega + \sigma \, . 
\ee
These are called  generalised vectors and transform in the ${\bf 27}$ of  $E_{6(6)}$. 

All geometric structures of conventional geometry on $M$, such as tensors, Lie derivative, connections etc,  admit an extension to $E$~\cite{Hull:2007zu,Pacheco:2008ps,Coimbra:2011ky}.
In particular,  generalised tensors are defined by considering bundles whose fibers transform in different representations of $E_{6(6)}$. 
We can define dual generalised vectors $Z$  as the sections of the dual tangent bundle
\be
 \label{dgentan} 
   E^* \,\simeq \, T^*M \oplus \Lambda^2TM \oplus \Lambda^5TM \, ,
\end{equation}
transforming in the ${\bf \overline{27}}$ of $\Ex{6}$. 
Locally the dual vectors are sums of a one-form $\hat{v}$, 
 a two-vector $ \hat{\omega} $ 
 and a five-vector $\hat{\sigma}$,
\be
Z = \hat{v} + \hat{\omega} + \hat{\sigma}  \, . 
\ee
The adjoint bundle transforms in  the ${\bf 1} + {\bf 78}$ of $E_{6(6)}$  and, in terms of $GL(6)$ tensors, is defined as 
\be
{\adj} F \,\simeq \, \bbR \oplus (TM \otimes T^* M) \oplus \Lambda^3 T^* M \oplus \Lambda^6 T^* M  \oplus \Lambda^3 T M \oplus \Lambda^6 T  M  \, , 
\ee
with sections
\be
R = l + r + a + \tilde a + \alpha + \tilde \alpha \, ,
\ee
where, locally,  $l \in \bbR$,  $r \in \End(TM)$, $a \in   \Lambda^3 T^* M$ is a three-form, $\tilde a \in   \Lambda^6 T^* M$  is a six-form and
$ \alpha \in   \Lambda^3 TM$ and  $ \tilde \alpha  \in  \Lambda^6 TM$. 
This bundle plays an important role as the components of the M-theory
three-form and six-form gauge potentials are embedded in ${\adj}
F$.

As we will see, the bosonic fields of eleven-dimensional supergravity can be
unified into generalised tensors. The supergravity spinors on the
other hand arrange into representations transforming under $\USp(8)$, 
the double cover of the maximal compact subgroup $\USp(8)/\bbZ_2$ of
$E_{6(6)}$. For example the supersymmetry parameters are section of the generalsied spinor bundle $\mathcal{S}$, transforming in the ${\bf 8}$ of 
$\USp(8)$. It will be this compact $\USp(8)$ or more generally a subgroup of
it, that determines the R-symmetry of the reduced five-dimensional theory. 

\bigskip

The manifold $M$  admits  a {\it generalised  structure}, $G_S\subset
\USp(8)/\bbZ_2$,  when the  structure group $E_{6(6)}$ is reduced to the
subgroup $G_S$. Typically this can be characterised by the existence of 
globally defined generalised tensors that are invariant under $G_S$.  
The amount of supersymmetry of the eleven-dimensional theory that is
preserved by the $G_S$ structure is given by  the number of $G_S$
singlets in the spinor bundle, $\mathcal{S}$.\footnote{Here we will assume that either $G_S$ is simply connected or is $U(1)$ so that it lifts to a $G_S$ subgroup of $\USp(8)$.} 

In this paper we are interested in structures preserving  $\mathcal{N}=2$ supersymmetry.  
The generic case is provided by what has been called an {\it HV structure}~\cite{Grana:2009im,Ashmore:2015joa,Ashmore:2016qvs}. It consists of 
a triplet of globally defined tensors in the adjoint bundle, $J_\alpha \in \Gamma(\adj F)$, with $\alpha =1,2,3$, satisfying 
\be 
\label{normJ}
[J_\alpha , J_\beta] \,=\,\ 2 \epsilon_{\alpha \beta \gamma} J_\gamma\,, \qquad \tr (J_\alpha J_\beta) \,=\,  - \delta_{\alpha \beta}  \, , 
\ee
together with a globally defined generalised vector $K\in \Gamma(E)$ having positive norm with respect to the $E_{6(6)}$ cubic invariant,
\be\label{eq:cKKK}
c(K,K,K) \,:=\, 6\,\kk^2 > 0  \, , 
\ee
where $\kk$ is a section\footnote{Recall that $\det T^*M$ is just a different notation for the top-form bundle $\Lambda^6T^*M$ that stresses that it is a real line bundle. Here we are assuming that the manifold is orientable and hence $\det T^*M$ is trivial and so we can define arbitrary powers $(\det T^*M)^p$ for any real $p$.} of $(\det T^*M)^{1/2}$, and satisfying the compatibility condition
\be
\begin{aligned}
\label{compJK}
 & J_\alpha  \cdot K \,=\, 0\,, 
\end{aligned}
\ee
where $\cdot$ denotes the adjoint action.\footnote{\label{J-conv} Note that we are using slightly different conventions for the $J_\alpha$ tensors compared with~\cite{Ashmore:2015joa}. In particular $J^\text{AW}_\alpha=\kk J^\text{here}_\alpha\in \Gamma({( \det T^*M)^{1/2}\otimes \rm ad}F)$.} See Appendix \ref{PreliminariesE66_Mth} for a definition 
of the cubic invariant and the other generalised geometry operations appearing in these formulae.
  

The HV structure $\{J_\alpha, K\}$ defines a reduction of the structure group to  $\USp(6) \subset E_{6(6)}$. Indeed the vector $K$ is stabilised by $F_{4(4)} \subset E_{6(6)}$, 
while the  $J_\alpha$ are invariant under the subgroup  $SU^*(6)$.  The compatible $K$ and $J_\alpha$ have $SU^*(6) \cap F_{4(4)} \simeq \USp(6)$ as a common stabiliser.
 The globally defined vector $K \in \Gamma(E)$ with positive norm  is called a  vector-multiplet structure, or {\it V structure} for short. A triplet of $J_\alpha \in \Gamma({\adj} F)$ that define the highest root $\frak{su}_2$ subalgebra of $\frak{e}_{6(6)}$ and satisfy the  conditions \eqref{normJ}  is called a  hypermultiplet structure, or {\it H structure}.  This justifies the name HV structure for the compatible pair $\{J_\alpha, K\}$.

It is easy to check that the amount of supersymmetry preserved by a HV structure is  $\mathcal{N}=2$. Under the breaking
\be
\label{Usp6br}  
\USp(8) \supset \USp(6) \times \SUH  \, . 
\ee 
the spinorial representation decomposes as  ${\bf 8} = ({\bf 6}, {\bf 1}) \oplus ({\bf 1}, {\bf 2})$ and  we see that the are
only two  $\USp(6)$ singlets.  The  $\SUH$ factor in  \eqref{Usp6br} is the R-symmetry of the reduced theory  so that the two singlets form
 an  R-symmetry doublet,  as expected for $\mathcal{N}=2$ supersymmetry parameters. 

A strict $\USp(6)$ structure is not the only option to obtain $\mathcal{N}=2$ supersymmetry. In fact, any subgroup $G_S$  that embeds in 
$\USp(6)$  in such a way that there are no extra singlets in the
decomposition of the spinorial representation of $\USp(8)$ does the
job. Although the number spinor singlets is unchanged, 
when the structure group is smaller than $\USp(6)$ in general one finds more $G_S$ singlets in the  decomposition of the ${\bf 27}$ and the  ${\bf 78}$ representations. 
Let us denote by 
\be
K_I\,, \qquad I= 0, \ldots, n_V\,,
\ee
the set of independent generalised vectors corresponding to $G_S$ singlets in the ${\bf 27}$, and by
\be
J_A\,, \qquad A = 1, \ldots, {\rm dim}\, \mathcal{H}\,,
\ee
 the set of independent sections of the adjoint bundle corresponding to $G_S$ singlets in the ${\bf 78}$ that also satisfy the condition\footnote{Note that there are  singlets  in the adjoint bundle that do not satisfy \eqref{eq:JdotK=0}. These are given by $K_I \times_{\adj} K^*_J$, where $K^*_J$  is the dual of the generalised vector $K_J$ and $\times_{\adj}$ is the projection onto the adjoint bundle, and will not play a relevant role in our construction.\label{foot:KtimesK*}}
\be\label{eq:JdotK=0}
J_A \cdot K_I \,=\, 0 \qquad \forall \, I \text{ and }\forall\, A \, .
\ee
The latter generate a subgroup $\mathcal{H}\subset\Com{G_S}{\Ex{6}}$, where $\Com{G_S}{\Ex{6}}$ is the commutant of $G_S$ in $E_{6(6)}$, so that
\be\label{commutorJAJB}
[ J_A, J_B ] = f_{AB}{}^C J_C \, ,
\ee
with $f_{AB}{}^C$ being the structure constants of $\mathcal{H}$. The generalised structure
$G_S \subseteq \USp(6)$  is fully characterised as the group
preserving the set 
\be
 \{ K_I, J_A  \}\,.
\ee
We can always normalise such that the $ n_V + 1$ generalised vectors
satisfy
\be\label{eq:CKKK_cond}
c(K_I, K_J, K_K) = 6\,\kk^2 C_{IJK} \,,
\ee
with $C_{IJK}$ a  symmetric, constant tensor and $\kk$ is a section of
$(\det T^*M)^{1/2}$ fixed by the structure. In addition we can
normalise the adjont singlets to satisfy
\be
\tr(J_A  J_B) = \,\eta_{AB} \,,
\ee
where $\eta_{AB}$ is a diagonal matrix with $-1$ and $+1$ entries in correspondence with compact and non-compact generators of $\mathcal{H}$, respectively.

Any generalised structure has an associated intrinsic torsion \cite{Coimbra:2014uxa}, which is defined as follows. 
Let $\tilde{D}$ be a generalised connection compatible with the
$\Gst$-structure, that is, sastisfying $\tilde{D}Q_i=0$ for all $i$,
where $Q_i$ is the set of invariant generalised tensors that define
the structure. Formally, the generalised torsion $T$ of $\tilde{D}$ is defined by, acting on any generalised tensor $\alpha$, 
\begin{equation}
   \label{eq:gen-torsion}
   \big(\Lgen^{\tilde{D}}_V - \Lgen_V\big)\, \alpha = T(V) \cdot \alpha \, , 
\end{equation}
where $\Lgen$ is the generalised Lie derivative, $\Lgen^{\tilde{D}}$ is the generalised Lie derivative calculated using $\tilde{D}$ and  $\cdot$ is the  adjoint action on $\alpha$.\footnote{
We view the torsion as a map $T:\Gs{E}\to\Gs{\adj F}$ where $\adj F$ is the $\Edd\times\bbR^+$ adjoint bundle.} The intrinsic torsion is  the component of $T$ that is independent of the choice of compatible connection $\tilde{D}$, and hence is fixed only by the choice of generalised structure. In general, one can decompose the intrinsic torsion into representations of $\Gst$. In particular, for a consistent truncation we will be interested in the case where only the singlet representations are non-zero.

\subsection{The generalised metric} 

An important ingredient to derive a consistent truncation is the generalised metric $G$ on $M$. This is a positive-definite, symmetric rank-2 tensor on the generalised tangent bundle, 
\begin{align}
G : E \otimes E &\to \bbR^+ \nn\\
 (V,V') &\mapsto G(V,V') = G_{MN}V^MV'{}^N\,,
\end{align}
that encodes the degrees of freedom of eleven-dimensional supergravity that correspond to scalars in the reduced theory. We provide the explicit relation between the generalised metric and the supergravity fields on $M$ in Eq.~\eqref{gen_metr_general}. 
The generalised metric is defined in analogy to the ordinary metric: a metric $g$ on $M$  can be seen as an $O(6)$ structure
on $TM$ that at each point on $M$ parameterises the coset $GL(6)/
O(6)$. Similarly, at each point $p\in M$ a choice of a generalised metric corresponds to an element of the coset
\be
 \left.G\right|_p \in \frac{ E_{6(6)} \times \mathbb{R}^+}{\USp(8)/\mathbb{Z}_2} \,  . 
\ee
 
Since a $G_S\subset\USp(8)/\bbZ_2$, the $G_S$ structure will determine
a $G_S$-invariant generalised metric, given in terms of the invariant tensors that are used to define the $G_S$ structure. The expression of $G_{MN}$ that is relevant for truncations preserving maximal and half-maximal supersymmetry was given in  \cite{Lee:2014mla,Cassani:2016ncu,Cassani:2019vcl}.  Here we will discuss the $\mathcal{N}=2$ case.  

Consider first the case of a generic $\USp(6)$ structure. As discussed in the previous section this is specified by an invariant generalised vector, $K$,  together with an 
 $\su(2)$ triplet of sections of the adjoint bundle $J_{\alpha}$, $\alpha=1,2,3$. These objects define a $\USp(6)$-invariant generalised metric through the formula
\be\label{USp6_gm}
G(V,V) \,=\, 3\left(3 \, \frac{c(K,K,V)^2}{c(K,K,K)^2}  - 2  \frac{c(K,V,V)}{c(K,K,K)} +  4 \, {\frac{c(K,J_3\cdot V,J_3\cdot V)}{c(K,K,K)}}\right)\ .
\ee
This formula can be motivated as follows.
The  globally defined  $K$ induces the splitting of the  $\rep{27}$ of $\Ex{6}$ into orthogonal subspaces 
 \be
 V  = V_{\rep{0}} + V_{\rep{26}}
 \ee
in the singlet and $\rep{26}$ representation of $F_{4(4)}$;
correspondingly, the $E_{6(6)}$ cubic invariant on the $\rep{26}$ reduces to the
symmetric invariant form of $F_{4(4)}$ 
\be
\label{genmet1}
c(K, V,V) = c(K, V_{\rep{0}}, V_{\rep{0}}) + c(K, V_{\rep{26}}, V_{\rep{26}} ) \, .
 \ee
This expression however is not positive definite, since the symmetric form of $F_{4(4)}$ has signature $(14 , 12)$ and overall  \eqref{genmet1} has signature 
$(14 , 13)$.  The first term in  \eqref{USp6_gm}  contains the contribution from the singlet component $V_{\bf 0}$ and makes the metric positive definite in the singlet.
To do the same in the $\rep{26}$ we need the full HV structure. Under $ \SUH \times \USp(6)$  the $\rep{27}$ decomposes  as 
 \be
 \rep{27} =  \rep{1} \oplus   (\rep{1}, \rep{14})   \oplus    (\rep{2}, \rep{6}) \, , 
\ee
and the action   $J_\alpha$ on  $V$ projects on the $({\bf 2,6})$ part, as the rest is an $\SUH$ singlet.  Then we can write the contribution to the metric in the $(\rep{2}, \rep{6})$
as 
\begin{equation}
\label{Jact}
c(K,J_3\cdot V,J_3\cdot V)  \, ,  
\end{equation}
and add it in  \eqref{USp6_gm} to make it positive definite. 
Note that \eqref{USp6_gm} only contains one element of the triplet
$J_\alpha$, that we chose to be $J_3$. This is because, for each
$J_\alpha$, 
\begin{equation}
c(K,J_\alpha\cdot V,J_\alpha\cdot V) = -c(K, V, (J_\alpha)^2\cdot V) = c(K,V_{(\rep{2}, \rep{6})},V_{(\rep{2}, \rep{6})})\,,
\end{equation}
where  there is no sum over $\alpha$ and in the last equality we have
used that $(J_\alpha)^2 = - \id$ in the $V_{(\rep{2}, \rep{6})}$
subspace. We see that the action of each of the $J_\alpha$ gives the
same result. This reflects the fact that the generalised metric is
independent of the action of the $\SUH$ supergravity R-symmetry. 

For the purpose of constructing the truncation ansatz by comparing with \eqref{gen_metr_general}, we will also need the inverse generalised metric. We can exploit the isomorphism between the generalised tangent bundle $E$ and its dual $E^*$ provided by the generalised metric to construct a $\USp(6)$ singlet $K^*\in \Gamma(E^*)$ as $K^*(V) = G(K,V)$, where $V$ is any generalised vector.
Then, denoting by $Z\in \Gamma(E^*)$ a generic dual vector, the inverse generalised metric is given by
\be
\label{USp6_igm}
G^{-1}(Z,Z) \,=\, 3\left( 3 \, \frac{c^*(K^*,K^*,Z)^2}{c^*(K^*,K^*,K^*)^2} - 2  \frac{c^*(K^*, Z, Z)}{c^*(K^*,K^*,K^*)}  + 4 \, \frac{c^*(K^*, J_3\cdot Z, J_3\cdot Z)}{c^*(K^*,K^*,K^*)}\right)\,,
\ee
where the action of the cubic invariant $c^*$ and of the adjoint elements $J_\alpha$ on the dual generalised vectors can be found in Appendix \ref{PreliminariesE66_Mth}.

\subsection{The HV structure moduli space and the intrinsic torsion}\label{sec:HVstr_moduli}

When the $G_S$ structure is a subgroup of $\USp(6)$ (and there is no
supersymmetry enhancement), it determines an $\USp(6)$ structure and
hence by definition defines a generalised metric. However, a given
$G_S$ structure can determine several different $\USp(6)$
structures. Thus one gets a family of generalised metrics that can be
obtained from the $G_S$-invariant tensors, depending on which
$\USp(6)$ structure one chooses. Concretely, we use the  $K_I$ and  $J_A$ tensors characterising the $G_S$ structure to construct a generalised vector $K$ and a triplet of $J_\alpha$  satisfying \eqref{normJ}--\eqref{compJK}, which then we use  to build the generalised metric as in \eqref{USp6_gm}.
 The parameterisation of $K$ and $J_\alpha$ in terms of $K_I$ and $J_A$ provides a set of deformations of a reference $\USp(6)$-invariant metric, that correspond to acting on  the structure with elements of  $\Ex{6}$ that commute with $G_S$,  modulo elements of $\USp(8)/\bbZ_2$ 
that commute with $G_S$. 
The resulting generalised metric thus parameterises the coset
 \begin{equation}
    \label{eq:GS-coset}
   \mathcal{M} \,=\, \frac{\Com{\Gst}{\Ex{6}}}{\Com{\Gst}{\USp(8)/\bbZ_2}} \,.
\end{equation}
This is the moduli space of our $G_S$ structure, namely the space of deformations of the reference $\USp(6)$ structure that preserve the $G_S$ structure. For the $\mathcal{N}=2$  structures of interest in this paper, this splits in the product
\be
\mathcal{M} \,=\, \mathcal{M}_{\rm V} \times \mathcal{M}_{\rm H}\,,
\ee
where $\mathcal{M}_{\rm V}$ is the V structure moduli space, corresponding to deformations of $K$ that leave $J_\alpha$ invariant, while $\mathcal{M}_{\rm H}$ is the H structure moduli space, which describes deformations of $J_\alpha$ that leave $K$ invariant. The fact that these deformations are independent follows from the requirement \eqref{eq:JdotK=0}.
 When given a dependence on the external spacetime coordinates these deformations provide the scalar fields in the truncated theory, with $\mathcal{M}_{\rm V}$ and $\mathcal{M}_{\rm H}$ being identified with the vector multiplet and the hypermultiplet scalar manifolds, respectively.

We next outline how to construct the V and H structure moduli spaces. The procedure will be further illustrated in Sections~\ref{sec:MN1section} and \ref{sec:BBBWsection}, where concrete examples will be discussed in detail.

\paragraph{The V structure moduli space.} A family of V structures is obtained by parameterising the generalised vector $K$ as the linear combination 
\be\label{eq:Kdressing}
K = h^I K_I\,,
\ee 
where $h^I$, $I= 0,\ldots, n_{\rm V}$, are real parameters, and imposing the property \eqref{eq:cKKK}. Using \eqref{eq:CKKK_cond}, this is equivalent to
\begin{equation}\label{constraint_Chhh}
C_{IJK}h^Ih^Jh^K=  1\,,
\end{equation}
showing that the $n_{\rm V}+1$ parameters $h^I$ are constrained by one real relation and thus define an $n_{\rm V}$-dimensional hypersurface,
\be
\mathcal{M}_{\rm V} = \{\,  h^I :\   C_{IJK}h^Ih^Jh^K=1 \,\}\,.
\ee
 This is our V structure moduli space. It will be identified with the vector multiplet scalar manifold in five-dimensional supergravity. The metric on $\mathcal{M}_{\rm V}$ is obtained by evaluating the generalised metric on the invariant generalised vectors,
 \be
 a_{IJ} = \tfrac{1}{3} \,G(K_I,K_J)\,.
 \ee
 Using \eqref{USp6_gm}, it is straightforward to see that this gives
 \be\label{aIJ_maintext}
 a_{IJ} = 3 h_I h_J - 2 C_{IJK}h^K\,,
 \ee
 where $h_I = C_{IKL}h^Kh^L$.
 Then $g_{\rm ambient} = \frac{3}{2}\,a_{IJ}\dd h^I \dd h^J$ gives the metric on the ambient space,\footnote{The normalisation is chosen so as to match standard conventions in $\mathcal{N}=2$ supergravity, see Appendix~\ref{app:sugra_review}.} and the metric on $\mathcal{M}_V$ is obtained as the induced metric on the hypersurface.

\paragraph{The H structure moduli space.}
A family of H structures is obtained by parameterising the possible
$\su_2$ subalgebras of the algebra spanned by the $J_A$. The fact that
we only have two singlet spinors means that
$\Com{\Gst}{\USp(8)/\bbZ_2}$ must contain an $\SUH$ factor (as in~\eqref{Usp6br}) that acts on the two singlet spinors. Furthermore, the corresponding $\su_2$ algebra must be generated by a highest root in $\ex6$. The Lie algebra $\mathfrak{h}=\Lie \mathcal{H}$ generated by the $J_A$ is the simple subalgebra of the Lie algebra of $\Com{\Gst}{\Ex6}$ that contains the $\su_2$ factor. Since $\mathfrak{h}\subset\ex6$ the $\su_2$ algebra is generated by a highest root in $\mathfrak{h}$. 

The H structure moduli space is the space of choices of such highest root $\su_2$ algebras in $\mathfrak{h}$, namely the symmetric space\footnote{Note the strictly the denominator group is not quite the product $\SUH\times \Com{\SUH}{\mathcal{H}}$ but generally involves modding out correctly by terms in the centre of each factor. Here we will ignore these subtleties.}
\be\label{Hstr_space_general}
\mathcal{M}_{\rm H} \,=\, \frac{\mathcal{H}}{\SUH\times \Com{\SUH}{\mathcal{H}}}\,.
\ee 
Such spaces are known as ``Wolf spaces'' and are all quaternionic--K\"ahler, as expected from the fact that $\mathcal{M}_{\rm H}$ is going to be identified with hyperscalar manifold in five-dimensional supergravity. Points in $\mathcal{M}_{\rm H}$ can be parameterised by starting from a reference subalgebra $\mathfrak{j} \simeq\su_2 \subset \mathfrak{h}$ and then acting on a basis $\{ j_1,j_2,j_3 \}$ of $\mathfrak{j}$ by the adjoint action of group elements $h \in \mathcal{H}$, defined as 
\be\label{adjoint_action_on_j}
 J_\alpha = 
 {\adj}_\mathcal{H} \, j_\alpha = h \, j_\alpha\, h^{-1} \,.
\ee 
Clearly, this action acts trivially on $\mathfrak{j}$ if $h\in \SUH\simeq \exp(\mathfrak{j})$, or if $h$ belongs to the commutant of this $\SUH$ in $\mathcal{H}$, that is $h\in \Com{\SUH}{\mathcal{H}}$. 
This way, we obtain a triplet of ``dressed'' generalised tensors $J_\alpha$, $\alpha=1,2,3$, which depend on the coset coordinates and parameterise our family of H structures. 

\paragraph{The intrinsic torsion.} This picture that the $G_S$-structure defines a family of HV structures also allows us to give a characterisation of the intrinsic torsion. As discussed in~\cite{Ashmore:2015joa}, the intrinsic torsion of an HV structure is encoded in the three quantities
\begin{equation}
   \label{eq:HV-intrinsic}
   L_KK , \quad
   L_K J_\alpha , \quad
   \mu_\alpha
\end{equation}
where, given a generalised vector $V\in\Gamma(E)$, one defines a triplet of functions\footnote{Recall the change of conventions from those of~\cite{Ashmore:2015joa} discussed in footnote~\ref{J-conv}.}
\begin{equation}
   \label{eq:moment-maps}
   \mu_\alpha(V) = - \tfrac12 \,\epsilon_{\alpha\beta\gamma}
       \int_M \kk^2 \tr \left( J_\beta (L_V J_\gamma ) \right)
\end{equation}
that formally are moment maps for the action of the generalised diffeomorphism group on the space of H structures.

In general, if $K_{G_S}$ is the space of $G_S$-compatible connections, then the definition~\eqref{eq:gen-torsion} defines a map $\tau: K_{G_S}\to W$
where now we view the generalised torsion $T$ as a section of $W\subset E^*\otimes \adj F$.\footnote{For $\Ex6$ generalised geometry $W$ transforms in the ${\bf \overline{27}}+{\bf 351}$ representation.} The $G_S$-intrinsic torsion is then an element of $W_{\text{int}}^{G_S}=W/W_{G_S}$ where $W_{G_S}=\im \tau$. Now let $p\in\mathcal{M}$ be a particular point in the family of HV structures~\eqref{eq:GS-coset} and $\USp(6)_p\subset\Ex6$ be the corresponding structure group. By construction, $G_S$ is the common subgroup of all the $\USp(6)_p$ subgroups. This means that
\begin{equation}
 K_{G_S} = \bigcap_p K_{\USp(6)_p} , 
\end{equation}
that is, only a $G_S$-compatible connection is compatible with every HV structure in the family. Hence $W_{G_S}=\bigcap_p W_{\USp(6)_p}$ and so
\begin{equation}
   W_{\text{int}}^{G_S} = \bigcup_p W_{\text{int}}^{\USp(6)_p} . 
\end{equation}
In other words, knowing the intrinsic torsion of every HV structure in the family fixes the intrinsic torsion of the $G_S$ structure.

Now, recall that each $K$ in the family of HV structures is a linear combination of $K_I$ (with constant coefficients), while each $J_\alpha$ is defined by the exponentiated adjoint action of a linear combination of $J_A$ (with constant coefficients) on a fixed reference $\su(2)$ algebra. Hence the intrinsic torsion components $L_KK$ and $L_KJ_\alpha$ for the whole family are determined by knowing
\begin{equation}
   \label{eq:GS-int-tor-1}
   L_{K_I}K_J , \qquad
   L_{K_I}J_A . 
\end{equation}
These also determine $\mu_\alpha(V)$ when $V$ has the form $V=V^IK_I$, even when the components $V^I$ are functions because of the condition~\eqref{eq:JdotK=0}. Thus the final components of the $G_S$ intrinsic torsion are determined by
\begin{equation}
   \label{eq:GS-int-tor-2}
   \int_M\kk^2 \tr(J_A(L_WJ_B))\, , 
\end{equation}
where we require $c(K_I,K_J,W)=0$, which defines a generalised vector that is orthogonal to those of the form $V=V^I K_I$. Note that the expressions~\eqref{eq:GS-int-tor-1} and~\eqref{eq:GS-int-tor-2} are in general not independent, but are sufficient to determine the intrinsic torsion.

\section{M-theory truncations to $\mathcal{N}=2$ supergravity in five dimensions}
\label{sec:gen_str_trunc_ansatz}

Any generalised $G_S$ structure on a manifold $M$ with only constant,
singlet intrinsic torsion gives rise to a consistent truncation of
eleven-dimensional or type II supergravity on $M$. While the general
ideas (along with the details for five-dimensional truncations
preserving half-maximal supersymmetry) were given in
\cite{Cassani:2019vcl}, here we focus on the specific case of
truncations of eleven-dimensional supergravity leading to
$\mathcal{N}=2$ supergravity in five dimensions, based on $G_S
\subseteq \USp(6)$ structures.

Although some of the formulae giving the truncation ansatz in terms of the
structure are necessarily quite involved, a great advantage is that they are
universal expressions good for any $\mathcal{N}=2$ consistent
truncation. One does not have to search for the correct set of
consistent modes on a case-by-case basis. All the particulars of the given truncations are encoded 
in terms of the given $G_S$ structure defined by the set of $K_I$ and
$J_A$ singlets. For example, following the discussion in the previous
section, the scalar matter content is determined by the commutant of
$G_S$ in $\Ex{6}$, giving $n_{\rm V}$ vector multiplets and $n_{\rm H}$
hypermultiplets, whose scalar manifolds are identified with the V
structure and H structure moduli spaces, respectively. 
The gauge interactions of the
truncated theory are determined by the torsion of the
$\Gst$-structure, which in turn depends only the generalised Lie
derivatives $L_{K_I}K_J$ and $L_{K_I}J_A$. Together this data completely specifies the full five-dimensional supergravity.

\subsection{The gauging}

The gauge interactions of the truncated theory are determined by the intrinsic torsion of the generalised structure $G_S$. As already emphasised, we assume that the intrinsic torsion takes values in the singlet representation of $G_S$, with components that are constant on~$M$.
As explained in~\cite{Cassani:2019vcl}, this means that 
the generalised Lie derivative  along the invariant vectors $K_I$ acting on any invariant tensor $Q_i$, is given by 
\begin{equation}
   \label{eq:LQ}
   \Lgen_{K_I}Q_i = - \Tint(K_I) \cdot Q_i  \, ,
\end{equation}
where $\Tint(K_{I})$  is a  $G_S$ singlet in the adjoint bundle.  This means that  $\Tint(K_{I})$ is in the Lie algebra of the commutant group $\mathcal{G}=\Com{\Gst}{\Edd}$. Thus $-\Tint$ defines an ``embedding tensor''~\cite{Samtleben:2008pe,Trigiante:2016mnt}, that is a linear map
\begin{equation}
   \label{eq:Theta}
   \Theta : {\rm span}(\{ K_I \}  )  \to \Lie\mathcal{G} \,.
\end{equation}
The image of this map defines the Lie algebra of the gauge group $\Ggauge$ of the truncated theory and also how it embeds $\Lie\Ggauge=\im \Theta\subseteq\Lie\mathcal{G}$, thus giving $\Ggauge$ as a subgroup of the commutant group 
\begin{equation}
   \label{eq:gauge}
 \Ggauge\subseteq \mathcal{G} =\Com{\Gst}{\Edd} \,.
\end{equation}

For the structures of interest in the present paper, the relevant invariant tensors are the vectors $K_I$ and the adjoint bundle singlets $J_A$ that generate $\mathcal{H}\subset \mathcal{G}$. The former are the generators of the gauge algebra with structure constants  $f_{[IJ]}{}^L$ given by 
\begin{equation}
\label{eq:K-alg}
   \Lgen_{K_I}K_J
   = \Theta_I\cdot K_J
   = \Theta_I{}^A(t_{A})_J{}^L K_L
   := f_{IJ}{}^L K_L\, , 
\end{equation}
where $(t_{A})_J{}^L$ are the representations of the generators of $\Lie\mathcal{G}$ acting on $\mathcal{V}$. For the $J_A$ singlets we have
\begin{align}\label{LieDerivativesForGauging}
   L_{K_I} J_A &= \Theta_I \cdot J_A = [J_{(K_I)}, J_A]
     = \Theta_I{}^B f_{BA}{}^C J_C := p_{IA}{}^B J_B \,,
\end{align}
where $f_{AB}{}^C$ are the  $\mathcal{H}$ structure constants, as in \eqref{commutorJAJB}. For convenience we have also defined the linear combination of the $J_A$ with constant coefficients,
\be\label{eq:JKI}
J_{(K_I)} \,:=\, \Theta_I{}^A J_A\,,
\ee
so that the action of the generalised vector $K_I$ on $J_A$ is represented by the adjoint action of $J_{(K_I)}$. Recall that generally the intrinsic torsion of the $G_S$ structure is captured by the expressions~\eqref{eq:GS-int-tor-1} and~\eqref{eq:GS-int-tor-2}. The condition that one has singlet, constant intrinsic torsion is thus that~\eqref{eq:K-alg} and~\eqref{LieDerivativesForGauging} are satisfied with constant $f_{IJ}{}^K$ and $p_{IA}{}^B$ and in addition that
\begin{equation}
   \label{eq:GS-singlet-int-tor}
   \int_M\kk^2 \tr(J_A(L_WJ_B)) = 0 \, , 
\end{equation}
where the generalised vector $W$ satisfies $c(K_I,K_J,W)=0$. The condition on $W$ implies it transforms non-trivially under $G_S$ and hence, since $J_A$ are singlets, the corresponding intrinsic torsion cannot be a singlet and so must vanish. Alternatively, recall from the discussion in Section~\ref{sec:HVstr_moduli} that \eqref{eq:GS-singlet-int-tor} is equivalent to the vanishing of the moment maps $\mu_\alpha(W)$ given in~\eqref{eq:moment-maps} for all H structures in the family of HV structures defined by the $G_S$ structure. Any one H structure is related to another by the action of $\mathcal{H}$, as in~\eqref{adjoint_action_on_j}. Furthermore it is straightforward to show that $\mu_\alpha(W)$ is invariant under this action. Hence if $\mu_\alpha(W)=0$ with $c(K_I,K_J,W)=0$ at any point in the family then it vanishes for all and \eqref{eq:GS-singlet-int-tor} holds. 

We now show how the singlet intrinsic torsion determines the gauging of the lower-dimensional $\mathcal{N}=2$ theory. The constants $f_{IJ}{}^L$ and $\Theta_I{}^A$, defined in \eqref{eq:K-alg} and \eqref{LieDerivativesForGauging} respectively, can be identified with the embedding tensor components that encode generic gaugings of five-dimensional $\mathcal{N}=2$ supergravity theories, including those involving vector fields that transform in non-adjoint representations of the gauge group, as well as antisymmetric rank-2 tensor fields.\footnote{The embedding tensor formalism  is most commonly used to describe the gauging of maximal and half-maximal supergravity \cite{Samtleben:2008pe,Trigiante:2016mnt}, see however \cite{deWit:2011gk,Louis:2012ux} for its use in an $\mathcal{N}=2$ context.} For simplicity, here we just discuss the case where \eqref{eq:K-alg} define the structure constants of a Lie algebra, implying that it is not necessary to introduce antisymmetric rank-2 tensor fields.
These determine the symmetries of the scalar manifold that are gauged, and hence all matter couplings of the $\mathcal{N}=2$ theory, completely fixing the five-dimensional Lagrangian (see Appendix~\ref{app:sugra_review} for a brief account of $\mathcal{N}=2$ supergravity in five dimensions).
In particular, the vector multiplet scalar covariant derivatives and the gauge field strengths are given by
\be\label{vectorm_scalar_cov_der_main_text}
\mathcal{D} h^I = \diff h^I + g\, f_{JK}{}^I \mathcal{A}^J \,h^K \,,
\ee
\be
\mathcal{F}^I \, =\,  \diff \mathcal{A}^I + \, \tfrac{1}{2}\,g\, f_{JK}{}^I \mathcal{A}^J\wedge \mathcal{A}^K  \, ,
\ee
where $g$ is the gauge coupling constant and $\mathcal{A}^I = \mathcal{A}_\mu^I \dd x^\mu$ are the five-dimensional gauge fields. In order to obtain the hyperscalar covariant derivatives, we need the  Killing vectors on the H structure moduli space \eqref{Hstr_space_general} that generate the gauged isometries. These can be constructed from \eqref{eq:JKI} using the standard formalism of coset spaces, see e.g.~\cite{Castellani:1983tb}.
 Given the left-action of a generator $J_{(K_I)}$ on the coset representative $L \in \mathcal{H}$, the corresponding Killing vector $k_I$ on $\mathcal{M}_{\rm H}$ is determined by the equation
\be\label{formula_for_k^X}
L^{-1} J_{(K_I)}\, L \,\sim\,   g \,  \iota_{k_I} \,(L^{-1}\diff L)\,,
\ee
where the symbol $\sim$ means that the equality holds up to an element
of the algebra one is modding out by, which in the present case is
$\SUH\times \Com{\SUH}{\mathcal{H}}$.\footnote{A similar
  construction could be made for the Killing vectors that gauge
  isometries in the V structure moduli space, starting from the
  sections of the adjoint bundle that generate $\mathcal{M}_{\rm V}$
  mentioned in Footnote~\ref{foot:KtimesK*}. However, this will not be
  needed for our purposes.} Writing $k_I = k_I^X \frac{\partial}{\partial q^X}$, where  $q^X$  denote the coordinates on $\mathcal{M}_{\rm H}$,
the hyperscalar covariant derivatives then read
\be\label{hyperscalar_cov_der_main_text}
\mathcal{D} q^X \,=\, \diff  q^X + g\, \mathcal{A}^I k^X_I \,.
\ee

From the Killing vectors $k_I$ we can then compute the triholomorphic
Killing prepotentials $P_I^\alpha$, $\alpha=1,2,3$, that determine the
fermionic shifts and the scalar potential of the $\mathcal{N}=2$
supergravity theory, see Appendix~\ref{app:sugra_review} for the
relevant formulae. These Killing prepotentials are moment maps of the isometries being gauged, and as such can be nicely computed  from the generalised geometry formalism. Recalling the definition of the moment map $\mu$ in \eqref{eq:moment-maps}, they are given by
\begin{equation}\label{KillingPrep_from_mommaps}
  \begin{aligned}
 g\, P_I^\alpha 
      \,&=\,  \tfrac 18    \, \epsilon^{\alpha\beta\gamma}
       \int_M \kk^2 \tr \left( J_\beta (L_{K_I} J_\gamma ) \right)
       \bigg/
       \int_M \kk^2 \\
      \,&=\,   \tfrac18 \, \epsilon^{\alpha\beta\gamma}
       \tr \left( J_\beta (L_{K_I} J_\gamma ) \right)\,.
     \end{aligned}
\end{equation}
In this formula, recall
that the $J_\alpha$ are the dressed triplet, hence the resulting moment
maps are function of the H structure moduli. In the second line, we
have used the fact that the singlet torsion components $\tr \left( J_\beta
  (L_{K_I}J_\gamma )\right)$ are constant on $M$ and hence the
integrals over $\kk^2$ cancel.

\subsection{The truncation ansatz}\label{sec:fielddec}

Our conventions for eleven-dimensional supergravity are as in \cite{Coimbra:2011ky}. 
The eleven-dimensional bosonic action is (we denote by a hat the 11d quantities)
\be
\hat{S} = \frac{1}{2} \int \left(\hat{R}\,\hat{*}\,1 - \tfrac{1}{2} \hat{F}\wedge * \hat{F} - \tfrac{1}{6} \hat{A}\wedge \hat{F}\wedge \hat{F} \right)\,,
\ee
where $\hat{F}=\diff \hat{A}$ and $\hat{A}$ is the three-form potential.
The equations of motion are
\begin{equation}
\begin{aligned}
\hat{R}_{\hat\mu\hat\nu} -\tfrac{1}{12}
\left(\hat{F}_{\hat\mu\hat\rho_1\hat\rho_2\hat\rho_3}\hat{F}_{\hat\nu}{}^{\hat\rho_1\hat\rho_2\hat\rho_3}
  - \tfrac{1}{12} \,\hat{g}_{\hat\mu\hat\nu} \hat{F}^2 \right) &=0\,, \\[1mm]
\diff \,\hat{*}\, \hat{F} + \tfrac{1}{2} \hat{F}\wedge \hat{F} &= 0\,.
\end{aligned}
\end{equation}
The six-form potential $\hat{\tilde{A}}$ dual to the three-form $\hat{A}$ may be introduced via the first-order relation
\be\label{dualityAtildeA}
\hat *\, \diff \hat A + \tfrac{1}{2}\,\hat A \wedge\diff \hat A \,=\, \diff \hat{\tilde A}\,,
\ee
whose exterior derivative gives the Maxwell equation.

As first step of the truncation procedure, we arrange the eleven-dimensional bosonic fields  into generalised tensors transforming in representations of $\GL(5,\bbR)\times E_{6(6)}$, where 
$\GL(5,\bbR)$  gives the tensorial structure of the fields in the five-dimensional theory obtained after reduction. Then we expand each  $E_{6(6)}$ representation in terms of the $G_S$ invariant tensors transforming in the same representation.
We separate the eleven-dimensional coordinates in coordinates $x^\mu$, $\mu=0,\ldots,4$, on the external spacetime $X$, and $z^m$, $m=1,\ldots,6$, on the internal manifold $M$. 

The bosonic fields of 
eleven-dimensional supergravity are decomposed as 
\begin{align}
\label{11d_metric_general}
\hat{g}  &= \rme^{2\Delta}\, g_{\mu\nu} \,\dd{x}^\mu\dd{x}^\nu + g_{mn} Dz^m Dz^n\ ,\nn \\[1mm]
\hat{A} &= \tfrac{1}{3!} A_{mnp} D{z}^{mnp} + \tfrac{1}{2}{A}_{\mu mn}
\dd{x}^\mu\wedge D{z}^{mn} + \tfrac{1}{2}\,{\bar{A}}_{\mu\nu
  m}\dd{x}^{\mu\nu} \wedge Dz^m +
\tfrac{1}{3!}\,{\bar{A}}_{\mu\nu\rho}\,\dd{x}^{\mu\nu\rho}  \, , \nn \\[1mm]
\hat{\tilde{A}} &= \tfrac{1}{6!} \tilde{A}_{m_1\ldots m_6}
Dz^{m_1\ldots m_6} + \tfrac{1}{5!} {\tilde A}_{\mu m_1\ldots m_5}
\dd{x}^\mu \!\wedge\! Dz^{m_1\ldots m_5}
+  \tfrac{1}{2\cdot 4!} \bar{{\tilde A}}_{\mu\nu m_1\ldots m_4} \dd{x}^{\mu\nu} \!\wedge\! Dz^{m_1\ldots m_4}\nn\\[1mm] & \quad  +\, \ldots  , 
\end{align}
where   $Dz^m \,=\, \dd{z}^m - h_\mu{}^m \dd{x}^\mu$, 
and all tensor field components may depend both on $x^\mu$ and $z^m$, except for the external metric, for which we assume a dependence on the external coordinates only, $g_{\mu\nu}= g_{\mu\nu}(x)$.

 The barred fields need to be redefined.  In  Appendix~\ref{sec:gauge_transf} we provide a justification for these redefinitions by studying the gauge transformations of the metric and three-form potential.
 For the three-form components we introduce the new fields $A_{\mu\nu m}$, $A_{\mu\nu\rho}$ via
 \be\label{redef_three_form}
 \bar{A}_{\mu\nu m} = A_{\mu\nu m} -\, h_{[\mu}{}^n A_{\nu] nm}\,,\qquad \bar{A}_{\mu\nu\rho} = A_{\mu\nu\rho} 
 +  h_{[\mu}{}^n h_\nu{}^p  A_{\rho] np}\,.
\ee
Similar redefinitions apply to the six-form components with at least two external indices, 
however we will not discuss them in detail here.

The supergravity fields having all components on the internal manifold $M$ arrange into the inverse generalised metric
\be
\label{gmscalars}
G^{MN} \ \leftrightarrow\ \{\Delta,\,g_{mn}, \, A_{mnp}, \, \tilde A_{m_1\ldots m_6}  \}\ ,
\ee
in the following way\footnote{This expression follows straightforwardly from the elements of the conformal split frame given in \cite{Coimbra:2011ky}.}
\begin{equation}
\label{gen_metr_general}
\begin{aligned}
 ( G^{-1})^{mn}  & = \rme^{2 \Delta} g^{mn} \\[1mm]
(G^{-1})^m{}_{n_1 n_2}  &  = \rme^{2 \Delta} g^{mp} A_{p n_1 n_2}  \\[1mm]
(G^{-1})^m{}_{n_1 \ldots  n_5}  & =   \rme^{2 \Delta} g^{mp}  ( A_{p [n_1 n_2}  A_{n_3 n_4 n_5]} + \tilde{A}_{p n_1 \ldots n_5} )  \\[1mm]
(G^{-1})_{m_1 m_2 \, n_1  n_2}  & =  \rme^{2 \Delta} (  g_{m_1 m_2, n_1 n_2} +  g^{pq}   A_{p m_1 m_2}  A_{q n_1 n_2]} )   \\[1mm]
(G^{-1})_{m_1 m_2 \,  n_1  \ldots n_5}  & =  \rme^{2 \Delta} [ g_{m_1
  m_2 , [n_1 n_2} A_{n_3 n_4 n_5]}
\\ & \qquad
{} + g^{pq}  ( A_{p m_1 m_2}  ( A_{q [n_1 n_2}  A_{n_3 n_4 n_5]} + \tilde{A}_{q n_1 \ldots n_5} ) ]  \\[1mm]
(G^{-1})_{m_1 \ldots m_5 \,  n_1  \ldots n_5}  & = 
\rme^{2 \Delta} [ g_{m_1 \ldots m_5 ,\,  n_1  \ldots n_5}
\\ & \qquad
{} + g^{pq}  ( A_{p [m_1 m_2} A_{m_3 m_4 m_5]} + \tilde{A}_{p m_1 \ldots m_5} ) 
( A_{q [n_1 n_2}  A_{n_3 n_4 n_5]} + \tilde{A}_{q n_1 \ldots n_5} ) ] \,,
\end{aligned}
\end{equation}
where $g_{m_1m_2,\,n_1n_2}= g_{m_1[n_1} g_{|m_2|n_2]}$, and similarly for $g_{m_1 \ldots m_5 ,\,  n_1  \ldots n_5}$.
 Since the generalised metric is a scalar on the external spacetime, after imposing our truncation ansatz it will provide the scalar fields of the reduced five-dimensional theory.
 
The density $\kk$ introduced in Section~\ref{sec:HVstr} when defining the HV structure is related to the determinant of the generalised metric and is an $\Ex{6}$ invariant. For eleven-dimensional metrics of the form \eqref{11d_metric_general}, this is given by  \cite{Coimbra:2011ky,Ashmore:2015joa}
\be\label{generalised_volume}
\kk^2 = \rme^{3\Delta}\sqrt{\det g_{mn}}\,.
\ee

 The tensors with one external leg arrange into a generalised vector $\mathcal{A}_\mu$ on $M$, with components
\be
\label{Avec}
\mathcal{A}_\mu{}^M = \{ h_\mu{}^m ,\, A_{\mu mn} , \tilde{A}_{\mu m_1\ldots m_5}\,     \}\, ,
\ee
and will provide the gauge potentials of the reduced theory. The tensors with two antisymmetrised external indices define a weighted  dual vector $\mathcal{B}_{\mu\nu}$ on $M$, which is a section of $\det T^*M \otimes E^*$, with components
\be
\label{Btens}
\mathcal{B}_{\mu\nu\,M} =\{  A_{\mu\nu m}, \,   \tilde{A}_{\mu\nu m_1\ldots m_4}  ,\, \tilde g_{\mu\nu m_1\ldots m_6,n} \} \,,
\ee
and will give the two-form fields of the reduced theory. 
The last term in \eqref{Btens} is  related to the dual graviton and we will not discuss it further here.
The tensors with three antisymmetrised external indices arrange into the generalised tensor
\be
\mathcal{C}_{\mu\nu\rho}{}^{\hat\alpha} = \{ A_{\mu\nu \rho}, \,\tilde{A}_{\mu\nu \rho m_1m_2 m_3},\, \tilde g_{\mu\nu\rho m_1\ldots m_5,n}\}\,,
\ee
which is a section of (a sub-bundle of) the weighted adjoint bundle $\det T^*M \otimes \adj F$, whose components are labeled by $\hat{\alpha}=1,\ldots,57$. See e.g.~\cite{Riccioni:2007ni,deWit:2008ta} for more details on this tensor hierarchy.

As discussed in \cite{Cassani:2019vcl}, the bosonic part of the
truncation ansatz is obtained by imposing that the generalised tensors
above are expanded in singlets of the $G_S$ structure. The generalised
metric is obtained by constructing the $K$ and $J_\alpha$
parameterising a family of HV structures as detailed in
Section~\ref{sec:HVstr_moduli}, and plugging these generalised tensors
in the formula \eqref{USp6_igm}. The resulting generalised metric
depends on the H and V structure moduli; when given a
dependence on the external coordinates $x^\mu$, these are then
identified with the hyperscalar and vector multiplet scalar fields of
the truncated $\mathcal{N}=2$ theory, respectively. Thus we have
\begin{equation}
  \left.\begin{aligned}
    K &= h^I(x) K_I  \\
    J_\alpha &= L(x) j_\alpha L(x)^{-1} 
  \end{aligned}\right\} \qquad 
  \text{giving $G^{MN}(x)$ from \eqref{USp6_igm}} \,,
\end{equation}
where $L$ is the representative of the coset $\mathcal{M}_{\rm H}$. Comparing the expression for the generalised metric with its general form \eqref{gen_metr_general}, we obtain the truncation ansatz for $\Delta$, $g_{mn}$, $A_{mnp}$ (as well as $\tilde A_{m_1\ldots m_6}$, whenever it is needed). Note that $\kk^2$ given in~\eqref{generalised_volume} is independent of the scalar fields $h^I(x)$ and $L(x)$, so it can be evaluated using any chosen representative of the family of HV structures defined by the $G_S$ structure. 

The gauge potentials $\mathcal{A}_\mu{}^I(x)$ on the external space-time are defined by taking
\be\label{ansatz_Avec}
\mathcal{A}_\mu \,=\, \mathcal{A}_\mu{}^I(x)\, K_I  \quad \in\, \Gs{T^*X}\otimes {\rm span}(\{K_I\})
\ee
where ${\rm span}(\{K_I\})\subset\Gs{E}$ is the  vector space spanned by the set of $G_S$ singlets $K_I$, $I=0,1,\ldots,n_{\rm V}$. 
Similarly the two-form fields are given by
\be
\label{ansatz-Bform}
\mathcal{B}_{\mu\nu} \,=\, \mathcal{B}_{\mu\nu\,I}(x)\, K^I_\flat   \quad \in\, \Gs{\Lambda^2T^*X}\otimes {\rm span}(\{K_{\flat}^I\})\,,
\ee
where ${\rm span}(\{K_\flat^I\})\subset \Gamma(\det T^*M \otimes E^*)$
is the vector space spanned by the weighted dual basis vectors
$K_\flat^I$, the latter being defined by $K_\flat^I(K_J)=3\kk^2\,\delta^I{}_J$. We also have
\be
\label{ansatz-Cform}
\mathcal{C}_{\mu\nu\rho} \,=\, \mathcal{C}_{\mu\nu\rho}{}^A(x)\, J^\flat_A   \quad \in\, \Gs{\Lambda^3T^*X}\otimes {\rm span}(\{J^\flat_A\})\,,
\ee
where ${\rm span}(\{J^\flat_A\})\subset \Gamma(\det T^*M \otimes
{\adj}(F))$ is spanned by the $G_S$ singlets in the weighted adjoint
bundle, here denoted by $J^\flat_A$ and given by $J^\flat_A=\kk^2 J_A$. In Appendix~\ref{sec:gauge_transf} we show that these expressions, together with the field redefinitions~\eqref{redef_three_form}, lead to the correct five-dimensional covariant objects, consistent with the expected gauge transformations.

\section{$\mathcal{N}=2$ truncations on Maldacena--Nu\~nez geometries}\label{sec:MN1section}

We now apply the above  formalism  to consistent truncations of eleven-dimensional supergravity based on generalised structures arising from M5-branes wrapping a Riemann surface.

We start with the $\mathcal{N}=2$ AdS$_5\times_{\rm w} M$ solution of Maldacena and Nu\~nez \cite{Maldacena:2000mw} and show that the manifold $M$ admits 
a generalised $\U(1)$ structure with singlet intrinsic torsion, and
therefore can be used to construct a consistent truncation. As we have
stressed above, once we identify the singlet $K_I$ and $J_A$ tensors defining
the structure it is straightforward to read off the form of the
$\mathcal{N}=2$ supergravity.  

We already observed in~\cite{Cassani:2019vcl} that this process yields
$\mathcal{N}=2$ supergravity with  one hypermultiplet and four vector multiplets. Here we give the details of the construction and derive the gauging, which defines an $\SO(3) \times \U(1)\times \mathbb{R}$ gauge group. Our truncation includes as a subtruncation the reduction to $\mathcal{N}=2$ supergravity with one vector multiplet, one hypermultiplet and $\U(1)\times \mathbb{R}$ gauging recently obtained in \cite{Faedo:2019cvr}.

\subsection{The MN1 solution}
\label{MNsol}

We are interested  in warped AdS$_5$ solutions to eleven-dimensional supergravity that describe the near-horizon region of M5-branes wrapping  supersymmetric cycles in a Calabi--Yau geometry. 
The amount of supersymmetry  of the solutions depends on how the cycle is embedded in the ambient geometry.  This corresponds to a topological twist of the 
 world-volume  $(0,2)$ theory on the M5-branes. 
The simplest examples are the solutions found by Maldacena and Nu\~nez~\cite{Maldacena:2000mw} describing the near-horizon geometry of M5-branes wrapped on 
a Riemann surface $\mathbb{\Sigma}$ of negative constant curvature.
The topological twist of the $(0,2)$ world-volume theory  is realised  by identifying the spin connection on $\mathbb{\Sigma}$ with a $\U(1)$ connection in the $SO(5)$ R-symmetry group of the M5-brane theory.  
The theory preserves  $\mathcal{N}=2$ or $\mathcal{N}=1$ superconformal symmetry in four dimensions, depending on how the $\U(1)$ is chosen inside $\SO(5)$.  
The corresponding supergravity solutions are  warped products of 
AdS$_5$ times a six-dimensional manifold, $M$, which is the fibration of a deformed $S^4$ over $\mathbb{\Sigma}$. The $SO(5)$ is realised via the action of the isometry group of the round $S^4$. The structure of the fibration reflects the twist of the world-volume theory and determines the amount of supersymmetry of the solutions, which in five-dimensional language is either
 $\mathcal{N}=4$ or  $\mathcal{N}=2$, respectively. 

In this paper we focus on  the $\mathcal{N}=2$ solution, which we call the  ``MN1 solution''  in the following.
The  eleven-dimensional metric is\footnote{We present the solution in a form similar to the one given in \cite[Sect.\:5]{Gauntlett:2004zh}. The precise dictionary with this reference is: $\alpha=\zeta$, $\nu=-\phi$, $\psi_{\rm GMSW}=\psi$, $\rme^{2\lambda}=\rme^{2\Delta}$, $m^{-1}=\ell_{\rm AdS_5}= \frac{3}{2}R$, where the variables on the left-hand side are those of \cite{Gauntlett:2004zh} while the variables on the 
right-hand side are those used here. The length scale $R$ that appears in our expressions is equal to the radius of $S^4$ in the related ${\rm AdS}_7\times S^4$ Freund-Rubin solution of eleven-dimensional supergravity. The four-form $\hat{F}$ in \eqref{F4MN1sol} has an overall opposite sign with respect to the one of \cite{Gauntlett:2004zh}, $\hat{F}= - F_{\rm GMSW}$; this sign does not affect the equations of motion, it just modifies the projection condition satisfied by the supersymmetry spinor parameter.}  
\be
\label{Msol}
\hat g \,=\, \rme^{2 \Delta } \,g_{\rm AdS_5}
    +  g_6  \, , 
\ee
where $g_{\rm AdS_5}$  is the Anti de Sitter metric with radius $\ell =\frac{3}{2}R$, $R$ being the length scale of the internal space $M$. The metric on $M$ takes the form
\be
\label{6dMN1}
g_6 \,=\,    R^2 \,\frac{3^{1/3}}{2^{4/3}}  \left(3 + \cos^2 \zeta\right)^{1/3} \left[ g_{\mathbb{\Sigma}} + \dd \zeta^2 +   \frac{ \sin^2 \zeta}{ 3 + \cos^2 \zeta }\left( \sigma_1^2 + \sigma_2^2 +
 ( \sigma_3  +\upsilon )^2 \right) \right] \, .
\ee
Here, $g_{\mathbb\Sigma}$ is the uniform metric on (a quotient of) the hyperbolic plane $\mathbb{\Sigma} =  H^2$, 
with Ricci scalar curvature $\mathcal{R}_{\mathbb{\Sigma}}=-2$, 
while $\upsilon$ is the spin connection on $\mathbb{\Sigma}$  satisfying
\be
\dd \upsilon = -\vol_{\mathbb{\Sigma}}\,,
\ee
with $\vol_{\mathbb{\Sigma}}$ the volume form on $\mathbb{\Sigma}$.\footnote{Choosing local coordinates $x,y$ on the hyperbolic plane, one can write $g_{\mathbb\Sigma} = \frac{ \dd x^2 + \dd y^2}{ y^2}$, $\vol_{\mathbb{\Sigma}} = \frac{\dd x\wedge \dd y}{y^2}$, and $\upsilon = - \frac{\diff x}{y} $.\label{foot:param_Riemann_surface}} The deformed $S^4$ is described as a foliation of a round $S^3$ over an interval, with the interval coordinate being $\zeta\in [0,\pi]$, while $\sigma_\alpha$, $\alpha=1,2,3$, are the standard $\SU(2)_{\rm left}$-invariant forms on $S^3$, expressed in terms of Euler angles $\{\theta,\phi,\psi\}$. 
Their explicit expression can be found in Appendix~\ref{app:param_and_frames_S4}, together with more details on the parameterisation of $S^4$. 
 
The warp factor is 
\be
\rme^{ 2\Delta} \,=\,  \left(\frac{2}{3}\right)^{2/3}  (3 + \cos^2 \zeta)^{1/3}  \, ,
\ee
while the four-form reads
\be
\begin{aligned}\label{F4MN1sol}
\hat F \,&=\,    \frac{R^3}{4} \left[  \,\frac{ 15 + \cos^2 \zeta}{ (3 + \cos^2 \zeta)^2} \sin^3\zeta\, \dd \zeta \wedge \sigma_1\wedge\sigma_2\wedge(\sigma_3 +\upsilon) \right. \\[1mm]
&\qquad \qquad \left.  
+\,\sin \zeta \left(- \dd\zeta  \wedge\sigma_3  + \frac{\sin(2\zeta)}{ 3 + \cos^2 \zeta} \, \sigma_1\wedge\sigma_2 \right) \wedge \vol_{\mathbb\Sigma} \right] \, .
\end{aligned} 
\ee
Note that the invariant volume form~\eqref{generalised_volume} is given by
\begin{equation}
\label{MN1-kappa}
  \kk^2 = R^2 \vol_{\mathbb{\Sigma}} \wedge \vol_4 \, , 
\end{equation}
where $\vol_4$ is the volume form of the round $S^4$ of radius $R$. 

The solution has $\SU(2)_{\rm left}\times \U(1)_{\rm right}$ symmetry, which embeds in the $ \SO(5)$ isometry group of a round $S^4$ as
\be
\label{decompose_SO5}
 \SO(5)\, \supset\, \SO(4) \,\simeq\, \SU(2)_{\rm left} \times \SU(2)_{\rm right} \,\supset\, \SU(2)_{\rm left}\times \U(1)_{\rm right} \,.
\ee
  This symmetry is manifest as the solution is given   in terms of the $\sigma_\alpha$. The globally-defined combination $(\sigma_3 + \upsilon)$ describes a fibration of $S^4$ over $\mathbb{\Sigma}$, such that the $\U(1)_{\rm right}$ action on $S^4$ is used to cancel the $\U(1)$ holonomy of $\mathbb{\Sigma}$.

The  $U(1)_{\rm right}$  factor provides the R-symmetry of the holographically dual $\mathcal{N}=1$ SCFT, while $\SU(2)_{\rm left}$ corresponds to a flavour symmetry. The dual $\mathcal{N}=1$ SCFT has been described in \cite{Benini:2009mz}.

\subsection{Generalised $\U(1)$ structure of the MN1 solution}
\label{sec:U1str}

The  solution reviewed above admits a  generalised $U(1)_S$ structure, which will be the basis for constructing our consistent truncation. In order to characterise it  we proceed in two steps. The first is purely group theoretical: it 
consists in embedding the relevant $U(1)_S$ in $E_{6(6)}$, computing its commutant and the corresponding decompositions of the generalised tangent and 
adjoint bundles. To this end, it is convenient to decompose $E_{6(6)}$ according to its maximal compact subgroup $\USp(8)/\bbZ_2$. Since the 
 $\usp(8)$ algebra can be given in terms of $\Cliff(6)$ gamma matrices (see  Appendix~\ref{Usp8SL6}), this reduces the problem to gamma matrix algebra.  The details of the derivation can be found in Appendix~\ref{app:U1MN}; here we will just give the results. 
Once the relevant $U(1)_S$ singlets are identified, the second step is to express them in terms of the geometry of the six-dimensional manifold $M$.  

The generalised $U(1)_S$  structure of the MN1 solution is the diagonal of the ordinary geometrical $\U(1)\simeq\SO(2)\subset\GL(2,\bbR)$ structure on the Riemann surface and a $\U(1)$  factor in the $\SO(5)\subset\SL(5,\bbR)\simeq E_{4(4)}$ generalised structure for the generalised tangent space of the four-sphere. In terms of the isometry group decomposition~\eqref{decompose_SO5} this can be identified with $U(1)_{\rm right}$. If we  denote by 1 to 4 the directions in $M$ along  $S^4$  and by 5,6 those along  $\mathbb{\Sigma}$, the generator of  $\U(1)_S$  can be written as a  $\usp(8)$ element as
\be 
\label{u1sgenb}
\mathfrak{u}(1)_S = \ii\,  \hat{\Gamma}_{56} -\frac{\ii}{2} (\hat{\Gamma}_{12} - \hat{\Gamma}_{34})  \, ,
\ee
where $\hat{\Gamma}_m$ are six-dimensional gamma matrices. The first term corresponds to the $\U(1)$ holonomy of $\mathbb{\Sigma}$ while the second one is the $\U(1)_{\rm right}$ in $\SO(5)$. By computing the commutators of \eqref{u1sgenb} in 
$\USp(8)$  we find that  the $\U(1)_S$ structure embeds in  $\USp(8)$ as\footnote{Here and below we give expressions ignoring subtleties involving the centres of each group; thus for instance we will not distinguish between embeddings in $\USp(8)$ and $\USp(8)/\bbZ_2$.}
\be
\label{sec:U1Usp8}
\USp(8) \supset \SU(2) \times \SUH \times  \U(1) \times \U(1)_S \, , 
\ee
where as above we distinguish the factor $\SUH$ that gives the R-symmetry of the five-dimensional supergravity theory. Under this splitting, the spinorial representation of $\USp(8)$ decomposes as
\be
{\bf 8} = ({\bf 1}, {\bf 2})_0 \oplus  ({\bf 2}, {\bf 1})_1 \oplus ({\bf 2}, {\bf 1})_{-1} \oplus ({\bf 1}, {\bf 1})_2 \oplus ({\bf 1}, {\bf 1})_{-2} \, ,
\ee
where the elements in the brackets denote the $\SU(2)\times\SUH$ representations and the subscript gives the $U(1)_S$ charge. 
We then see that there are only two spinors that are singlets under $\U(1)_S$ and that transform as a doublet of $\SUH$ as required by $\mathcal{N}=2$ supersymmetry.

The embedding of the $U(1)_S$ structure in the full $E_{6(6)}$  is obtained in a similar way (see Appendix~\ref{app:U1str} for details)
\be
\label{E6dec}
E_{6(6)} \supset \Com{U(1)_S}{E_{6(6)}}
= \mathbb{R}^+ \times \Spin(3,1) \times \SU(2,1) \times  U(1)_S  \, , 
\ee
where  $\Com{U(1)_S}{E_{6(6)}}$ is the commutant of $U(1)_S$ in $E_{6(6)}$. We can now determine how many generalised vectors and adjoint elements are $U(1)_S$ singlets. Under \eqref{E6dec} the ${\bf 27}$
decomposes as
\be
\label{27dec}
\begin{aligned}
{\bf 27} \,&=\,   ({\bf 1},{\bf 1} )_{(0,8)} \oplus   ({\bf 4},{\bf 1} )_{(0,-4)} \oplus  ({\bf 2},{\bf 1} )_{(3,-2)}  \oplus ({\bf  \bar{2}}, {\bf 1} )_{(-3,-2)}  \\ 
\,&\,   \oplus   ({\bf 1},{\bf 3} )_{(2,-4)} \oplus   ({\bf 1},{\bf \bar 3} )_{(-2,-4)}    \oplus   ({\bf \bar 2},{\bf 3} )_{(1,2)} \oplus   ({\bf 2},{\bf \bar 3} )_{(-1, 2)}\,,
 \end{aligned} 
\ee
where the first subscript denotes the $U(1)_S$ charge and the second one the $\mathbb{R}^+$ charge. We see that there are five singlets $K_I$,
$I=0,1, \ldots, 4$, where
\be
K_0 \in  ({\bf 1},{\bf 1} )_{(0,8)}
\ee
is only charged under the $\mathbb{R}^+$, while 
\be
 \{ K_1, K_2, K_3, K_4 \} \in  ({\bf 4},{\bf 1} )_{(0,-4)}
 \ee
form a vector of $ SO(3,1)$. 

The singlets in the ${\bf 78}$ adjoint representation are the generators of the commutant $C_{E_{6(6)}}(U(1)_S)$.  However only the generators of the $\SU(2,1)$ subgroup are relevant for the structure.
Indeed,   \eqref{27dec}  shows that the generators of $\mathbb{R}^+ \times \SO(3,1)$  do not  leave the singlet vectors invariant, and therefore, as discussed in Section \ref{sec:HVstr}, do not contribute to the truncation. As shown in  \eqref{app:so31genprod} and \eqref{app:Rpgenprod}, they can be  obtained as products $K_I \times_{\adj} K^*_J$.
 We denote by $J_A$, $A=1,\ldots,8$, the elements of the adjoint bundle generating $\su_{2,1}$. Four of them are in the ${\bf 36}$ of $\USp(8)$ and generate the compact subalgebra $\su_2\oplus \uni{1}$, and four more are  in the ${\bf 42}$ of   $\USp(8)$  and generate the rest of $\su_{2,1}$.

The  $U(1)_S$ structure is then defined by 
\be\label{eq:singlets_MN1}
\{ K_I , J_A \} \, ,\qquad  I=0, \ldots, 4\,,  \quad A= 1, \ldots, 8\,.
\ee 

The derivation of the  explicit expressions for these generalised tensors relies on the way the solution  of~\cite{Maldacena:2000mw} is constructed by 
deforming the AdS$_7 \times S^4$ background dual to flat M5-branes so as to describe their backreaction when wrapping a Riemann surface $\mathbb{\Sigma}$. The world-volume theory on the wrapped M5-branes is made supersymmetric by
a topological twist, where the spin connection on the Riemann surface is cancelled by switching on a background gauge field for a $U(1)$ subgroup of the $SO(5)$ R-symmetry. On the dual background the topological twist implies that $M$ is an $S^4$ fibration over $\mathbb{\Sigma}$
 \begin{equation}
\label{eq:S4-fibration}
   \begin{tikzcd} 
      S^4 \arrow[r,"i"] & M \arrow[d,"\pi"] \\
      & \mathbb{\Sigma}
   \end{tikzcd} 
\end{equation}

The generalised tangent bundle for $S^4$ is given by 
\be
 \label{sec:S4sets}
 E_4 \,\simeq\,  TS^4 \oplus \Lambda^2T^*S^4\,,
\ee
and transforms under $\SL(5,\bbR)\simeq\Ex{4}$. It is generalised parallelisable, meaning it  admits a globally defined frame \cite{Lee:2014mla}. 
The idea is then to consider first  the direct product $\mathbb{\Sigma}\times S^4$, express the $E_{6(6)}$  generalised tensors on this manifold in terms of the frame on $\mathbb{\Sigma}$ and the parallelisation on $S^4$, and then implement the twist of $S^4$ over $\mathbb{\Sigma}$ so as to make globally well-defined objects. 
In the decomposition
\be
\Ex{6} \supset \GL(2,\bbR)\times\SL(5,\bbR) \, ,
\ee
where  $\GL(2,\bbR)$ is the structure group of the conventional tangent bundle on $\mathbb{\Sigma}$ and $\SL(5,\bbR)\simeq\Ex{4}$ is the structure group of the generalised tangent bundle on $S^4$,  the $E_{6(6)}$ generalised tangent bundle on $\mathbb{\Sigma}\times S^4$ decomposes as
\be
\begin{aligned}
\label{eq:SL5-vecdecomp}
 E &\,\simeq\, T\mathbb{\Sigma} \oplus (T^*\mathbb{\Sigma} \otimes N_4)
      \oplus (\Lambda^2 T^*\mathbb{\Sigma} \otimes N_4') \oplus E_4\,, \\
\end{aligned}
\ee
and the adjoint bundle as
\be
\label{eq:SL5-addecomp}
\begin{aligned}
{\adj} F   \, & \simeq\,  {\adj} F_4 \oplus  (T \Sigma \otimes T^\ast \Sigma)  \oplus  (  T^\ast \Sigma \otimes E_4) \\ & \qquad \qquad 
  \oplus ( \Lambda^2  T^\ast \Sigma \otimes N_4) 
  \oplus (  T \Sigma \otimes E^\ast_4) \oplus ( \Lambda^2  T \Sigma \otimes N^\ast_4)  \, .
\end{aligned}
\ee
In the expressions above $E_4$ is  the generalised tangent bundle on $S^4$ introduced in \eqref{sec:S4sets}, $\adj F_4 $ is the adjoint bundle on $S^4$,
\be
\adj F_4  \simeq \mathbb{R} \oplus  (T S^4 \otimes  T^\ast S^4)  \oplus \Lambda^3  T^\ast S^4 \oplus  \Lambda^3 T S^4  \,,
\ee
and  $N_4$ and   $N_4' $ are the following bundles on $S^4$, 
\begin{equation}
\begin{aligned}
   N_4 &\,\simeq\, T^*S^4 \oplus \Lambda^4 T^*S^4\,, \\
   N_4' &\,\simeq\, \mathbb{R} \oplus \Lambda^3 T^*S^4\, . 
\end{aligned}
\end{equation}

The  bundles $E_4$, $N_4$ and   $N_4'$ admit the globally defined  generalised  frames 
\be
E_{ij}\in\, \Gamma(E_4)\,,\qquad E_i \in\, \Gamma(N_4)\,,\qquad E'_i \in\, \Gamma(N'_4)\,, \qquad i,j=1,\ldots,5\,,
\ee
see Appendix~\ref{app:param_and_frames_S4} for their expression in a coordinate basis and note that they include a contribution from the three-form gauge potential $A_{S^4}$ of the flux on the $S^4$. Geometrically this defines a generalised identity structure on $S^4$. Given the way $\U(1)_S$ is embedded in $\USp(8)$, we will find it useful to also introduce the following linear combinations of the generalised frame elements $E_{ij}$ on $S^4$,
 \begin{align}
 \label{sec:Xi_tildeXi}
&\Xi_1 = E_{13}+E_{24}\,,\quad \Xi_2 = E_{14}-E_{23} \,,\quad \Xi_3 = E_{12}-E_{34}\,,\nn\\
&\ti \Xi_1 = E_{13}-E_{24}\,,\quad \ti \Xi_2 = E_{14}+E_{23} \,,\quad \ti \Xi_3 = E_{12}+E_{34}\, . 
\end{align}
Since their restriction to $TM$ corresponds to the Killing vectors generating the $\SU(2)_{\rm left}\times\SU(2)_{\rm right}\simeq\SO(4)\subset\SO(5)$ isometries of $S^4$ (again see Appendix~\ref{app:param_and_frames_S4} for their explicit expression), $\Xi_\alpha$ and $\ti{\Xi}_\alpha$, $\alpha=1,2,3$, may be seen as generalised Killing vectors generating the corresponding generalised isometries.

As for the Riemann surface $\mathbb{\Sigma}$, it can be (a quotient of) the hyperbolic plane $H^2$ as in the MN1 solution reviewed in Section~\ref{MNsol}, but we can also take a torus $T^2$, or a sphere $S^2$. We introduce orthonormal co-frame one-forms  $e_1$, $e_2$ on $\mathbb{\Sigma}$, such that  the constant curvature metric and the compatible volume form on $\mathbb{\Sigma}$  are given by 
\be
\label{gSigmavolSigma}
g_{\mathbb{\Sigma}} = (e_1)^2+(e_2)^2\,,\qquad \vol_\mathbb{\Sigma} = e_1\wedge e_2 \, . 
\ee
The metric is normalised so that the Ricci scalar curvature is $\mathcal{R}_{\mathbb{\Sigma}}=2\kappa$, where $\kappa =+1$ for $S^2$, $\kappa =0$ for $T^2$ and $\kappa = -1$ for $H^2$ (and quotients thereof).
We also define the $U(1)$ spin connection, $\upsilon$,  on  $\mathbb{\Sigma}$  as 
\begin{equation}
   \label{d_frame_Sigma}
\dd (e_1+\ii\,e_2) \,=\, \ii\, \upsilon \wedge (e_1+\ii\,e_2)\,, \qquad \dd \upsilon\, =\, \kappa \,\vol_\mathbb{\Sigma}\,.
\end{equation}

The decompositions \eqref{eq:SL5-vecdecomp} and \eqref{eq:SL5-addecomp} allow us to express the $U(1)_S$ invariant generalised tensors in terms of tensors on $\mathbb{\Sigma}$ and the $S^4$ generalised frames introduced above. We provide the derivation in Appendix~\ref{app:U1MN} and here just present the resulting expressions. 
Let us first focus on the singlet generalised vectors $K_I$. These can be written as
 \be\label{KI_before_twist}
K_0\, = \,\rme^\Upsilon \cdot (R^2\,\text{vol}_{\mathbb{\Sigma}}\wedge \, E'_5) \,,  \qquad 
K_{1,2,3} \,  =\, \rme^\Upsilon \cdot  \ti{\Xi}_{1,2,3}\,, \qquad  K _4 \, = \, \rme^\Upsilon \cdot \Xi_3    \,  ,
\ee
where $\Upsilon$ is a section of the adjoint bundle implementing the twist of $S^4$ over $\mathbb{\Sigma}$ as in \eqref{eq:S4-fibration}, ensuring that these are globally defined objects on the six-dimensional manifold. 
 Recall that in the MN1 solution, the $U(1)$ that is used to twist the four-sphere and compensate the spin connection $\upsilon$ on $\mathbb{\Sigma}$ is the Cartan of $\SU(2)_{\rm right}\subset \SO(5)$.   
  The $E_{6(6)}$ twist element $\Upsilon$ is constructed in a way similar to the one used in~\cite{Cassani:2019vcl}, albeit with a different choice of $\U(1)$ in $\SO(5)$. We embed  the connection one-form $\upsilon$ in a generalised dual vector, the  Killing vector generating the Cartan of $\SU(2)_{\rm right}$ in the generalised vector $\Xi_3$ introduced above, and we project their product onto the adjoint of $E_{6(6)}$. That is,
\begin{align} 
\label{twistel}
\Upsilon \,&=\,  -\frac{R}{2} \, \upsilon \times_{\adj}\Xi_3 \,,
\end{align} 
where   $\times_{\adj}$ denotes the projection onto the adjoint and again $R$ is the radius of $S^4$. 
Evaluating its action in \eqref{KI_before_twist}, we find that this is trivial for all the $K_I$'s except for $K_4$, and we obtain  our final expressions
\be
\label{sec:Ksing}
K_0\,=\,  R^2\,\text{vol}_{\mathbb{\Sigma}}\wedge \, E'_5   \,,\qquad K_{1,2,3} \,=\,   \ti{\Xi}_{1,2,3}  \,,\qquad
K_4\,=\,   \Xi_3 - R\,\upsilon\wedge E_5   \,.
\ee
A similar procedure applies to the singlets $J_A$, $A = 1, \dots, 8$, in the adjoint bundle. In this way we obtain (see Appendix~\ref{app:U1MN}  for the derivation)
\be
\label{sec:Jsing}
\begin{aligned}
J_1 \,&=\,  \tfrac{1}{2}\, \rme^{\Upsilon}\cdot\, \big( - R\,e_1\times_{\text{ad}} \Xi_1  - R\,e_2\times_{\text{ad}} \Xi_2  + R^{-1}\,\Xi_1^* \times_{\text{ad}} \hat{e}_1  + R^{-1}\,\Xi_2^*  \times_{\text{ad}} \hat{e}_2 \big),\\[1mm]
J_2 \,&=\,   \tfrac{1}{2}\,\rme^{\Upsilon}\cdot \,\big( R\,e_1\times_{\text{ad}} \Xi_2    -  R\,e_2\times_{\text{ad}} \Xi_1  - R^{-1}\,\Xi_2^*  \times_{\text{ad}} \hat{e}_1 + R^{-1}\, \Xi_1^* \times_{\text{ad}} \hat{e}_2  \big) \, , \\[1mm] 
J_3 \,&=\,   \tfrac{1}{2}\,\rme^{\Upsilon}\cdot\, \big( \hat{e}_1\otimes e_2 - \hat{e}_2\otimes e_1 - R\, e_{2}  \times_{\adj} \Psi_{15} + R^{-1}\, \Psi^*_{15} \times_{\adj} \hat{e}_{2} \\
&\,\qquad\ \qquad - E_{5[1}^*\times_{\text{ad}}E_{2]5} + E_{5[3}^*\times_{\text{ad}}E_{4]5} \big)\,, \\[1mm]
J_4 \,&=\,  \tfrac{1}{2}\,\rme^{\Upsilon}\cdot\, \big( R\, e_1\times_{\text{ad}} \Xi_2  -R\, e_2\times_{\text{ad}} \Xi_1 +R^{-1}\, \Xi_2^*  \times_{\text{ad}} \hat{e}_1 - R^{-1}\, \Xi_1^*  \times_{\text{ad}} \hat{e}_2   \big) \, ,\\[1mm]
J_5 \,&=\,  \tfrac{1}{2}\,\rme^{\Upsilon}\cdot\, \big( R\, e_1\times_{\text{ad}} \Xi_1    + R\, e_2\times_{\text{ad}} \Xi_2  + R^{-1}\, \Xi_1^*  \times_{\text{ad}} \hat{e}_1 + R^{-1}\, \Xi_2^*  \times_{\text{ad}} \hat{e}_2 \big) \, ,\\
J_6 \,&=\, -\tfrac{1}{3}\,\rme^{\Upsilon}\cdot\,\big(\hat{e}_1\otimes e_1+\hat{e}_2\otimes e_2+\sum_{i=1}^4 E_{i5}^*\times_{\text{ad}}E_{i5} +2 \big)\,, \\ 
J_7 \,&=\, \rme^{\Upsilon}\cdot \,\big(R\, e_{2}  \times_{\adj} \Psi_{15} + R^{-1}\, \Psi^*_{15} \times_{\adj} \hat{e}_{2}\big) \, ,\\[1mm]
J_8 \,&=\, \tfrac{1}{2\sqrt{3}}\,\rme^{\Upsilon}\cdot\, \big(  \hat{e}_1\otimes e_2-\hat{e}_2\otimes e_1-3R\, e_{2}  \times_{\adj} \Psi_{15} +3R^{-1}\, \Psi^*_{15} \times_{\adj} \hat{e}_{2} \\[1mm]
&\, \qquad\qquad \qquad - E_{5[1}^*\times_{\text{ad}}E_{2]5} + E_{5[3}^*\times_{\text{ad}}E_{4]5}   \big)\,,
\end{aligned}
\ee
where the superscript $*$ denotes dual generalised vectors, transforming in the ${\bf \overline{27}}$, and we introduced
$  \Psi_{1i} = R\, e_1 \wedge E_i$  and $  \Psi_{2i} = R\, e_2 \wedge E_i$.  The adjoint action of $\rme^{\Upsilon}$ is evaluated using the formula \eqref{adjcomm}; we do  
not show  the resulting expressions as   they are rather lengthy. 
Evaluating the commutators $[J_A,J_B]$ using again \eqref{adjcomm}, we checked that the $J_A$ satisfy precisely the $\SU(2,1)$ commutation relations (see  \eqref{comm_su21_generators} for our choice of $\SU(2,1)$ structure constants).

\subsection{The V and H structure moduli spaces}

We now construct the V structure and H structure moduli spaces. Applying the general discussion of Section~\ref{sec:HVstr_moduli} we have 
\begin{equation}
\label{eq:MN1ms}
   \mathcal{M}_{\rm V} \times \mathcal{M}_{\rm H} \,=\,
       \frac{\Com{\Gst}{\Ex{6}}}{\Com{\Gst}{\USp(8)/\bbZ_2}}
       \,=\, \bbR^+ \times \frac{\Spin(3,1)}{\SU(2)} 
           \times \frac{\SU(2,1)}{\SUH\times\U(1)}
   \,,
\end{equation}
As we now show the first two factors give the V structure moduli space and the last factor the H structure moduli space.  

\subsubsection*{The V structure}

Evaluating \eqref{eq:CKKK_cond} for the $K_I$ constructed above we obtain the constant, symmetric tensor $C_{IJK}$. Using the invariant volume~\eqref{MN1-kappa},
we find that the non-vanishing components of $C_{IJK}$ are given by
\begin{equation}\label{cubicinv_on_K}
C_{0IJ} = C_{I0J} = C_{IJ0}= \,\tfrac{1}{3}\, \eta_{IJ}\,,\qquad \text{for}\ I,J=1,\ldots,4\,,
\end{equation}
where 
\be
\eta = {\rm diag}(-1,-1,-1,1)\,.
\ee 
A family of V structures is then obtained by defining $K$ as the linear combination~\eqref{eq:Kdressing} and imposing the condition \eqref{constraint_Chhh}. It follows that our V structure moduli space is the hypersurface
\begin{equation}\label{towards_hyperboloid}
C_{IJK}h^Ih^Jh^K \,= \,   h^0 \left(  - (h^1)^2- (h^2)^2- (h^3)^2 + (h^4)^2\right) \,=\, 1\,.
\end{equation}
It will be convenient to redefine the $h^I$ in terms of the parameters
\be
\{\Sigma,H^1,H^2,H^3,H^4\}
\ee
as
\begin{align}\label{from_h_to_H}
 h^0 &= \Sigma^{-2}\,,\nn\\[1mm]
h^I &= -\Sigma \, H^I \,,\quad I = 1,\ldots,4\,,
\end{align}
so that
\be\label{eq:final_param_K}
K \,=\, \Sigma^{-2}K_0 - \Sigma \, \left( H^1 K_1 + H^2 K_2 + H^3 K_3 + H^4 K_4 \right)\,.
\ee
From \eqref{towards_hyperboloid} we see that $H^I$ are coordinates on the unit hyperboloid $\frac{\SO(3,1)}{\SO(3)}$,
\be
  - (H^1)^2 - (H^2)^2 - (H^3)^2 + (H^4)^2 = 1\,,
\ee
 while $\Sigma$ (that we assume strictly positive) is a coordinate on $\mathbb{R}^+$, whose powers in \eqref{from_h_to_H} are dictated by the weight of the $K_I$'s under the action of the $\mathbb{R}^+$ that commutes with the generalised structure.
 The resulting V structure moduli space thus is 
\be
\mathcal{M}_{\rm V} \,=\,\mathbb{R}^+\times \frac{\SO(3,1)}{\SO(3)}\,,
\ee
and will determine $n_{\rm V}=4$ vector multiplets in five-dimensional $\mathcal{N}=2$ supergravity. Note that by identifying $\SU(2)\simeq\Spin(3)$ this matches the first two factors in~\eqref{eq:MN1ms}. The isometry group is $\SO(3,1)$ because the $h^I$ form a vector rather than a spinor representation of $\Spin(3,1)$.

\subsubsection*{The H structure}

We next turn to the H structure moduli space, again following the general discussion given in  Section~\ref{sec:HVstr_moduli}.
Since the commutant of $\SUH$ in $\SU(2,1)$ is $\U(1)$, from \eqref{Hstr_space_general} we obtain that the H structure moduli space is\footnote{More precisely one has $\mathcal{M}_{\rm H} = \SU(2,1)/S(\U(2)\times\U(1)).$}
\be\label{MH_moduli_space_MN1}
\mathcal{M}_{\rm H} = \frac{\SU(2,1)}{\SUH\times\U(1)}\,.
\ee
This is a simple quaternionic-K\"ahler manifold of quaternionic dimension $n_{\rm H}=1$.
 We will denote by 
 \be
 \{\varphi,\xi,\theta_1,\theta_2\}
 \ee the coordinates on this space. 
In Appendix~\ref{app:param_Hstructure}  we give the explicit parameterisation chosen for the coset space as well as the explicit form of the ``dressed'' $\su(2)$ elements $J_\alpha$, depending on $\{\varphi,\xi,\theta_1,\theta_2\}$, in terms of $\su(2,1)$ elements. 
 Below we will use this dressed triplet to construct the generalised metric. In Appendix~\ref{app:param_Hstructure} we also give the $\SU(2,1)$ invariant metric on $\mathcal{M}_H$, which will provide the hyperscalar kinetic term in the five-dimensional theory.

\subsection{Intrinsic torsion and gauging}\label{sec:gauging_MN1_trun}

For the $U(1)_S$ structure constructed in the previous section to lead to a consistent truncation, it must be  checked that its intrinsic torsion only contains $U(1)_S$ singlets, and that these are constant. In particular we need to show that  equations~\eqref{eq:K-alg}, \eqref{LieDerivativesForGauging} and~\eqref{eq:GS-singlet-int-tor} hold. For the first two conditions we  evaluate the generalised Lie derivatives of the tensors  $K_I$ and $J_A$ in \eqref{sec:Ksing} and \eqref{sec:Jsing}, using the action of generalised Lie derivative on a generalised vector and on sections of the adjoint bundle given in Appendix \ref{PreliminariesE66_Mth}. 

Consider first the algebra of the generalised vectors  \eqref{sec:Ksing}. Using the fact that,  under the generalised Lie derivative, the $S^4$ frames $E_{ij}$ generate an
 $\so(5)$ algebra
\begin{equation}
   L_{E_{ij}} E_{kl} = - R^{-1} \left(
      \delta_{ik}E_{jl} -\delta_{il}E_{jk} + \delta_{jl} E_{ik}
      - \delta_{jk} E_{il} \right) \,,
\end{equation}
one can show that the only non-zero  Lie derivatives are 
\be\label{Lie_der_K_alpha}
L_{K_\alpha} K_\beta \,=\, -\tfrac{2}{R} \,\epsilon_{\alpha\beta\gamma} K_\gamma\,,\qquad\qquad \alpha,\beta,\gamma=1,2,3  \, ,
\ee
so that the vectors $K_\alpha$, $\alpha=1,2,3$, lead to an $SO(3)$ factor in the gauge group in the five-dimensional supergravity.\footnote{For simplicity, we use the indices $\alpha,\beta=1,2,3$ both to label the generators of the $\SUH$ entering in the definition of the H structure and the generators of the $\SU(2)$ in the V structure, although these are different subgroups of $\Ex{6}$. } This embeds in the  $SO(3,1)$ factor of the global symmetry group of the ungauged theory in the obvious way. Hence \eqref{Lie_der_K_alpha} determines the components of the embedding tensor acting on the vector multiplet sector of the five-dimensional supergravity theory.

We thus have that the non-vanishing structure constants are $f_{\alpha\beta\gamma}= -2\,\epsilon_{\alpha\beta\gamma}$ and the gauge coupling constant is $g = \frac{1}{R}$.
Recalling \eqref{vectorm_scalar_cov_der_main_text},  the non-trivial vector multiplet scalar covariant derivatives are
\be
\mathcal{D} H^\alpha = \diff H^\alpha - \tfrac{2}{R}\, \epsilon^\alpha{}_{\beta\gamma} \mathcal{A}^\beta \,H^\gamma \,,
\ee
while the gauge field strengths read
\be\label{gauge_field_strengths}
\mathcal{F}^0= \dd \mathcal{A}^0\,,\qquad  \mathcal{F}^\alpha = \diff\mathcal{A}^\alpha - \tfrac{1}{R}
\, \epsilon^\alpha{}_{\beta\gamma} \mathcal{A}^\beta\wedge\mathcal{A}^\gamma\,,\qquad \mathcal{F}^4 = \diff\mathcal{A}^4 \,.
 \ee
 
In order to determine the gauging in the hypersector we also need to compute the Lie derivative of the $J_A$ along the generalised vectors $K_I$. We find that the $J_A$ are neutral under the action of the $\SO(3)$ generators $K_\alpha$,
\be
L_{K_\alpha}J_A =0  \,, \qquad A =1, \ldots 8 \, , 
\ee
consistently with the fact that the gauging in the vector multiplet sector does not affect the hypersector.  On the other hand, the remaining generalised vectors $K_0$ and $K_4$ act non-trivially on the $J_A$, and determine an abelian gauging of the $\SU(2,1)$ generators. In detail, the generalised Lie derivative of the $J_A$ along $K_0$ gives
\begin{equation}
\label{K0Jac}
\begin{aligned}
L_{K_0} (J_1-J_5)&=0\,,\\[1mm]
L_{K_0} (J_1+J_5) &=\tfrac{1}{R}(J_2+J_4)\,,\\[1mm]
L_{K_0} (J_2+J_4)&=0\,,\\[1mm]
L_{K_0} (J_2-J_4)&= -\tfrac{1}{R}(J_1-J_5)\,,
\end{aligned}
\hspace{2cm}
\begin{aligned}
L_{K_0} J_3&=-\tfrac{1}{2R}J_6\,,\\[1mm]
L_{K_0} J_6&=-\tfrac{1}{2R}\big(J_3+2J_7-\sqrt{3}J_8 \big)\,,\\[1mm]
L_{K_0} J_7&=\tfrac{1}{R}J_6\,,\\[1mm]
L_{K_0} J_{8}&=\tfrac{\sqrt{3}}{2R}J_6\,,
\end{aligned}
\end{equation}
while the one along $K_4$ yields
\begin{equation}
\label{K4Jac}
\begin{aligned}
L_{K_4}{(J_1-J_5)}&=-\tfrac{2}{R}(J_2+ J_4)\,,\\[1mm]
L_{K_4}{(J_1+J_5)}&=-\tfrac{2}{R}(J_2-J_4)-\tfrac{\kappa}{R}\, (J_2+ J_4)\,,\\[1mm]
L_{K_4}{(J_2+J_4)}&=\tfrac{2}{R}(J_1- J_5)\,,\\[1mm]
L_{K_4}{(J_2-J_4)}&= \tfrac{2}{R}(J_1+J_5)+\tfrac{\kappa}{R} (J_1- J_5)\,,
\end{aligned}
\hspace{1.5cm}
\begin{aligned}
L_{K_4} J_3&=\tfrac{\kappa}{2R}J_6\,,\\[1mm]
L_{K_4} J_6&=\tfrac{\kappa}{2R}\big(J_3+2J_7-\sqrt{3}J_8\big)\,,\\[1mm]
L_{K_4} J_7&= -\tfrac{\kappa}{R} J_6\,,\\[1mm]
L_{K_4} J_8&= -\tfrac{\sqrt{3}\,\kappa}{2R}J_6\,.
\end{aligned}
\end{equation}
The actions \eqref{K0Jac} and \eqref{K4Jac} 
can equivalently be expressed in terms of an adjoint action as
\be
L_{K_0} J_A = [J_{(K_0)}, J_A]\,,\qquad L_{K_4} J_A = [J_{(K_4)}, J_A]\,, \qquad  A=1, \dots, 8 \, , 
\ee
where the sections of the adjoint bundle
\begin{align}
J_{(K_0)} &= \tfrac{1}{4R}\,\big(J_3 + 2J_7 - \sqrt{3} J_8 \big)\,,\nn\\[1mm]
J_{(K_4)} &= -\tfrac{\kappa}{4R}\,\big(J_3 + 2J_7 - \sqrt{3} J_8 \big)   -  \tfrac{1}{R}\,\big(  J_3 + \tfrac{1}{\sqrt3} J_8  \big)\,
\end{align} 
correspond to $\SU(2,1)$ generators acting on the H-structure moduli space \eqref{MH_moduli_space_MN1} as isometries. We denote by $k_0$ and $k_4$ the corresponding Killing vectors on $\mathcal{M}_{\rm H}$. These can be calculated applying \eqref{formula_for_k^X} to  the coset representative $L$ given in Appendix~\ref{app:param_Hstructure}, and read
\begin{align}\label{KillingVectHyper}
k_0 \,&=\, \partial_\xi\,,\nn\\
k_4 \,&=\, - \kappa\, \partial_\xi + 2\left(\theta_2 \partial_{\theta_1} - \theta_1 \partial_{\theta_2} \right) \,.
\end{align}
These Killing vectors specify the isometries of $\mathcal{M}_{\rm H}$ that are gauged in the five-dimensional supergravity.
The hyperscalar covariant derivatives \eqref{hyperscalar_cov_der_main_text} are determined as 
 \begin{align}\label{scal_cov_der}
 \mathcal{D}(\theta_1 + \ii\, \theta_2) \,&=\, \dd (\theta_1 + \ii\, \theta_2) - \tfrac{2}{R}\,\ii \,  \mathcal{A}^4\,(\theta_1 +\ii\,\theta_2)\,,\nn\\[1mm]
 \mathcal{D} \xi \,&=\, \dd \xi + \tfrac{1}{R} \mathcal{A}^0 - \tfrac{\kappa}{R} \mathcal{A}^4\,.
 \end{align}

The triholomorphic Killing prepotentials $P_I^\alpha$ obtained by evaluating the moment maps \eqref{KillingPrep_from_mommaps} read 
\begin{align}\label{KillingPrepMN1}
P^\alpha_0 \,& =\, \big\{0\,,\,0\,,\,\tfrac{1}{4}\,\rme^{2\varphi} \big\} \,,\nn\\[1mm]
P^\alpha_4 \,& =\, \big\{\sqrt{2}\,\rme^{\varphi}\theta_1\,,\,\sqrt{2}\,\rme^{\varphi}\theta_2\,,-1+\tfrac{1}{4}\,\rme^{2\varphi} \big(2\theta_1^2 + 2\theta_2^2-\kappa\big)\,    \big\}\,,
\end{align}
with $P_1^\alpha=P_2^\alpha=P_3^\alpha=0$.

The information above completely characterises the five-dimensional $\mathcal{N}=2$ supergravity obtained upon reduction on $M$. This will be discussed in Section~\ref{sec:5d_theory_MN1}. Before coming to that, we provide the explicit bosonic truncation ansatz.

\subsection{The truncation ansatz}

The truncation ansatz is built following the general procedure described in Section~\ref{sec:fielddec}. We compute the inverse generalised metric \eqref{USp6_igm} out of the $\U(1)_S$ invariant generalised tensors. This depends on  the V structure moduli $\{\Sigma,H^1,H^2,H^3,H^4\}$ and on the H structure moduli $ \{\varphi,\xi,\theta_1,\theta_2\}$, which are now promoted to scalar fields in the external, five-dimensional spacetime.  
Then we evaluate the generalised tensors $\mathcal{A}_\mu,\mathcal{B}_{\mu\nu},\mathcal{C}_{\mu\nu\rho}$ using \eqref{ansatz_Avec}--\eqref{ansatz-Cform}. Separating the components of these tensors as described in Section~\ref{sec:fielddec}, we obtain the ansatz for the eleven-dimensional metric $\hat{g}$ and three-form potential $\hat{A}$.

We start from the covariantised differentials $Dz^m \,=\, \diff z^m-h_\mu{}^m \diff x^\mu$ of the coordinates on $M$, that appear in \eqref{11d_metric_general}. From \eqref{Avec} and \eqref{ansatz_Avec} we see that $h_\mu = h_\mu{}^m \partial_m$ is given by
\be
h_\mu \,=\, \mathcal{A}_\mu^I\, K_I |_{TM}\,,
\ee
where $K_I |_{TM}$ is the restriction of $K_I$ to the tangent bundle of $M$.
Evaluating the right hand side using the explicit form \eqref{sec:Ksing} of the generalised vectors $K_I$, we obtain
\be\label{hmu_MN1trunc}
h_\mu \,=\, \tfrac{2}{R}\, \big( \mathcal{A}_\mu^\alpha\, \tilde{\xi}_\alpha + \mathcal{A}^4_\mu\,\xi_3 \big) \,,
\ee
where we recall that $\xi_\alpha$, $\tilde\xi_{\alpha}$, $\alpha = 1,2,3$, are the pull-back to $TM$ of the $\SU(2)_{\rm left}$- and $\SU(2)_{\rm right}$-invariant vectors on $S^3$, respectively, whose coordinate expression is given in \eqref{left_inv_Killing_vectors} and \eqref{right_inv_Killing_vectors}. 
It follows that $Dz^m$, and thus both the eleven-dimensional metric and three-form, contain the five-dimensional gauge potentials $\mathcal{A}^\alpha$, $\mathcal{A}^4$, gauging the $\SU(2)_{\rm left}\times \U(1)_{\rm right}$ isometries of $S^3$ in $M$. Notice that $\mathcal{A}^0$ does not appear in \eqref{hmu_MN1trunc} as $K_0$ does not have a component in $TM$, hence it will not enter in the eleven-dimensional metric. However $K_0$  will appear in the ansatz for the three-form, as it does have a component in $\Lambda^2 T^*M$.

In order to express our ansatz in a more compact way, it will be convenient to introduce new one-forms $\Omega_\alpha$ and $\tOmega_\alpha$, $\alpha=1,2,3$, adapted to the symmetries of the problem, that  incorporate the covariantised differentials above but also include some more terms. 
Recall that we describe $S^4$ as a foliation of $S^3$, parameterised by Euler angles $\{\theta,\phi,\psi\}$, over an interval, parameterised by $\zeta$. We define
\begin{align}
\Omega_1 &= \cos\psi \,{\sf D}\theta +\sin\psi \sin\theta\,{\sf D}\phi\,, \qquad &\tOmega_1 = \cos\phi \,{\sf D}\theta +\sin\phi \sin\theta\,{\sf D}\psi  \,,\nn\\
\Omega_2 &= -\sin\psi\,{\sf D}\theta + \cos\psi\sin\theta\,{\sf D}\phi\,,\qquad &\tOmega_2 =  \sin\phi\,{\sf D}\theta - \cos\phi\sin\theta\,{\sf D}\psi \,, \nn\\
\Omega_3 &= {\sf D}\psi + \cos\theta\,{\sf D}\phi\,,\qquad &\tOmega_3 =  {\sf D}\phi + \cos\theta\,{\sf D}\psi \,,
\end{align}
which are analogous to the left- and right-invariant forms $\sigma_\alpha,\tilde{\sigma}_\alpha$ given in \eqref{leftinv_oneforms} and \eqref{rightinv_oneforms}, but with the ordinary differential of the coordinates being replaced by the new covariantised differential ${\sf D}$. This extends the differential $D$ given above and is defined as
\be\label{eq:AtildeAindz}
{\sf D} z^m = \diff z^m - \tfrac{2}{R} \big( {\sf A}^\alpha \xi^m_\alpha +\tilde{\sf A}^\alpha \tilde{\xi}^m_\alpha \big)\,,
\ee
with
\begin{align}\label{sevend_gauge_fields}
& {\sf A}^1 = \tfrac{R}{\sqrt{2}}\left(\theta_2 e_1 - \theta_1 e_2\right)   ,\qquad {\sf A}^2 = \tfrac{R}{\sqrt{2}}\left( \theta_1 e_1 + \theta_2 e_2 \right) ,\qquad {\sf A}^3= -\tfrac{R}{2}\,\upsilon + \mathcal{A}^4\,,\nn\\
& \tilde{{\sf A}}^\alpha = \mathcal{A}^\alpha\,,\qquad \alpha=1,2,3\,,
\end{align}
where the five-dimensional scalars $\theta_1,\theta_2$ are two of the H structure moduli, and we recall that $e_1,e_2$ are the vielbeine on the Riemann surface $\mathbb{\Sigma}$ while $\upsilon$ is the connection on $\mathbb{\Sigma}$.
The local one-forms $\tilde{\sf A}^\alpha,{\sf A}^\alpha$ gauge all the left- and right- isometries of $S^3$, respectively, and would correspond to $\SO(4)\simeq \SU(2)_{\rm left}\times \SU(2)_{\rm right}$ gauge potentials in the reduction of eleven-dimensional supergravity on $S^4$ down to seven-dimensional supergravity. However, in the further reduction on $\mathbb{\Sigma}$ of interest here only $ \mathcal{A}^\alpha, \mathcal{A}^4$ become five-dimensional gauge fields, while the rest of \eqref{sevend_gauge_fields} implements the twist on the Riemann surface and introduces the five-dimensional scalars $\theta_1,\theta_2$.

We are now in the position to give the truncation ansatz for the eleven-dimensional metric
\be
\hat{g} \,=\,  \rme^{2\Delta} g_{\mu\nu}\dd x^\mu\dd x^\nu + g_{mn} Dz^m Dz^n \,.
\ee
The warp factor is
\be
\rme^{2\Delta}\, =\, \bar\Delta^{1/3}  \left(\rme^{\varphi} \Sigma\right)^{4/5}\,,
\ee
while the part with at least one internal leg reads
\begin{align}
g_{mn} Dz^m D z^n &=   R^2 \bar\Delta^{1/3}\! \left(\rme^{\varphi} \Sigma\right)^{-6/5}g_{\mathbb\Sigma}  + R^2\bar\Delta^{-2/3}\rme^{2\varphi/5}\Sigma^{-3/5} \Big[\! \left(\rme^{-2\varphi} \Sigma^3 \sin^2\zeta + H_- \cos^2\zeta \right)\! \diff\zeta^2 \nn\\[1mm]
& + \tfrac{1}{4}H_+\sin^2\zeta\, \delta^{\alpha\beta}\Omega_\alpha \otimes\Omega_\beta -  \tfrac{1}{2}\sin^2\zeta\, H^{\alpha}\, \tOmega_\alpha   \otimes_s\Omega_3  - \cos\zeta\sin\zeta\,\diff\zeta \otimes_s\diff_6 H_+  \Big]\,,
\end{align}
where $\otimes_s$ is the symmetrised tensor product, defined as $\Omega\otimes_s \ti\Omega = \frac{1}{2}(\Omega\otimes \ti\Omega +\ti\Omega\otimes \Omega)$. In these expressions we introduced the function
\begin{align}
\bar\Delta \,=\, \big(\rme^{-2\varphi}\Sigma^{3}\big)^{-4/5} \cos^2\zeta + \big(\rme^{-2\varphi}\Sigma^{3}\big)^{1/5} H_+\sin^2\zeta\,,
\end{align}
as well as
\begin{align}
H_{\pm} \,&=\, H^4 \pm \left( H^1 \sin\theta\sin\phi -H^2 \sin\theta\cos\phi  + H^3 \cos\theta \right)\,,\nn\\
\diff_{6} H_+ \,&=\,  H^1\, \diff(\sin\theta\sin\phi) -H^2\, \diff(\sin\theta\cos\phi)  + H^3\, \diff\cos\theta \,.
\end{align}
Note that in the last expression the exterior derivative acts on the internal coordinates and not on the scalars $H^I$, which only depend on the external coordinates.

We next come to the eleven-dimensional three-form potential $\hat{A}$. We first give our result and then make some comments. The ansatz for $\hat{A}$ reads
\begin{align}\label{our_3form_ansatz}
\hat{A} &= - \tfrac{1}{8}R^3 \cos\zeta\,\big[ 2 + \sin^2\zeta  \,\bar\Delta^{-1}(\rme^{-2\varphi}\Sigma^{3})^{-4/5} \, \big]\, \Omega_1\wedge\Omega_2\wedge\Omega_3 \nn\\[1mm]
&\quad +\tfrac{1}{4}R^3 \sin^3\zeta\,\bar\Delta^{-1} (\rme^{-2\varphi}\Sigma^{3})^{1/5} \,\diff\zeta \wedge H^{\alpha}\, \tOmega_\alpha\wedge\Omega_3\nn\\[1mm]
&\quad + R^3\, \cos\zeta\, \big(\mathcal{D}\xi - \theta_1 \mathcal{D}\theta_2 + \theta_2 \mathcal{D}\theta_1\big) \wedge\vol_{\mathbb{\Sigma}} +\tfrac{1}{4}R^3\cos\zeta\left( 2\theta_1^2 + 2\theta_2^2 - \kappa\right)\vol_{\mathbb{\Sigma}}\wedge\Omega_3 \nn\\[1mm]
&\quad +   \tfrac{1}{2}R^2\cos\zeta \,\big( \mathcal{F}^4\wedge \Omega_3 -\mathcal{F}^\alpha\wedge \ti\Omega_\alpha\big)  + R\,\cos\zeta\,\Sigma^4 *_5  \mathcal{F}^0   \nn\\[1mm]
&\quad + \tfrac{1}{2\sqrt2}\,R^3\cos\zeta \left[ \left( \mathcal{D}\theta_2\wedge  e^1 - \mathcal{D}\theta_1\wedge  e^2\right)\wedge \Omega_1 + \left( \mathcal{D}\theta_1\wedge  e^1 + \mathcal{D}\theta_2\wedge  e^2\right)\wedge \Omega_2   \right] \,,
\end{align}
 where the five-dimensional gauge field strengths, $\mathcal{F}$,  and the covariant derivatives, $\mathcal{D}$, of the five-dimensional scalars were given in~\eqref{gauge_field_strengths} and~\eqref{scal_cov_der}, respectively.

Equation \eqref{our_3form_ansatz} has been obtained by first computing $\hat{A}$ through the general procedure of Section~\ref{sec:fielddec}, then implementing a gauge shift by an exact three-form so as to obtain a nicer expression (this is why derivatives of the external fields appear), and finally dualising away the five-dimensional two- and three-forms, so that the only five-dimensional degrees of freedom contained in the ansatz are scalar and vector fields, in addition to the metric $g_{\mu\nu}$. Let us outline how this dualisation is performed.
Evaluating \eqref{ansatz-Bform} and \eqref{ansatz-Cform}, we find that only one external two-form $\mathcal{B}$ and one external three-form $\mathcal{C}$ appear in the ansatz for $\hat{A}$. These are paired up with the generalised tensors $E_5$ and $E'_5$ on $S^4$, 
which,  as generalised tensors on $M$, are sections of $\det T^*M \otimes E^*$ and $\det T^*M \otimes \adj F$, respectively. The combination entering in $\hat{A}$ is
\begin{align}
\left[\mathcal{B} E_5 + \mathcal{C} E_5' \right]_3 \,&=\, R\,\mathcal{B}\wedge \dd\cos\zeta + R\,\mathcal{C} \cos\zeta \,=\, (\mathcal{C} - \diff \mathcal{B})\,R\cos\zeta + \diff \left( \mathcal{B} \,R\cos\zeta\right)\,,
\end{align}
where the subscript on the left-hand side indicates the restriction  to the three-form part, and the last term in the expression is removable via a gauge transformation of $\hat A$. 
Hence $\mathcal{B}$ and $\mathcal{C}$ only appear in the combination 
$(\mathcal{C} - \diff \mathcal{B})$. This means that the two-form gets eaten by the three-form via the Stuckelberg mechanism, giving it a mass. While a massless three-form in five-dimensions is dual to a scalar field, here we dualise the two-form at the same time and also obtain a vector field. 
The duality relation is obtained considering the duality between the eleven-dimensional three-form $\hat{A}$ and six-form $\hat{\tilde{A}}$ given in~\eqref{dualityAtildeA}, and looking at the relevant terms with three external indices. In this way we find that
\be\label{duality_Cthree}
\mathcal{C}-\dd \mathcal{B} \,=\,  \Sigma^4 *_5 \dd \mathcal{A}^0 - \mathcal{A}^4\wedge \dd \mathcal{A}^4 +   \mathcal{A}^\alpha\wedge \dd\mathcal{A}^\alpha -\tfrac{1}{3R}\, \epsilon_{\alpha\beta\gamma}\mathcal{A}^\alpha\wedge\mathcal{A}^\beta \wedge\mathcal{A}^\gamma   \,.
\ee
We have used this expression to eliminate $(\mathcal{C}-\dd \mathcal{B})$ completely from the truncation ansatz. This explains the $*_5  \mathcal{F}^0$ term appearing in \eqref{our_3form_ansatz}.

\medskip

Our truncation ansatz reproduces the Maldacena--Nu\~nez solution given in Section \ref{MNsol} upon taking $\kappa=-1$ and setting the scalars to
\be\label{values_scalars_MN1}
H^1=H^2=H^3= \theta_1=\theta_2=\xi=0\,,\quad H^4= \Sigma=1\,, \quad \varphi=\tfrac{1}{2}\log\tfrac{4}{3}\,.
\ee

The consistent truncation of \cite{Faedo:2019cvr} is recovered as a subtruncation that projects out the fields transforming under $\SU(2)_{\rm left}$, that is setting $\tilde{\mathcal{A}}^\alpha = H^\alpha = 0$, $\alpha=1,2,3$, which also implies $H^4=1$.\footnote{Then the one-forms $\Omega_\alpha$ essentially reduce to those in \cite{Faedo:2019cvr}, up to slightly different conventions, while $\tOmega_\alpha$ drop out of the ansatz. When comparing our truncation ansatz with the one given in Section~4.1 of \cite{Faedo:2019cvr}, one should take into account that $\hat{A}^{\rm here}=-\hat{A}^{\rm FNR}$ (this is seen from comparing our 11d Maxwell equation with the one in \cite{Lu:1999bc}, which provides the 7d to 11d uplift formulae used in \cite{Faedo:2019cvr}). Moreover $\zeta^{\rm here}=\zeta^{\rm FNR} + \pi/2\,,$
 $\mathcal{A}^4 \propto \mathcal{A}^{\rm FNR}$, $\mathcal{A}^0 \propto \chi_1^{\rm FNR}$, $\Sigma = 2^{1/3}\Sigma^{\rm FNR}$, $\rme^{2\varphi}= 2(\rme^{2\varphi})^{\rm FNR}$, $|\theta_{1,2}| = \frac{1}{\sqrt2}|\theta_{1,2}|^{\rm FNR}$, $\xi = \frac{1}{2}\xi^{\rm FNR}$.
} The further truncation to minimal gauged supergravity is obtained by setting the scalars to their AdS value \eqref{values_scalars_MN1} and taking $\mathcal{A}^0 = - \mathcal{A}^4$.

One can obtain a slightly larger subtruncation by projecting out only the modes charged under $\U(1)_{\rm left}$, rather than $\SU(2)_{\rm left}$, namely setting $\tilde{\mathcal{A}}^1=\tilde{\mathcal{A}}^2=H^1=H^2=0$. This leaves us with two vector multiplets, one hypermultiplet and just the abelian gauging generated by the Killing vectors \eqref{KillingVectHyper}, which is the same as the one in the truncation of \cite{Faedo:2019cvr}.\footnote{Curiously, this five-dimensional supergravity with two vector multiplets and one hypermultiplet looks closely related to the $\mathcal{N}=2$ ``Betti-vector'' model obtained in \cite[Section~7]{Cassani:2010na} as a consistent truncation of IIB supergravity on $T^{1,1}$. The two models are not the same though, as the details of the couplings between the fields are different. 
} A notable generalisation of this subtruncation will be discussed in Section~\ref{sec:BBBWsection}.

The truncation of \cite{Faedo:2019cvr} was obtained via a reduction of gauged seven-dimensional supergravity on the Riemann surface $\mathbb{\Sigma}$. Similarly, we can obtain our truncation ansatz by combining the well-known truncation of eleven-dimensional supergravity on $S^4$ \cite{Nastase:1999kf}, leading to seven-dimensional maximal $\SO(5)$ supergravity, with a further truncation reducing the seven-dimensional theory on $\mathbb{\Sigma}$. Starting from the convenient form of the bosonic truncation ansatz on $S^4$ given in~\cite{Cvetic:2000ah}, we have explicitly checked that this procedure works out as expected and reproduces the ansatz above.

\subsection{The five-dimensional theory}\label{sec:5d_theory_MN1}

We now put together the ingredients defining the truncated five-dimensional theory and discuss it in more detail. This is an $\mathcal{N}=2$ gauged supergravity coupled to four vector multiplets and one hypermultiplet. The vector multiplet scalar manifold is 
\be
\mathcal{M}_{\rm V} \,=\, \mathbb{R}^+\times \frac{\SO(3,1)}{\SO(3)}\,,
\ee
while the hypermultiplet scalar manifold is
\be
\mathcal{M}_{\rm H} \,=\, \frac{\SU(2,1)}{\SUH\times\U(1)}\,.
\ee 
As discussed before, these have a geometric origin as the V and H structure moduli spaces of the internal manifold.
At the bosonic level, the vector multiplets are made of gauge fields $\mathcal{A}^I$ and constrained scalar fields $h^I$, $I=0,1,\ldots,4$, which we have parameterised in terms of $\Sigma$ and $H^I$, $I=1,\ldots,4$, in \eqref{from_h_to_H}. The latter scalars satisfy the constraint $\eta_{IJ}H^I H^J=1$, with $\eta = {\rm diag}(-1,-1,-1,1)$. 
 We have also found that the non-vanishing components of the symmetric tensor $C_{IJK}$ are given by
\be
C_{0IJ} = C_{I0J} = C_{IJ0}= \,\tfrac{1}{3}\, \eta_{IJ}\,,\qquad \quad I,J=1,\ldots,4\,.
\ee
The kinetic terms in the vector multiplet sector are controlled by the matrix $a_{IJ}$, given by the general formula~\eqref{aIJ_maintext}, which in our case reads
\begin{align}\label{kin_matr_MN1}
a_{00}&= \tfrac{1}{3}\, \Sigma^{4}\,, \nn\\[1mm]
a_{0J}&=0  \,,\nn\\[1mm] 
a_{IJ}&= \tfrac{2}{3}\,\Sigma^{-2} \left( 2\eta_{IK}H^K \eta_{JL} H^L - \eta_{IJ} \right)\,, \qquad\quad I,J =1,\ldots,4 \,.
\end{align}

The hypermultiplet comprises the scalars $q^X = \{\varphi,\xi,\theta_1,\theta_2\}$, and the kinetic term is given by the quaternionic-K\"ahler metric on $\mathcal{M}_{\rm H}$ that we derived in Appendix~\ref{app:param_Hstructure},
\be\label{Quaternionic_metric}
g_{XY}\dd q^X\dd q^Y \,=\, 2\, \diff \varphi^2 + \rme^{2\varphi} \left( \diff \theta_1^2 + \diff \theta_2^2 \right) + \tfrac12\,\rme^{4\varphi} \left( \diff \xi -\theta_1 \diff \theta_2 + \theta_2 \diff \theta_1\right)^2\,.
\ee

The gauge group is $\SO(3)\times \U(1)\times \mathbb{R}$.
The symmetries being gauged are the $\SO(3)\subset\SO(3,1)$ isometries of $\mathcal{M}_{\rm V}$ and two abelian isometries in $\mathcal{M}_{\rm H}$, generated by the Killing vectors~\eqref{KillingVectHyper}. Note that the $\der_\xi$ term generates the non-compact $\bbR$ factor and the $\theta_2\der_{\theta_1}-\theta_1\der_{\theta_2}$ term generates the compact $U(1)$. 

We recall for convenience the gauge field strengths 
\be
\mathcal{F}^0= \dd \mathcal{A}^0\,,\qquad  \mathcal{F}^\alpha = \diff\mathcal{A}^\alpha - g
\, \epsilon^\alpha{}_{\beta\gamma} \mathcal{A}^\beta\wedge\mathcal{A}^\gamma\,,\qquad \mathcal{F}^4 = \diff\mathcal{A}^4 \,,\quad \alpha =1,2,3\,,
 \ee
 and the covariant derivatives of the charged scalars,
 \begin{align}
\mathcal{D} H^\alpha \,&=\, \diff H^\alpha - \tfrac{2}{R}\, \epsilon^\alpha{}_{\beta\gamma} \mathcal{A}^\beta \,H^\gamma \,,\nn\\[1mm]
 \mathcal{D}(\theta_1 + \ii\, \theta_2) \,&=\, \dd (\theta_1 + \ii\, \theta_2) - \tfrac{2}{R}\,\ii \,  \mathcal{A}^4\,(\theta_1 +\ii\,\theta_2)\,,\nn\\[1mm]
 \mathcal{D} \xi \,&=\, \dd \xi + \tfrac{1}{R} \mathcal{A}^0 - \tfrac{\kappa}{R}\, \mathcal{A}^4\,,
 \end{align}
where the gauge coupling constant is given by the inverse $S^4$ radius, $g=\frac{1}{R}$. The scalars $\Sigma$, $H^4$ and $\varphi$ remain uncharged. The gauging in the hypersector is the same as in \cite{Faedo:2019cvr}, while the gauging in the vector multiplet sector is a novel feature of our truncation.

Plugging these data in the general form of the $\mathcal{N}=2$ supergravity action given in Appendix~\ref{app:sugra_review}, we obtain the bosonic action for our model,
\begin{align} 
S \, &= \, \frac{1}{16\pi G_5}\int \, \Big[ \left( \mathcal{R} -2\mathcal{V}\right)*1  -  \tfrac{1}{2}\,\Sigma^{4} \mathcal{F}^{0}\wedge*  \mathcal{F}^{0} -  \tfrac{3}{2}\sum_{I,J=1}^4 a_{IJ} \mathcal{F}^{I}\wedge*  \mathcal{F}^{J} - 2\Sigma^{-2}  \dd \Sigma \wedge* \dd \Sigma \nn\\[-1mm]
&\qquad\   - \tfrac{3}{2}\sum_{I,J=1}^4 a_{IJ}  \mathcal{D} (\Sigma H^I) \wedge* \mathcal{D} (\Sigma H^J)  -  g_{XY}  \mathcal{D} q^{X} \wedge * \mathcal{D} q^{Y} + \sum_{I,J=1}^4 \eta_{IJ}\mathcal{A}^0\wedge \mathcal{F}^I\wedge \mathcal{F}^J \Big] \, ,
\end{align}
where $G_5$ is the five-dimensional Newton constant.\footnote{As discussed in \cite{Ashmore:2016qvs}, the five-dimensional Newton constant is given by $(G_5)^{-1} \propto \int_M \rme^{3\Delta}\vol_6 = \int_M \kk^2$. In the present case, $\int_M \kk^2 = R^2\,{\rm Vol}_{\mathbb{\Sigma}}\,{\rm Vol}_4$, where 
${\rm Vol}_{\mathbb{\Sigma}}=\frac{4\pi(1-g)}{\kappa}$ is the standard volume of a Riemann surface of genus $g$ and ${\rm Vol}_4=\frac{8\pi^3}{3}R^4$ is the volume of a round $S^4$ with radius $R$.}
The scalar potential $\mathcal{V}$ is obtained from the Killing prepotentials of the gauged isometries as summarised in Appendix~\ref{app:sugra_review}.
The Killing prepotentials were already given in~\eqref{KillingPrepMN1}. We can check this expression by starting from the Killing vectors \eqref{KillingVectHyper} and evaluating~\eqref{KillingPrepGeneral} using a standard parameterisation for the universal hypermultiplet; we have verified that indeed the same result is  obtained. 
Then \eqref{ScalPotGeneralFinal} gives for the scalar potential
\begin{align}\label{eq:ScalPotMN1}
\mathcal{V} \,&=\, \frac{1}{R^2}\,\bigg\{ \frac{\rme^{4\varphi}}{4\Sigma^4} - \frac{2\,H^4 \,\rme^{2\varphi}}{\Sigma}  + \Sigma^2 \Big[ -2  + \rme^{2\varphi} \left( 2\big((H^4)^2-1\big)\big(\theta_1^2+\theta_2^2\big) - \kappa  \right) \nn\\[1mm]
\,&\ \quad\qquad + \frac{1}{8}\,\rme^{4\varphi}\big(2 (H^4)^2-1\big) \big(2\theta_1^2+2\theta_2^2-\kappa\big)^2 \Big]  \bigg\}\,.
\end{align}

The supersymmetric AdS vacuum conditions summarised in Eq.~\eqref{vanishing_shifts} are easily solved and give  the scalar field values
\be\label{values_scalars_MN1_from5d}
H^1=H^2=H^3= \theta_1=\theta_2=0\,,\quad H^4= \Sigma=1\,, \quad \varphi=\tfrac{1}{2}\log\tfrac{4}{3}\,,
\ee
that is precisely the values~\eqref{values_scalars_MN1} that reproduce the MN1 solution reviewed in Section~\ref{MNsol}. The negative curvature $\kappa = -1$ for the Riemann surface arises as a positivity condition for the scalars $\Sigma$ and $\rme^{2\varphi}$. The critical value of the scalar potential yields the cosmological constant $\Lambda \equiv \mathcal{V} = - \frac{8}{3R^2}$, corresponding to an AdS$_5$ radius $\ell = \frac{3}{2}R$, again in harmony with the solution in Section~\ref{MNsol}.

By extremising the scalar potential \eqref{eq:ScalPotMN1} we can search for further AdS$_5$ vacua within our truncation. Then, by analysing the mass matrix of the scalar field fluctuations around the extrema we can test their perturbative stability. In the following we discuss the outcome of this analysis for the three extrema that we have found. 
\begin{itemize}
\item  We recover the supersymmetric vacuum~\eqref{values_scalars_MN1_from5d}. Being supersymmetric, this is stable. The supergravity field fluctuations source $\SU(2,2|1)$ superconformal multiplets in the dual $\mathcal{N}=1$ SCFT \cite{Benini:2009mz}, with the supergravity mass eigenvalues providing the conformal dimension $\Delta$ of the operators in the multiplets. The field fluctuations that were also considered in~\cite{Faedo:2019cvr} correspond to the energy-momentum multiplet (containing the energy-momentum tensor with $\Delta=4$ and the R-current with $\Delta=3$) and to a long vector multiplet of conformal dimension $\Delta=1+\sqrt{7}$ (see~\cite{Faedo:2019cvr} for more details). The additional $\SO(3)$ vector multiplet included in this paper sources a conserved $\SO(3)$ flavour current multiplet in the dual SCFT. The three scalar operators in this multiplet have conformal dimension $\Delta= 2$ (once) and $\Delta = 4$ (twice), while the $\SO(3)$ flavour current has conformal dimension $\Delta=3$, as required for a conserved current.
Another piece of information about the dual SCFT is given by the Weyl anomaly coefficients; these are obtained from the five-dimensional Newton constant $G_5$ and the AdS$_5$ radius $\ell$ through the formula $a = c = \frac{\pi \ell^3}{8 G_5}$.

\item When $\kappa=-1$ we also recover the non-supersymmetric vacuum discussed in~\cite{Faedo:2019cvr}, that was originally found in \cite{Gauntlett:2002rv}. The analysis of the scalar mass matrix shows that the fluctuation of $H^4$ has a mass squared $m^2 \ell^2 \simeq -4.46 $, which is below the Breitenlohner--Freedman bound $\ell^2 m_{\rm BF}^2= -4$. We thus establish that this vacuum is perturbatively unstable. Note that the unstable mode lies outside the truncation of~\cite{Faedo:2019cvr}.

\item For $\kappa = +1$, we find a non-supersymmetric vacuum with non zero value of the $H$-scalars, given by
\be
\Sigma = \frac{2^{1/3}}{5^{1/6}}\,,\qquad \rme^{2\varphi} = \frac{8}{3}\,,\qquad H^4 = \frac{3\sqrt{5}}{4}\,,\qquad\theta_1 = \theta_2 = 0\,,\qquad \ell = 3\,\frac{ 2^{1/6}}{5^{5/6}}R\,,
\ee
where $\ell$ is the AdS radius. This appears to be a new solution. It represents an $\SO(3)$ worth of vacua really, since the scalars $H^\alpha$, $\alpha=1,2,3$, can take any value such that $\sqrt{(H^1)^2+(H^2)^2+(H^3)^2} = \sqrt{(H^4)^2 -1} = \frac{\sqrt{29}}{4}$.
We find that a linear combination of the fluctuations of $\Sigma,\varphi$ and $H^4$ has mass squared $m^2\ell^2\simeq -5.86 < m^2_{\rm BF}\ell^2$, hence this vacuum is perturbatively unstable. Nevertheless, it allowed us to perform a non-trivial check of our truncation ansatz for non-vanishing $H$-fields, as we have verified that its uplift does satisfy the equations of motion of eleven-dimensional supergravity.
\end{itemize}

\section{Truncations for more general wrapped M5-branes}\label{sec:BBBWsection}

The $\mathcal{N}=2$ and $\mathcal{N}=4$ Maldacena--Nu\~nez solutions are special cases of an infinite family of  $\mathcal{N}=2$ solutions~\cite{Bah:2011vv, Bah:2012dg},\footnote{See also~\cite{Cucu:2003bm}, where a subset of the solutions was previously found.}
describing M5-branes wrapping a Riemann surface in a Calabi--Yau geometry.
These solutions, which we will denote as BBBW solutions, have the same general features of the MN1 solution. In particular, they all admit a generalised $U(1)_S$ structure, which we use to derive the most general consistent truncation to $\mathcal{N}=2$ gauged supergravity in five dimensions associated with such backgrounds. As  we will see, the truncated theory 
has two vector multiplets, one hypermultiplet and gauge group $\U(1) \times \mathbb{R}$. It generalises the  $\U(1)_{\rm right}$ invariant subtruncation of the truncation presented in the previous section: the matter content is the same and the gauging is deformed by one (discrete) parameter. Our systematic approach allows us to complete the consistent truncation derived from seven-dimensional maximal $\SO(5)$ supergravity on $\mathbb{\Sigma}$ previously presented in \cite{Szepietowski:2012tb} by including all scalar fields in the hypermultiplet and directly deriving the gauging.\footnote{We thank Nikolay Bobev and Alberto Zaffaroni for pointing out this reference.}

\subsection{The BBBW solutions}\label{sec:BBBW_sol}

The BBBW solutions describe M5-branes wrapped on a Riemann surface $\mathbb{\Sigma}$, such that the $(2,0)$ theory on the branes has a twisting over $\mathbb{\Sigma}$ depending on two integer parameters $p$ and $q$. 
The way the Riemann surface is embedded in the ambient space determines the local structure of the latter. 
The authors of \cite{Bah:2011vv,Bah:2012dg} showed that there is an infinite family of allowed geometries, 
corresponding to  the fibration $\mathcal{L}_1 \oplus \mathcal{L}_2  \hookrightarrow \mathbb{\Sigma}$ 
 of two complex line bundles over the Riemann surface, so that the total space is Calabi--Yau. 
 The degrees of these line bundles are identified with the integers that parameterise the twist of the M5 world-volume theory, $p= {\rm deg}\, \mathcal{L}_1$ and   $q= {\rm deg} \,\mathcal{L}_2$.
By the Calabi--Yau condition $p$ and $q$ must  satisfy $p+q = 2g - 2$, with $g$ the genus of $\mathbb{\Sigma}$.  
 In this setup, the $\mathcal{N}=1$ and $\mathcal{N}=2$ twistings considered in~\cite{Maldacena:2000mw} arise from setting $p=q$ and $q=0$ (or $p=0$), respectively. 
 
The corresponding AdS$_5 \times_{\rm w}M$ supergravity solutions are generalisation of the MN1 solution reviewed in Section~\ref{MNsol}.  
The  eleven-dimensional metric is a warped product 
\be
\hat{g} = \rme^{2 \Delta} g_{{\rm AdS}_5} + g_6  \, ,
\ee
with warp factor
\be
 \rme^{2 \Delta}\, \ell^2 \,=\,  \rme^{2 f_0} \bar{\Delta}^{1/3}  \, .
\ee
where $\ell$ is the AdS radius.
The six-dimensional manifold $M$ is still a fibration of a squashed four-sphere over the Riemann surface, with metric
\be
g_6 =    \bar{\Delta}^{1/3}  \rme^{2 g_0}  g_{\mathbb{\Sigma}}  + \tfrac{1}{4} \, \bar{\Delta}^{-2/3}  g_{4}   \, , 
\ee
where the Riemann surface metric $g_{\mathbb{\Sigma}}$ satisfies~\eqref{gSigmavolSigma}, \eqref{d_frame_Sigma},
 and the metric on the squashed and fibered $S^4$ is
\begin{align}
\label{4dmet}  
g_{4} =  X_0^{-1} \dd \mu_0^2 + \sum_{i=1,2} X^{-1}_i  \,\big(\dd \mu_i^2 + \mu_i^2 (\dd \varphi_i  + A^{(i)})^2\big) \, . 
\end{align} 
The angles $\varphi_1,\varphi_2$ vary in  $[0, 2 \pi]$,\footnote{They are related to the angles of Section \ref{MNsol} by $\varphi_1 = - ( \phi + \psi)/2$ and  $\varphi_2 =   (\phi - \psi)/2$.} and 
\be
\mu_0 = \cos \zeta \, , \qquad  \mu_1 = \sin \zeta \cos \tfrac{\theta}{2}   \, , \qquad  \mu_2 = \sin \zeta \sin   \tfrac{\theta}{2} \, , 
\ee
with $\zeta, \theta \in [0,\pi]$.  The  two circles $\varphi_1$ and $\varphi_2$ are independently fibered over the Riemann surface, with connections 
\be
 A^{(1)} =  -\frac{1+z}{2}\, \upsilon \qquad   A^{(2)} =  -\frac{1-z}{2}\, \upsilon \, ,
\ee
where $\upsilon$ is again the connection on $\mathbb{\Sigma}$ and the discrete parameter $z$ is related to the integers $p$ and $q$ as  
\be\label{defz_param}
z \,=\, \frac{p- q}{p+q}\,.
\ee
The warping function $\bar{\Delta}$ and the constants $f_0$, $g_0$ depend on $z$ and on the curvature $\kappa$  of the Riemann surface as
\be
 \bar{\Delta} = \sum_{I = 0}^2 X_I \mu_I^2\,, \qquad 
\rme^{f_0} =  X_0^{-1}\,,  \qquad 
\rme^{2g_0}  = -\tfrac{1}{8}\,\kappa \, X_1 X_2\, [(1-z)  X_1 + (1+z)  X_2 ]   \, ,
\ee
with
\be
\begin{aligned}
& X_0 =  ( X_1 X_2)^{-2}\,, \\[1mm]
& X_1 X_2^{-1} = \frac{1+z}{2 z -\kappa  \sqrt{1+3 z^2}}\,, \\[1mm]
& X^5_1 = \frac{1 + 7 z + 7 z^2 + 33 z^3 +\kappa  (1 + 4 z + 19 z^2) \sqrt{1 + 3 z^2}}{4 z (1-z)^2} \,.
\end{aligned}
\ee
The four-form flux is given by  
\begin{align}
\hat F =&  - \frac{1}{4} \, \bar\Delta^{-5/2}  \,\Big[ \sum_{I=0}^2(X_I^2\mu_I^2- \bar\Delta X_I)  + 2   \bar\Delta X_0 \Big] \vol_{4} \nn \\
& +  \frac{1}{16} \bar\Delta^{-1/2} \sum_{i=1}^2X_i^{-2}  \ast_{4} \big[\dd(\mu_i^2)\wedge(\dd \varphi_i+A^{(i)}) \big] \wedge \dd A^{(i)}\,,
\end{align}
where the Hodge star $*_4$ is computed using the metric \eqref{4dmet}. 

The solution has two $U(1)$ isometries corresponding to shifts of the angles $\varphi_1,\varphi_2$ that parameterise the two diagonal combinations of the $U(1)_{\rm right}$ and  $U(1)_{\rm left}$ subgroups of $SO(5)$.  It turns out that neither of them corresponds to the superconformal R-symmetry of the dual $\mathcal{N}=1$ SCFT, which is given by a linear combination involving $X_1,X_2$~\cite{Bah:2011vv, Bah:2012dg}.

\subsection{Generalised $\U(1)_S$ structure}

The construction of the generalised structure associated to the BBBW solutions follows the same lines as for the MN1 solution. 
We first embed the ordinary $U(1)$ structure  in $E_{6(6)}$ and then look for the invariant generalised tensors. 
The  generalised $U(1)_S$ structure of the solutions is determined by the topological twist of the M5 world-volume theory, as a linear combination of the   $U(1)$   holonomy of $\mathbb{\Sigma}$ and the 
  $U(1)_{\rm  right}$ and  $U(1)_{\rm  left}$  subgroups of the $SO(5)$ R-symmetry group 
\be
\label{BBBWU1a}
U(1)_S \,\sim \,  U(1)_\mathbb{\Sigma} -   U(1)_{\rm right} - z\, U(1)_{\rm left} \, . 
\ee
This embeds in  $E_{6(6)}$ as an element  of its compact subgroup  $\USp(8)$ with generator
\be 
\label{u1sgen}
\mathfrak{u}(1)_S \,=\, \ii\,\hat\Gamma_{56} -\frac{\ii}{p+q} \big(p \,\hat\Gamma_{12} - q\,\hat\Gamma_{34}\big) \,,
\ee
where  $\hat\Gamma_{56}$ is the $\mathfrak{usp}_8$ element generating  $U(1)_\mathbb{\Sigma}$ and $\frac{1}{2}(\hat\Gamma_{12} \pm \hat\Gamma_{34})$ generate  $U(1)_{\rm left/right}$.  When
$p=q$ we recover the $\U(1)_S$ structure group of the MN1 solution, whereas   $q=0$  (or  $p=0$) gives the MN2 structure considered in \cite{Cassani:2019vcl}. Below we assume that $p,q$ are generic, and do not fulfill these special conditions which as we have seen lead to a larger truncation.

By looking at the singlets under  $\mathfrak{u}(1)_S$ in the $\rep{27}$ and $\rep{78}$ representations of  $E_{6(6)}$, we find that the  $\U(1)_S$ structure is defined by  eight $J_A$, $A=1,\ldots,8$, in the adjoint bundle and three generalised vectors $K_I$, $I=0,1,2$.
The singlets in the adjoint bundle have the  same form \eqref{sec:Jsing} as for the  MN1 solution, while the 
three singlet generalised vectors take the same form as a subset of the MN1 generalised vectors,\footnote{Before acting with $\Upsilon$, the singlets for the BBBW solutions are related to those used for the MN1 solutions as
\begin{equation}
K_0=K^{\rm MN1}_0\,, \quad  K_1=K^{\rm MN1}_3\,,\quad  K_2=K^{\rm MN1}_4\,, \nn
\end{equation}
and to the structure of the MN2 solution in \cite{Cassani:2019vcl} as 
\be\nn
K_0 = \tfrac{1}{2}(K^{\rm MN2}_5- K^{\rm MN2}_8)\,,\quad 
\begin{array}{l}
K_1 = K^{\rm MN2}_0+\tfrac{1}{2}(K^{\rm MN2}_5+K^{\rm MN2}_8)\,,\\
K_2 = K^{\rm MN2}_0-\tfrac{1}{2}(K^{\rm MN2}_5+K^{\rm MN2}_8)\, .
\end{array}
\ee
}
\be
\begin{aligned}
K_0 \,&=\,  \rme^\Upsilon \cdot (R^2  \vol_{\mathbb\Sigma} \wedge E^\prime_5) \, , \\
K_1 \,&=\,  \rme^\Upsilon \cdot\ti \Xi_3  \, , \\
K_2\, &= \, \rme^\Upsilon \cdot\Xi_3\, . \\
\end{aligned}
\ee
However now the twisting element $\Upsilon$ has a more general form dictated by the embedding \eqref{u1sgen}, that is
\be
\label{twistel_BBBW}
\Upsilon \,=\, - \frac{R}{p+q}\, \upsilon \times_{\adj}(p\, E_{12}-q \, E_{34})\,. 
\ee
This makes our generalised tensors globally well-defined. We emphasise that these depend on the integers $p,q$ only through \eqref{twistel_BBBW}.

\subsection{Features of the truncation} 

The number of $U(1)_S$ singlets in the $\rep{27}$ and $\rep{78}$ implies that the truncated  supergravity theory contains two vector multiplets and one hypermultiplet. 
The H structure moduli space is the same as for the MN1 case,
\be
\mathcal{M}_{\rm H} \,=\, \frac{\SU(2,1)}{\SUH\times U(1)}\,.
\ee 
As before, this is parameterised by real coordinates $q^X= \{\varphi,\xi,\theta_1,\theta_2\}$ and the metric is given by Eq.~\eqref{Quaternionic_metric}.
The V structure moduli space is determined again following our discussion in Section~\ref{sec:genN2}, and is a subspace of the one for the MN1 truncation. Evaluating the cubic invariant on the singlets $K_I$ as in \eqref{eq:CKKK_cond}, we obtain that the non-zero components of the $C_{IJK}$ tensor are
\begin{equation}
C_{0IJ} = C_{I0J} = C_{IJ0}= \,\tfrac{1}{3}\, \eta_{IJ}\,,\qquad \text{for}\ I,J=1,2\,,
\end{equation}
with $\eta = {\rm diag}(-1,1)\,.$
Parameterising the V structure moduli as in \eqref{from_h_to_H}, with $I=1,2$, the constraint~\eqref{constraint_Chhh} gives the equation of the unit hyperboloid  $SO(1,1)$, 
\be
-(H^1)^2 + (H^2)^2 =1   \, ,
\ee 
while again $\Sigma$ parameterises $\mathbb{R}^+$.
Thus the V structure moduli space is 
\be
\mathcal{M}_{\rm V} \,=\, \mathbb{R}^+\times \SO(1,1)\,.
\ee 
The kinetic matrix $a_{IJ}$ then takes the same form \eqref{kin_matr_MN1}, that is
\begin{align}
a_{00}&= \tfrac{1}{3}\, \Sigma^{4}\,, \nn\\[1mm]
a_{01}&= a_{02}=0  \,,\nn\\[1mm] 
a_{IJ}&= \tfrac{2}{3}\,\Sigma^{-2}  \begin{pmatrix} 2(H^1)^2+ 1 & -2H^1H^2 \\[1mm] -2H^1H^2 & 2(H^2)^2- 1\end{pmatrix} \,,\qquad I,J =1,2 \,.
\end{align}

The gauging of the reduced theory is obtained from the generalised Lie derivative $L_{K_I}$ acting on the  $K_J$ and the $J_A$. The Lie derivatives among vectors are now trivial,
\be
\label{BLievec} 
L_{K_I} K_J = 0 \,,\qquad   I,J = 0,1,2 \, . 
\ee
As discussed in~Section~\ref{sec:genN2}, the Lie derivatives $L_{K_I}J_A$ are conveniently expressed as the adjoint action of 
$SU(2,1)$ generators,
\be
\label{BLieJ}
L_{K_0} J_A = [J_{(K_0)}, J_A]\,,\qquad L_{K_1} J_A = [J_{(K_1)}, J_A]\,,\qquad L_{K_2} J_A = [J_{(K_2)}, J_A]\,.
\ee
Evaluating the generalised Lie derivatives we find
\begin{align}\label{generators_gauging_BBBW}
J_{(K_0)} \,&=\, \tfrac{1}{4R}\,\big(J_3 + 2J_7 - \sqrt{3} J_8 \big)\,,\nn\\[1mm]
J_{(K_1)} \,&=\, \tfrac{1}{4R}\,\kappa\,z\,\big(J_3 + 2J_7 - \sqrt{3} J_8 \big)\,,\nn\\[1mm]
J_{(K_2)} &= -\tfrac{1}{4R}\,\kappa\,\big(J_3 + 2J_7 - \sqrt{3} J_8 \big)   -  \tfrac{1}{R}\,\big(  J_3 + \tfrac{1}{\sqrt3} J_8  \big)\,.
\end{align} 
Eq.~\eqref{BLievec} implies that the vector multiplet sector is not gauged, so the field strengths are all abelian, 
\be
\mathcal{F}^I = \dd\mathcal{A}^I\,,
\ee while~\eqref{generators_gauging_BBBW} specifies the gauging in the hypermultiplet sector in terms of $\kappa$ and $z$. The $\SU(2,1)$ generators act as isometries on $\mathcal{M}_{\rm H}$; the corresponding Killing vectors can again be computed using \eqref{formula_for_k^X} and read
\begin{align}\label{KillingVectHyperBBBW}
k_0 \,&=\, \partial_\xi\,,\nn\\[1mm]
k_1 \,&=\, \kappa\,z\,\partial_\xi\,,\nn\\[1mm]
k_2 \,&=\, - \kappa\, \partial_\xi + 2\left(\theta_2 \partial_{\theta_1} - \theta_1 \partial_{\theta_2} \right) \,.
\end{align}
It follows that the covariant derivatives of the charged scalars are
 \begin{align}
 \mathcal{D}(\theta_1 + \ii\, \theta_2) \,&=\, \dd (\theta_1 + \ii\, \theta_2) - \tfrac{2}{R}\,\ii  \mathcal{A}^2\,(\theta_1 +\ii\,\theta_2)\,,\nn\\[1mm]
 \mathcal{D} \xi \,&=\, \dd \xi + \tfrac{1}{R}\, \mathcal{A}^0 + \tfrac{1}{R}\,\kappa  \left(z \mathcal{A}^1  - \mathcal{A}^2 \right)\,,
 \end{align}
where again the inverse $S^4$ radius $\frac{1}{R}$ plays the role of the gauge coupling constant.
The Killing prepotentials can be computed either from \eqref{KillingPrep_from_mommaps} or from \eqref{KillingPrepGeneral}, and read 
\begin{align}\label{KillingPrepBBBW}
P^\alpha_0 \,& =\, \big\{0\,,\,0\,,\,\tfrac{1}{4}\,\rme^{2\varphi} \big\} \,,\nn\\[1mm]
P^\alpha_1 \,& =\, \big\{0\,,\,0\,,\,\tfrac{1}{4}\,\kappa\,z\,\rme^{2\varphi} \big\} \,,\nn\\[1mm]
P^\alpha_2 \,& =\, \big\{\sqrt{2}\,\rme^{\varphi}\theta_1\,,\,\sqrt{2}\,\rme^{\varphi}\theta_2\,,-1+\tfrac{1}{4}\,\rme^{2\varphi}\, \big(2\theta_1^2 + 2\theta_2^2-\kappa\big)\,    \big\}\,.
\end{align}
Notice that for $z=0$ (that is $p=q$), the quantities above reduce to those obtained for the MN1 structure in Section \ref{sec:gauging_MN1_trun}. 

The five-dimensional bosonic action is then determined to be
\begin{align} 
S \, &= \, \frac{1}{16\pi G_5} \int \, \Big[\left( \mathcal{R} -2\mathcal{V}\right)*1  -  \tfrac{1}{2}\,\Sigma^{4} \mathcal{F}^{0}\wedge*  \mathcal{F}^{0} -  \tfrac{3}{2}\sum_{I,J=1}^2 a_{IJ} \mathcal{F}^{I}\wedge*  \mathcal{F}^{J} - 2\Sigma^{-2}  \dd \Sigma \wedge* \dd \Sigma  \nn\\[-1mm]
&\quad   - \tfrac{3}{2}\sum_{I,J=1}^2 a_{IJ} \, \dd (\Sigma H^I) \wedge* \dd (\Sigma H^J)  -   g_{XY}  \mathcal{D} q^{X} \wedge * \mathcal{D} q^{Y}  -  \mathcal{A}^0\wedge \big(\mathcal{F}^1\wedge \mathcal{F}^1 - \mathcal{F}^2\wedge \mathcal{F}^2 \big ) \Big] \, ,
\end{align}
where the scalar potential reads
\begin{align}
\mathcal{V} \,&=\, \frac{1}{R^2}\,\bigg\{ \frac{\rme^{4\varphi}}{4\Sigma^4} - \frac{2 \,\rme^{2\varphi}H^2}{\Sigma}  + \Sigma^2 \Big[ -2  + \rme^{2\varphi} \left( 2(H^1)^2\big(\theta_1^2+\theta_2^2\big) - \kappa  \right) \nn\\[1mm]
\,&\ \quad\qquad + \frac{1}{8}\,\rme^{4\varphi}\big((H^1)^2 + (H^2)^2\big) \big(2\theta_1^2+2\theta_2^2-\kappa\big)^2   \nn\\[1mm]
\,&\ \quad\qquad  +  z\,\kappa \left( z\kappa \, (H^1)^2 + z\kappa\, (H^2)^2  +4H^1H^2 \big(2\theta_1^2 + \theta_2^2 - \kappa \big) \right) \Big]  \bigg\}\,.
\end{align}

It is straightforward to analyse the supersymmetric AdS$_5$ vacuum conditions~\eqref{vanishing_shifts}. The  hyperino equation gives
\begin{align}\label{hyperino_BBBW}
&\theta_1 = \theta_2 = 0\,,\nn\\
&\Sigma^{-3} = \kappa\left(  z H^1 - H^2 \right) \,,
\end{align}
where we assume $\kappa = \pm1$ (hence leaving aside the case $\kappa = 0$).
The gaugino equation gives
\begin{align}
&2 \Sigma^{-3} \,P_0^\alpha + H^1 P^\alpha_1 + H^2 P^\alpha_2 = 0\,,\nn\\[1mm]
&H^2 P^\alpha_1 + H^1 P^\alpha_2 = 0\,.
\end{align}
Plugging the Killing prepotentials \eqref{KillingPrepBBBW} and using \eqref{hyperino_BBBW} we obtain
\begin{align}
 3\kappa \,\rme^{2\varphi}\left( z H^1 - H^2 \right)  - 4H^2  \,=\, 0\,,\nn\\[1mm]
\kappa \,\rme^{2\varphi} (z\, H^2 - H^1) - 4H^1 \,=\,0\,.
\end{align}
 Taking into account the allowed range of the scalar fields, 
 the solution to these equations is
\be
\frac{H^1}{H^2} = \frac{1+\kappa\,\sqrt{1+3z^2}}{3z}\,,\qquad \rme^{2\varphi} = \frac{4}{\sqrt{1+3z^2}-2\kappa}\,.
\ee
For $\kappa=1$, well-definiteness of the fields requires $|z|>1$, as in \cite{Bah:2012dg}, while $z$ can be generic for $\kappa=-1$. The MN1 case $z= 0$ is recovered as a limiting case after fixing $\kappa=-1$. The critical value of the scalar potential determines the AdS radius $\ell$ as
\be
\ell \,=\, \bigg(\frac{\kappa - 9\kappa z^2 + (1+3z^2)^{3/2}}{4z^2}\bigg)^{1/3}\,R\,. 
\ee

Although we do not present the uplift formulae for this truncation, we have checked that the supersymmetric vacuum identified above matches the BBBW solution summarised in Section~\ref{sec:BBBW_sol}. To do so, we have computed the inverse generalised metric $G^{-1}$ associated with the $\U(1)_S$ structure under consideration; this depends on the V structure and H structure parameters. From the generalised metric we have reconstructed the ordinary metric $g_6$ and the three-form potential on $M$, as well as the warp factor $\rme^{2\Delta}$. Substituting the values for the scalars found above, we find agreement with the solution in Section~\ref{sec:BBBW_sol} upon fixing the $S^4$ radius as $R = \frac{1}{2}$ and implementing the following dictionary:
\begin{equation}\begin{gathered}
\rme^{2\varphi} = \tfrac{1}{4}\,\rme^{-2g_0-\frac{1}{2}f_0} \,,\\
\Sigma^3\,=\,  \tfrac{1}{4}\,\rme^{-2g_0+\frac{3}{4}f_0} \,, \\
H^1\,=\,\tfrac{1}{2}X_0^{\frac{1}{4}}(X_1-X_2)\,, \\
H^2\,=\,\tfrac{1}{2}X_0^{\frac{1}{4}}(X_1+X_2)\,,
\end{gathered}\end{equation}
with our AdS radius being given in terms of the quantities appearing there as
\be
\ell \,=\, 2^{2/3}\,\rme^{f_0 + \frac{2}{3}g_0} R\,.
\ee

By extremising the scalar potential\footnote{To do so, it is convenient to parameterise $H^1 = \sinh\alpha$, $H^2 = \cosh\alpha$, and extremise with respect to $\alpha$.} we recover the supersymmetric vacuum and also find new non-supersymmetric vacua, where the scalar field values are rather complicated functions of the parameter $z$. As an example, we give the numerical values for one chosen value of $z$, that we take $z=\frac{1}{2}$. When $\kappa=-1$ we find a new extremum of the potential at
\be
\Sigma \simeq 0.9388\,,\quad \varphi \simeq 0.1109\,,\quad H^2 \simeq 1.0217\,,\quad \theta_1 = \theta_2 = 0\,,\qquad \ell \simeq 1.5276\, R\,,
\ee
while when $\kappa=1$ we find an extremum at
\be
\Sigma \simeq 0.8631\,,\quad \varphi \simeq 0.2812\,,\quad H^2 \simeq 1.5506\,,\quad \theta_1 = \theta_2 = 0\,,\qquad \ell \simeq 1.0644\, R\,,
\ee
and another one at
\be
\Sigma \simeq 1.1580\,,\quad \varphi \simeq 0.8455\,,\quad H^2 \simeq 1.9847\,,\quad \theta_1 = \theta_2 = 0\,,\qquad \ell \simeq 0.6198\, R\,,
\ee
where for each solution we have also indicated the corresponding AdS radius $\ell$.

\section{Conclusions}\label{sec:Conclusions}

In this paper we have illustrated the Exceptional Generalised Geometry approach to $\mathcal{N}=2$ consistent truncations of eleven-dimensional supergravity on a six-dimensional manifold~$M$. We have argued that for the truncation to go through, $M$ must admit a generalised \hbox{$G_S\subseteq \USp(6)$} structure with constant singlet intrinsic torsion, and we have explained how this completely determines the resulting five-dimensional supergravity theory. We have also given an algorithm to construct the full bosonic truncation ansatz. This formalism provides a geometric understanding of the origin of the truncations, in particular those that are not based on invariants of conventional $G$-structures on the tangent bundle. It also sidesteps the need to reduce the equations of motion in order to uncover the matter content and couplings of the truncated theory.

The main technically involved part of this formalism is deriving the truncation ansatz. However a significant advantage is that once this is done the relevant expressions can be used to derive the uplift formulae for any $\mathcal{N}=2$ consistent truncation. One does not have postulate the set of consistent modes case-by-case. Furthermore the structure of the resulting gauged supergravity is then simply determined by the generalised structure. 

To demonstrate the concrete effectiveness of the formalism we worked out the full bosonic truncation ansatz on Maldacena--Nu\~nez geometries, leading to five-dimensional $\mathcal{N}=2$ supergravity with four vector multiplets, one hypermultiplet and a non-abelian gauging, having the \hbox{$\mathcal{N}=2$} AdS$_5$ solution of \cite{Maldacena:2000mw} as a vacuum solution. This extends the truncation of~\cite{Faedo:2019cvr} by $\SO(3)$ vector multiplets. For the BBBW geometries \cite{Bah:2011vv,Bah:2012dg}, we obtained a truncation featuring two vector multiplets, one hypermultiplet and an abelian gauging, completing the truncation obtained in \cite{Szepietowski:2012tb}. This can be seen as a one-parameter deformation of the truncation obtained from the one on Maldacena--Nu\~nez geometry by imposing invariance under the Cartan of $\SO(3)$. Although in this case we did not give all details of the truncation ansatz, it should be clear that it can be obtained by following precisely the same steps presented for the case of Maldacena--Nu\~nez geometry. Since the generalised geometry tensors on $S^4$ used in these trucations are a subset of those appearing in the reduction of eleven-dimensional supergravity to maximal $\SO(5)$ supergravity in seven dimensions, it should also be clear that our consistent truncations can equivalently be obtained as truncations of maximal $\SO(5)$ supergravity on a Riemann surface.

Together with the half-maximal truncation presented in \cite{Cheung:2019pge,Cassani:2019vcl}, which is based on the $\mathcal{N}=4$ solution of \cite{Maldacena:2000mw}, this work provides the largest possible consistent truncations of eleven-dimensional supergravity that have as seed known AdS$_5\times_{\rm w}M$ supersymmetric solutions describing M5-branes wrapped on a Riemann surface (larger truncations may be possible by including degrees of freedom that go beyond eleven-dimensional supergravity, such as membrane degrees of freedom).

It would be interesting to explore further the relatively simple five-dimensional supergravity models obtained in this paper and construct new solutions thereof. These would have an automatic uplift to eleven dimensions, and may have an interpretation in the dual SCFT. For the subtruncation with no $\SO(3)$ vector multiplet, solutions of holographic interest have been discussed in~\cite{Faedo:2019cvr}. Our larger consistent truncation may offer the possibility to obtain solutions where non-abelian gauge fields are activated, which are quite rare in holography. For instance, constructing a supersymmetric, asymptotically AdS$_5$ black hole with non-abelian hair would represent a qualitatively new type of solutions.

It will be natural to adapt our construction to truncations of eleven-dimensional supergravity on a seven-dimensional manifold, leading to four-dimensional gauged $\mathcal{N}=2$ supergravity. This uses $G_S\subseteq\SU(6)$ structures in $\Ex{7}$ generalised geometry, and would allow one to derive new consistent truncations based on the generalised structures underlying the AdS$_4\times_{\rm w} M_7$ solutions of \cite{Gauntlett:2006ux,Gabella:2012rc}, which in terms of ordinary $G$-structures only admit a local $\SU(2)$ structure. The solutions of \cite{Gauntlett:2006ux} are the most general $\mathcal{N}=2$ AdS$_4$ solutions to eleven-dimensional supergravity supported by purely magnetic four-form flux; they represent the near-horizon region of M5-branes wrapping a special lagrangian three-cycle in $M_7$. The solutions of \cite{Gabella:2012rc} have both electric and magnetic flux, and should arise from M2-M5 brane systems.
Analysing the respective generalised structure it will become possible to enhance the truncation to minimal gauged supergravity obtained in \cite{Gauntlett:2007ma} and \cite{Larios:2019lxq} (for the solutions of \cite{Gauntlett:2006ux} and \cite{Gabella:2012rc}, respectively) by adding matter multiplets. One example of this construction has been given in~\cite{Donos:2010ax}.

It will also be useful to extend our formalism to $\mathcal{N}=2$ truncations of type II supergravity. Minimally supersymmetric AdS$_5$ solutions of type IIB and massive type IIA supergravity were classified in~\cite{Gauntlett:2005ww} and~\cite{Apruzzi:2015zna}, respectively. It would be useful to reformulate the classification of explicit solutions in terms of generalised $G_S \subseteq \USp(6)$ structures; this would be a first step towards constructing consistent truncations to five-dimensional supergravity using our approach. One concrete application would be to check if the IIB solution of \cite{Pilch:2000ej}, given by a warped product of AdS$_5$ and a deformed $S^5$, admits a consistent truncation to five-dimensional supergravity including (massive) KK modes that do not belong to the well-known IIB truncation leading to maximal $\SO(6)$ gauged supergravity. This would be somewhat analogous to the IIB consistent truncation on Sasaki--Einstein structures \cite{Cassani:2010uw,Gauntlett:2010vu}, where only a subset of the retained KK modes are also captured by $\SO(6)$ gauged supergravity.

A more challenging generalisation of our formalism would be the one to truncations preserving only $\mathcal{N}=1$ supersymmetry in four dimensions. 
Although a considerable amount of work remains to be done, it should be clear that the generalised structure approach to consistent truncations has the potential to  classify all possible consistent truncations of higher-dimensional supergravity that preserve any given amount of supersymmetry.

\section*{Acknowledgments}

We would like to thank Nikolay Bobev, Gianguido Dall'Agata, Ant\'on Faedo, Gianluca Inverso, Carlos Nu\~nez Chris Rosen and Alberto Zaffaroni for useful discussions and comments. DW is supported in part by the STFC Consolidated Grants ST/P000762/1 and ST/T000791/1. We acknowledge the Mainz Institute for Theoretical Physics (MITP) of the Cluster of Excellence PRISMA+ (Project ID 39083149) for hospitality and support during part of this work.


\appendix

\section{ $\mathbf{E}_{6(6)}$ generalised geometry for M-theory}
\label{PreliminariesE66_Mth}

In this section we briefly recall the main features of  the generalised geometry of M-theory compactifications  on a six-dimensional manifold $M$. 
For a more detailed discussion we refer to \cite{Coimbra:2011ky} and~\cite[App.~E]{Ashmore:2015joa}. 

We use the following conventions for  wedges and contractions among tensors on $M$
\begin{align}
( v \wedge u)^{a_1 \ldots a_{p+p^\prime}} & = \frac{(p+p^\prime)!}{p!\, p^\prime!}\, v^{[a_1 \ldots a_p}  u^{a_{p+1} \ldots a_{p+p^\prime} ]}  , \nn \\
( \lambda \wedge \rho)_{a_1 \ldots a_{q+q^\prime}}  & = \frac{(q+q^\prime)!}{q! \,q^\prime!}\, \lambda_{[a_1 \ldots a_q}  \rho_{a_{q+1} \ldots a_{q+q^\prime} ]} , \nn \\
(v \,\lrcorner\, \lambda)_{a_1 \ldots a_{q-p}}  & = \frac{1}{p!} v^{b_1 \ldots b_p} \lambda_{b_1 \ldots b_p a_1 \ldots a_{q-p}}  \quad {\rm if}\  p \leq q , \nn \\
(v \,\lrcorner\, \lambda)^{a_1 \ldots a_{p-q} } & = \frac{1}{q!} v^{a_1 \ldots a_{p-q}b_1\ldots b_q} \lambda_{b_1 \ldots b_q }  \quad  {\rm if}\  p \geq q , \nn \\
(j v \,\lrcorner\, j  \lambda)^a{}_b   &  = \frac{1}{(p-1)!} v^{a c_1 \ldots c_{p-1}} \lambda_{b  c_1 \ldots c_{p-1}} ,  \nn \\
\left(j \lambda \wedge \rho\right)_{a,\,a_1\ldots a_d} &= \frac{d!}{(q-1)!(d+1-q)!}\,\lambda_{a[a_1\ldots a_{q-1}}\rho_{a_q\ldots a_d]}\ .
\end{align} 
We will denote by  $\cdot$ the $\mathfrak{gl}(6)$ action on tensors:  given a frame $\{ \hat{e}_a \}$ for $TM$ and a co-frame  $\{ e_a \}$ for $T^*M$, $a=1, \ldots, 6$, the action,  for 
instance, on a vector and a two-form is 
\be
\label{app:gl6ac}
(r \cdot  v )^a = r^a{}_b v^b \qquad  \quad ( r \cdot  \omega)_{ab} = - r^c{}_a \omega_{c b} - r^c{}_b \omega_{a c } \, . 
\ee

\vspace{0.2cm}

For M-theory on a six-dimensional manifold we use $E_{6(6)} \times \bbR^+$ generalised geometry.  
 The generalised  tangent bundle $E$ is 
\begin{equation} 
\label{app:gentan}
   E \,\simeq \, TM \oplus \Lambda^2T^*M \oplus \Lambda^5T^*M \, ,
\end{equation}
where, as customary,  we decompose  the various   bundles in representations of 
 $GL(6)$, the geometric  subgroup of  $E_{6(6)}$. 
The sections of $E$, the  generalised vectors, transform in the  ${\bf 27}$ of $E_{6(6)}$ and can be written as 
\begin{equation}
\label{app:genvec}
V = v + \omega + \sigma\,,
\end{equation}
where $v$ is an ordinary vector field, $\omega$ is a two-form and  $\sigma$  is a five-form.\footnote{The generalised tangent bundle $E$  has a non-trivial  structure that  takes into account the non-trivial gauge potentials of M-theory.  
To be more precise the sections of $E$ are defined as 
\be
\label{twist_Mth}
 V \,=\, \rme^{A + \tilde A}  \cdot \check{V} \, , 
\ee 
where $A + \tilde A$ is an element of the adjoint bundle,  $\check{V}  = v + \omega + \sigma$, with $v \in \Gamma(TM)$ are vectors, $\omega\in \Gamma(\Lambda^2 T^*M)$ and $\sigma\in \Gamma(\Lambda^5T^*M)$, and $\cdot$ defines the adjoint action defined in \eqref{adj27}.   The patching condition on the overlaps $U_{\alpha} \cap U_{\beta}$ is 
\be
V_{(\alpha)} \,=\, \rme^{\dd \Lambda_{(\alpha \beta)} + \dd \tilde \Lambda_ {(\alpha \beta)}} \cdot V_{(\beta)} \, , 
\ee
where $\Lambda_{(\alpha \beta)}$ and $ \tilde\Lambda_ {(\alpha \beta)}$ are a two- and five-form, respectively. This corresponds to the gauge-transformation of  the three- and six-form potentials in~\eqref{twist_Mth} as 
\begin{align}
A_{(\alpha)} \,&=\, A_{(\beta)} + \dd \Lambda_{(\alpha \beta)}\ , \nn \\
\tilde A_{(\alpha)} \,&=\, \tilde A_{(\beta)} + \dd \tilde \Lambda_{(\alpha \beta)}  -\frac{1}{2}   \dd \Lambda_{(\alpha \beta)}  \wedge A_{(\beta)} \ .
\end{align}
The respective gauge-invariant field-strengths reproduce the supergravity ones:
\begin{align}
F \,&=\, \dd A \, ,  \nn \\
\tilde F \,&=\, \dd \tilde A - \frac{1}{2} A \wedge F \, .
\end{align}
}

The dual bundle $E^*$ is defined as 
\begin{equation}
   E^* \,\simeq \, T^* M \oplus \Lambda^2T M \oplus \Lambda^5T M \, ,
\end{equation}
with sections
\begin{equation}
Z =  \hat{v} +  \hat{\omega} +  \hat{\sigma} \,,
\end{equation}
where $\hat{v}$ is one-form,  $\hat{\omega}$ is a two-vector and  $ \hat{\sigma}$ is a five-vector.  Generalised vectors and dual generalised vectors have a natural pairing given by
\be
\label{pairing_vector_dualvector}
\GM{Z}{V} = \hat v_m v^m + \tfrac{1}{2}\,  \hat\omega^{mn} \omega_{m n} +  \tfrac{1}{5!}\, \hat\sigma^{mnpqr}\sigma_{mnpqr}  \, . 
\ee

We will also need the bundle  $N\simeq \det T^*M \otimes E^*$. In terms of $\GL(6)$ tensors, $N$ decomposes as 
\be
N \simeq T^*M \oplus \Lambda^4T^*M \oplus  (T^*M \otimes  \Lambda^6 T^*M )  \,,
\ee
and correspondingly its sections $Z_{\flat}$ decompose as
\be
Z_{\flat}  = \lambda  +   \rho  + \tau  \, . 
\ee
The bundle $N$ is  obtained from the symmetric product of two generalised vectors via the  map $\otimes_{N}: E \otimes E \to N$  with
\be
\begin{aligned}
\label{N'prod_Mth}
\lambda \,&= \, v \,\lrcorner\, \omega' + v'\,\lrcorner\,\omega\,,  \\
\rho \,&=\, v \,\lrcorner\, \sigma' + v' \,\lrcorner\, \sigma - \omega \wedge \omega' \,,\\
\tau \,&=\,    j \omega \wedge \sigma' + j \omega' \wedge \sigma   \,.
\end{aligned}
\ee

The $E_{6(6)}$ cubic invariant is defined  on $E$ and $E^*$as\footnote{This is 6 times the cubic invariant given in \cite{Ashmore:2015joa}. Because of this, we introduced a compensating factor of 6 in the formulae \eqref{eq:cKKK} and \eqref{compJK}.}
\begin{align}
c(V,V,V)  &= - \, 6\,\iota_v\, \omega \wedge \sigma -  \omega\wedge \omega\wedge\omega \, , \nn\\
c^*(Z,Z,Z) &= - \, 6\,\iota_{\hat v}\,\hat \omega \wedge \hat\sigma -  \hat\omega\wedge \hat\omega\wedge\hat\omega\,.
\end{align}

The adjoint bundle is defined as 
\be
{\adj} F \,\simeq \, \bbR \oplus (TM \otimes T^* M) \oplus \Lambda^3 T^* M \oplus \Lambda^6 T^* M  \oplus \Lambda^3 T M \oplus \Lambda^6 T  M  \, , 
\ee
with sections
\be
\label{sec78}
R = l + r + a + \tilde a + \alpha + \tilde \alpha \, ,
\ee
where locally $l \in \bbR$,  $r \in \End(TM)$, $a \in   \Lambda^3 T^* M$, etc.  The  $\mathfrak{e}_{d(d)}$ sub-algebra 
is obtained by fixing the factor $l$  in terms of the trace of  $r$ as $ l =  \frac{1}{3}  \tr r$. This choice 
 fixes the weight of the generalised tensors under the $\bbR^+$ factor. In particular it implies that a scalar of weight $k$ is a section of 
$(\det T^* M)^{k/3}$: $ \mathbb{1}_k \in \Gamma ((\det T^* M)^{k/3})$. 

It is also useful to introduce the weighted adjoint bundle 
\be 
(\det T^*M) \otimes {\adj} \,  F    \,\supset\,  \bbR  \oplus  \Lambda^3 T^* M    \oplus (TM \otimes \Lambda^5 T  M)   \, , 
\ee
whose sections are  locally given by the sum 
\be
R_{\flat} =   \tilde{ \phi} + \phi     + \psi    \, , 
\ee
where $ \tilde{ \phi} $, $\phi$   and $\psi$  are obtained from  the adjoint elements $r \in TM \otimes T^*M$,  $\alpha \in  \Lambda^3 T M$,  $\tilde \alpha \in  \Lambda^3 T M$ as 
\be
\tilde  \phi  = \tilde \alpha \lrcorner {\rm vol}_6  \qquad 
 \phi   =  \alpha \lrcorner {\rm vol}_6  \qquad 
  \psi    = r  \cdot {\rm vol}_6   \, , 
\ee
where $\vol_6$ is a reference volume form.

The action of an adjoint element $R$ on another adjoint element $R^\prime$ is given by the commutator, $R^{\prime \prime} = [ R , R^\prime]$. In components, $R^{\prime \prime}$ reads 
\be
\label{adjcomm}
\begin{aligned}
l^{\prime \prime} & = \tfrac{1}{3} (\alpha\, \lrcorner \,a^\prime - \alpha^\prime \,\lrcorner\, a )  +\tfrac{2}{3}(\tilde{\alpha}'\,\lrcorner\, \tilde{a} - \tilde{\alpha}\,\lrcorner\,\tilde{a}') \, ,\\ 
r^{\prime \prime} &= [r, r^\prime] + j \alpha \,\lrcorner\, j a^\prime  -  j \alpha^\prime \,\lrcorner\, j a - \tfrac{1}{3} ( \alpha\, \lrcorner\,  a^\prime  -  \alpha^\prime\, \lrcorner\, a )\,\id   \, , \\
& \quad  +  j  \tilde \alpha^\prime \,\lrcorner\, j  \tilde a  -  j \tilde  \alpha \,\lrcorner\, j \tilde  a^\prime  - \tfrac{2}{3} ( \tilde \alpha^\prime \,\lrcorner\,  \tilde a -  \tilde  \alpha \,\lrcorner\,  \tilde  a^\prime  )\, \id  \, ,\\
a^{\prime \prime} & = r \cdot a^\prime - r^\prime \cdot a + \alpha^\prime \,\lrcorner\, \tilde a - \alpha  \,\lrcorner\, \tilde a^\prime     \, ,  \\ 
\tilde a^{\prime \prime} & = r \cdot \tilde a^\prime - r^\prime \cdot  \tilde a - a \wedge a^\prime    \, ,  \\
\alpha^{\prime \prime} & = r \cdot \alpha^\prime - r^\prime \cdot \alpha + \tilde  \alpha^\prime \,\lrcorner\, a -  \tilde \alpha  \,\lrcorner\,  a^\prime    \, , \\
\tilde \alpha^{\prime \prime} & = r \cdot \tilde \alpha^\prime - r^\prime \cdot  \tilde \alpha - \alpha \wedge \alpha^\prime  \, , 
\end{aligned} 
\ee
where   $\cdot$ denotes the $\mathfrak{gl}(6)$ action defined in \eqref{app:gl6ac}.

The action of an adjoint element  $R$ on a generalised vector $V \in \Gamma( E)$  and on a dual generalised vector $Z$ is also denoted by $\cdot$ and is defined as
\be
V^\prime = R \cdot V \qquad \quad  Z^\prime = R \cdot Z \, ,
\ee
where the components of $V^\prime$ are 
\be
\label{adj27}
\begin{aligned}
v^{\prime} & =  l v +  r \cdot v + \alpha \,\lrcorner\, \omega - \tilde \alpha \,\lrcorner\, \sigma   \, ,  \\ 
\omega^{\prime} &=  l \omega +  r \cdot \omega  + v \,\lrcorner\, a  +  \alpha \,\lrcorner\, \sigma   \, ,  \\ 
\sigma^{\prime} & = l \sigma +  r \cdot \sigma  + v \,\lrcorner\, \tilde a  + a \wedge \omega    \, ,  
\end{aligned} 
\ee
and those  of $Z^\prime$ are
\be
\label{adj27b}
\begin{aligned}
\hat v^{\prime} & = -  l  \hat v +  r \cdot  \hat v - \hat \omega  \,\lrcorner\, a  + \hat  \sigma \,\lrcorner\, \tilde  a    \, ,   \\ 
\hat \omega^{\prime} &= -  l \hat\omega +  r \cdot  \hat \omega  - \alpha \,\lrcorner\, \hat{v}   -  \hat  \sigma \,\lrcorner\, a    \, ,  \\ 
\hat \sigma^{\prime} & = - l  \hat \sigma +  r \cdot  \hat \sigma  - \tilde \alpha  \,\lrcorner\, \hat v   - \alpha  \wedge  \hat \omega     \, . 
\end{aligned} 
\ee

The $\mathfrak{e}_{6(6)}$ Killing form on two elements of the adjoint bundle is given by
\be
{\rm tr}(R,R') \,=\, \tfrac{1}{2} \left( \tfrac{1}{3}\,{\rm tr}(r){\rm tr}(r') + {\rm tr}(rr') + \alpha \,\lrcorner\, a' + \alpha' \,\lrcorner\, a -\tilde\alpha \,\lrcorner\, \tilde a' - \tilde \alpha' \,\lrcorner\, \tilde a \right)\,.
\ee

The combination of diffeomorphisms and gauge transformations by the three-form and six-form potentials defines the generalised diffeomorphisms. 
The action of an infinitesimal generalised diffeomorphism is generated by the generalised Lie (or Dorfman) derivative along a generalised vector. The  Lie derivative  between two ordinary vectors $v$ and $v'$ on $TM$ 
can be written in components as a $\mathfrak{gl}(6)$ action 
\be
( \mathcal{L}_v v')^m \,=\,  v^n \,\partial_n v'^{\,m} - (\partial \times v)^m{}_n \,v'^{\,n}  \, ,
\ee
where the symbol $\times$ denotes the projection onto the adjoint of the product of the fundamental and dual representation of $\GL(6)$. 
The generalised Lie derivative is defined in an  analogous way;  we introduce  the operators $\partial_M = \partial_m$ as sections of the dual tangent bundle and we define the generalised Lie derivative as
\be 
\label{eq:Liedefgapp}
(L_V V')^M \,=\,  V^N \partial_N  V'^M - (\partial \times_{\adj} V)^M{}_N V'^N \, , 
\ee
where  $V^M$, $M=1,\ldots,27$, are the components of $V$ in a standard coordinate basis, and $ \times_{\adj}$ is the projection onto the adjoint bundle,
\be
 \times_{\adj} \, : \, E^* \otimes E \rightarrow  {\adj} F\, , 
\ee
whose explicit expression can be found in~\cite[Eq.$\:$(C.13)]{Coimbra:2011ky}.
In terms of  $\GL(6)$  tensors, \eqref{eq:Liedefgapp} becomes
\begin{equation}
\label{dorf7}
L_{V} V' = \mathcal{L}_{v} v' + \left(\mathcal{L}_{v}  \omega^{\prime} -\iota_{v^\prime}\mathrm{d} \omega\right) + \left(\mathcal{L}_{v} \sigma' -\iota_{v^\prime}\mathrm{d}\sigma -  \omega^{\prime}\wedge \mathrm{d} \omega\right) .
\end{equation}
 The action of the generalised Lie  derivative on a section of the adjoint bundle  \eqref{sec78} is 
\begin{align}
\label{dor78} 
L_{V} R  &=  (\mathcal{L}_v r + j \alpha \,\lrcorner\, j {\rm d} \omega - \tfrac{1}{3}\, \mathbb{1}  \alpha \,\lrcorner\, {\rm d} \omega - 
 j \tilde \alpha \,\lrcorner\, j {\rm d} \sigma + \tfrac{2}{3}\, \mathbb{1} \tilde  \alpha \,\lrcorner\, {\rm d} \sigma) + (\mathcal{L}_v a + r \cdot {\rm d} \omega - \alpha \,\lrcorner\,  {\rm d} \sigma) \nonumber \\
 &\quad +  (\mathcal{L}_v  \tilde a +  r \cdot {\rm d} \sigma +  {\rm d} \omega  \wedge  a ) +    (\mathcal{L}_v  \alpha  - \tilde \alpha \,\lrcorner\,  {\rm d} \omega ) +  \mathcal{L}_v  \tilde  \alpha   \, . 
\end{align}

We will also need the action of $L_V$ on the elements of the bundle $N$. Given a section $Z_\flat = \lambda + \rho + \tau$ of $N$, its Lie derivative along the 
 generalised vector $V = v + \omega + \sigma$ is 
\be
\label{dorN} 
L_V Z_\flat = \mathcal{L}_v \lambda + (  \mathcal{L}_v  \rho -\lambda  \wedge \dd \omega ) +  (  \mathcal{L}_v  \tau - j  \rho  \wedge \dd \omega + j \lambda \wedge \dd \sigma ) \,.
\ee
 Since  $Z_\flat = V^\prime \otimes_N V^{\prime \prime}$,   this is easily obtained by applying the Leibniz rule for $L_V$.
\be
L_V (Z_\flat) =   L_V V^\prime   \otimes_N   V^{\prime \prime} + V^\prime \otimes_N  L_V V^{\prime \prime} \,.
\ee
It is  also  straightforward to verify that 
\be
\label{derid}
\dd Z_\flat = L_V  V^\prime + L_{V^\prime}  V \, ,
\ee
for any element $Z_\flat =V \otimes_{N} V'   \in N$.

\section{Five-dimensional $\mathcal{N}=2$ gauged supergravity}\label{app:sugra_review}

In this appendix we summarise some essential features of matter-coupled five-dimensional $\mathcal{N}=2$ gauged supergravity \cite{Gunaydin:1999zx,Ceresole:2000jd,Bergshoeff:2004kh}, following the conventions of \cite{Bergshoeff:2004kh}.\footnote{However, in order to match the normalisations defined by our truncation ansatz, we rescale the gauge fields appearing in~\cite{Bergshoeff:2004kh} as  $\mathcal{A}^I_{\rm here} = -\sqrt{\frac{2}{3}} A^I_{\rm there}$. Since we maintain the same form of the covariant derivatives, it follows that the gauge coupling constant $g$ is rescaled as $g_{\rm here}= -\sqrt{\frac{3}{2}} \,g_{\rm there}$. This implies that the expression for the scalar potential given in \eqref{scal_pot_fermi_shift} below acquires a multiplicative 2/3 factor compared to the one in \cite{Bergshoeff:2004kh}.} We limit ourselves to the bosonic sector and  only consider gaugings that do not require the introduction of two-form fields, as these are enough to describe our examples in Sections~\ref{sec:MN1section} and~\ref{sec:BBBWsection}.

The fields of five-dimensional $\mathcal{N}=2$ supergravity arrange into the gravity multiplet,  $n_{\rm V}$  vector multiplets and $n_{\rm H}$ hypermultiplets. The bosonic content consists of the vielbein $e^a_\mu$, $n_{\rm V}+1$ vector fields $\mathcal{A}^I_\mu$, $I=0,\ldots,n_{\rm V}$, together with $n_{\rm V}$ vector multiplet scalars $\phi^x$, $x=1,\ldots,n_{\rm V}$, and $4n_{\rm H}$ hypermultiplet scalars $q^{X}$, $X=1,\ldots,4n_{\rm H}$.
The $\phi^x$ parameterise a `very special real' manifold  $\mathcal{M}_{\rm V}$, with metric $g_{xy}$, while the $q^X$ parameterise a quaternionic-K\"ahler manifold $\mathcal{M}_{\rm H}$, with metric $g_{XY}$. All together, the  scalar  manifold of the theory is the direct product 
\be
\mathcal{M} = \mathcal{M}_{\rm V}\times  \mathcal{M}_{\rm H} \, .
\ee

A very special real manifold  $\mathcal{M}_{\rm V}$ is a hypersurface that is conveniently described in terms of $n_{\rm V}+1$ embedding functions $h^I(\phi)$, $I=0,\ldots,n_{\rm V}$, satisfying the constraint
\be
C_{IJK} h^{I}  h^{J} h^{K} = 1 \, ,
\ee
where $C_{IJK} $ is a completely symmetric constant tensor.
The metric on $\mathcal{M}_{\rm V}$ is given by
\be\label{from_gxy_to_aIJ}
g_{xy} = h_x^I h_y^J \,a_{IJ}\,,
\ee
where  
\be
h_x^I = - \sqrt{\tfrac{3}{2}}\, \partial_x h^I\,,
\ee
and
\be\label{aIJ_general_formula}
a_{IJ} =  3 h_Ih_J -2C_{IJK} h^K \,,
\ee
with the lower-index functions being
\be
h_I = C_{IKL}h^Kh^L = a_{IK}h^K\,.
\ee
The matrix $a_{IJ}$ is assumed invertible, and also controls the gauge kinetic terms.

In the gauged theory, a subgroup of the isometries of the scalar manifold $\mathcal{M}$, which are global symmetries of the Lagrangian, is promoted to  a gauge group. The gauge generators $t_I$ satisfy $[t_I,t_J]=-f_{IJ}{}^K t_K$, with the structure constants $f_{IJ}{}^K$ obeying  $f_{I(J}{}^H C_{KL)H}=0$.
The gauge covariant derivatives of the scalars are given by
\be
\begin{aligned}
\mathcal{D}_\mu \phi^{x} & = \partial_\mu  \phi^{x} + g\, k^{x}_I \mathcal{A}_\mu^I  \,,\\[1mm]
\mathcal{D}_\mu q^X & = \partial_\mu  q^X + g \, k^X_I \mathcal{A}_\mu^I \,,\\
\end{aligned}
\ee
where $ k^{x}_I(\phi)$ and $k^X_I(q)$  are the Killing vector fields generating the gauged isometries in the vector multiplet and hypermultiplet scalar manifolds, respectively.
Equivalently, the vector multiplet scalar covariant derivatives can be expressed in terms of the embedding functions $h^I$ as
\be\label{eq:covderhI}
\mathcal{D}_\mu h^I = \partial_\mu h^I + g\, f_{JK}{}^I \mathcal{A}_\mu^J \,h^K 
=  \partial_xh^I  \mathcal{D}_\mu\phi^x \,.
\ee

One also has the gauge field-strengths
\be
\mathcal{F}^I_{\mu\nu} \, =\,  2\partial_{[\mu}\mathcal{A}^I_{\nu]} + \, g\, f_{JK}{}^I \mathcal{A}^J_\mu \mathcal{A}^K_\nu  \, .
\ee

We now have all the elements to write down the bosonic Lagrangian. This reads
\begin{align} 
e^{-1} \mathcal{L} \, &= \,  \tfrac{1}{2} \,\mathcal{R} -  \mathcal{V} -  \tfrac{3}{8}\, a_{IJ} \mathcal{F}^{I}_{\mu \nu}  \mathcal{F}^{J \mu \nu}    - \tfrac{1}{2}\, g_{xy}  \mathcal{D}_\mu \phi^{x}  \mathcal{D}^\mu \phi^{y}  - \tfrac{1}{2}\,  g_{XY}  \mathcal{D}_\mu q^{X}  \mathcal{D}^\mu q^{Y} \nonumber \\[2mm]
&\quad - \tfrac{1}{ 8}\,e^{-1}\epsilon^{\mu \nu \lambda\rho \sigma}\, C_{IJK}   \mathcal{A}^I_\mu \left[ \mathcal{F}^J_{\nu\lambda} \mathcal{F}^K_{\rho \sigma} + g f_{MN}{}^J \mathcal{A}_\nu^M\mathcal{A}_\lambda^N \left(  - \tfrac{1}{2}\, \mathcal{F}_{\rho\sigma}^K + \tfrac{1}{10}\,g f_{HL}{}^K \mathcal{A}_\rho^H \mathcal{A}_\sigma^L \right) \right]   \, .
\end{align}
The vector multiplet scalar kinetic term can also be written in terms of the constrained scalars $h^I$ using the identity
\be
g_{xy}  \mathcal{D}_\mu \phi^{x}  \mathcal{D}^\mu \phi^{y}\, =\, \tfrac{3}{2}\,a_{IJ}  \mathcal{D}_\mu h^I  \mathcal{D}^\mu h^J\,.
\ee
Using a differential form notation, the action reads
\begin{align} 
S \, &= \,  \int \tfrac{1}{2} \,(\mathcal{R}-  2\mathcal{V}) *1  -  \tfrac{3}{4}\, a_{IJ} \mathcal{F}^{I}\wedge*  \mathcal{F}^{J}    - \tfrac{3}{4}\,a_{IJ}  \mathcal{D} h^I \wedge* \mathcal{D} h^J  - \tfrac{1}{2}\,  g_{XY}  \mathcal{D} q^{X} \wedge * \mathcal{D} q^{Y}  \nonumber \\[2mm]
&\quad + \tfrac{1}{ 8}\, C_{IJK}   \mathcal{A}^I \wedge \left[ 4\mathcal{F}^J\wedge \mathcal{F}^K + g \,f_{MN}{}^J \mathcal{A}^M \wedge \mathcal{A}^N \wedge \left(  -  \mathcal{F}^K + \tfrac{1}{10}\,g f_{HL}{}^K \mathcal{A}^H \wedge \mathcal{A}^L \right) \right]   \, .
\end{align}
The scalar potential $ \mathcal{V}$ is given as a sum of squares as
\be\label{scal_pot_fermi_shift}
\mathcal{V} \,=\,\tfrac{4}{3}\, g^2 \left(   -  2 \vec{P} \cdot \vec{P} +  g^{xy}\vec{P}_{x}\cdot \vec{P}_{y}  
+  \mathcal{N}_{i A} \mathcal{N}^{i A} \right) \, ,
\ee
where
\begin{align}
\vec{P} & = h^I \vec{P}_I\,,\nn\\ 
\vec{P}_{x}  & =  h^I_x \vec{P}_I \,,\nn\\
\mathcal{N}^{i A} &= \tfrac{\sqrt{6}}{4} h^I k_I^X f_X^{iA}\,,
\end{align}
are the fermionic shifts, also appearing in the supersymmetry variations of the fermion fields: $\vec{P}$ is the gravitino shift, $\vec{P}_x$ is the gaugino shift, and $\mathcal{N}^{iA}$ is the hyperino shift. Here,
the arrow symbol denotes a triplet of the $\SUH$ R-symmetry,
and $f_X^{iA}$ are the quaternionic vielbeins, satisfying $f_X^{iA}f_{YiA}=g_{XY}$. The Killing prepotentials $\vec{P}_I$ on $\mathcal{M}_{\rm H}$ are defined for $n_{\rm H}\neq0$ by
\be\label{KillingPrepGeneral}
4n_{\rm H} \vec{P}_I \,=\,  \vec{J}_X{}^Y \nabla_Y k_I^X\,,
\ee
where $\vec{J}_X{}^Y$ is the triplet of almost complex structures defined on any quaternionic-K\"ahler manifold. Plugging these expressions in \eqref{scal_pot_fermi_shift} and using the identity 
(cf.~\cite[App.$\:$C]{Bergshoeff:2004kh})
\be
g^{xy}h_x^Ih_y^J \,=\, a^{IJ}- h^Ih^J\,,
\ee
we can express the scalar potential as
\be\label{ScalPotGeneralFinal}
\mathcal{V}\, =\, \tfrac{2}{3}\,g^2 \left[   \left(2a^{IJ} -  6 h^Ih^J  \right) \vec{P}_I\cdot \vec{P}_J  + \tfrac{3}{4}\,  g_{XY}k_I^X  k_J^Y  h^I h^J \right] \, . 
\ee
Notice that the Killing vectors $k_I^x$ on $\mathcal{M}_{\rm V}$ do not appear here, i.e.~the gauging in the vector multiplet sector does not contribute to the scalar potential. This is true as long as we restrict to gaugings that do not require the introduction of two-form fields.

Supersymmetric AdS$_5$ vacua are obtained by setting all gauge fields to zero, all scalar fields to constant, and imposing that the gaugino and hyperino shifts vanish,
\be\label{vanishing_shifts}
h^I_x \vec{P}_I  = 0\,,\qquad h^I k_I^X = 0\,.
\ee
Then the gravitino shift gives the AdS cosmological constant via
\be
\Lambda \,\equiv\, \mathcal{V} \,=\, - \tfrac{8}{3}\,g^2 \,\vec{P}\cdot \vec{P}\,.
\ee

\section{Gauge transformations}\label{sec:gauge_transf}

In this appendix, we study the reduction  gauge transformations of eleven-dimensional supergravity to five dimensions.  We first repackage them in terms of generalised geometric objects and then use our truncation ansatz to derive the gauge transformations of five-dimensional  $\mathcal{N}=2$ supergravity. 

The infinitesimal gauge transformations of the eleven-dimensional metric and three- and six-form potentials are 
\begin{align}
\label{gauge_var_full}
\delta\hat{g} \,&=\, \hat{\mathcal{L}}_{\hat{v}}\, \hat{g} \,,\nn\\[1mm]
\delta \hat{A} \,&=\, \hat{\mathcal{L}}_{\hat{v}} \hat{A} -    \hat\dd \hat{\lambda} \, ,\nn\\[1mm]
\delta \hat{\tilde A} \,&=\, \hat{\mathcal{L}}_{\hat{v}} \hat{ \tilde A}  -   \hat\dd \hat{ \tilde \lambda} +  \frac{1}{2} \dd \lambda \wedge A \,,
\end{align}
where $\hat v$ is a vector field, $\hat\lambda$ a two-form and $\hat{ \tilde \lambda}$ a five-form.  The hat on the Lie and exterior derivative operators emphasises that the derivatives are taken with respect to all the eleven-dimensional coordinates.  
The  fields $g$, $\hat{A}$ and $\hat{\tilde{A}}$ are decomposed as in  \eqref{11d_metric_general}, while  the  gauge parameters are expanded as 
\begin{align}
\label{app:gaugepexp}
{\hat{v}} \,&=\, v \,=\, v^m \frac{\partial}{\partial z^m}\,,\nn\\[1mm]
\hat{\lambda} \,&=\, \lambda  -  \bar{\lambda}_{\mu}  \dd{x}^\mu  + \tfrac{1}{2}\,\bar{\lambda}_{\mu\nu} \dd{x}^{\mu\nu} \, , \nn\\[1mm]
\hat{ \tilde \lambda}   \,&=\,  \tilde{\lambda} + \bar{ \tilde{\lambda}}_{\mu} \dd x^\mu   + \tfrac{1}{2}  \bar{ \tilde{\lambda}}_{\mu \nu }\dd x^{\mu \nu}  +  \tfrac{1}{3!}  \bar{ \tilde{\lambda}}_{\mu \nu \rho}\dd x^{\mu \nu \rho } + \ldots \, , 
\end{align}
where, in the last line we omitted the terms that are not relevant in what follows.  
We only consider internal diffeomorphisms, as the external ones have the standard action dictated by the tensorial structure of the field.  That is why the vector $\hat{v}$ has only components on $M$. 
As in \eqref{11d_metric_general}, we do not impose any restriction on the dependence of the fields on the coordinates $\{x^\mu,z^m\}$. 
 
 In \eqref{app:gaugepexp} and the rest of this section we use a notation that manifestly displays the external indices and always contracts the internal ones. For the metric components we define
\be
g = g_{mn} \dd z^m \dd z^n\,,\qquad h_\mu = h_\mu{}^m \frac{\partial}{\partial z^m}\,, 
\ee
and for a generic $p$-form $\omega$ 
\begin{align}
 \omega  & = \tfrac{1}{p!}\, \omega_{m_1 \ldots m_p} \dd z^{m_1 \dots m_p} \,,  \nn\\[1mm]  
\omega_\mu & = \tfrac{1}{(p-1)! }\,\omega_{\mu m_1 \ldots m_{p-1}} \dd z^{m_1 \ldots m_{p-1}}  \,,  \nn\\[1mm] 
\omega_{\mu\nu}  &=  \tfrac{1}{(p-2)! } \omega_{\mu \nu m_1 \ldots m_{p-2} } \dd z^{m_1 \ldots m_{p-2}}\,,  
\end{align}

We already  mentioned in Section \ref{sec:fielddec}  that the barred components of the three- and six-form must be redefined as\footnote{The contractions are defined as follows
\be
\iota_{h_[\mu} A_{\nu]} =  h_{[\mu}{}^m A_{\nu] m n}  \dd z^n  \qquad \iota_{h_[\mu} A_{\nu \rho] }  =  h_{[\mu}{}^m A_{\nu \rho] m}   \qquad 
 \iota_{h_[\mu} \iota_{h_\nu}  A_{\rho]}  =  h_{[\mu}{}^m   h_{\nu}{}^n  A_{\rho]  n m  }  
\ee}  
\be 
\begin{aligned} 
\bar{A}_{\mu\nu}  &= A_{\mu\nu} -  \iota_{h_{[\mu}} A_{\nu] }\,, \\
\bar{A}_{\mu\nu\rho} &= A_{\mu\nu\rho}  - \,  \iota_{h_{[\mu}} \iota_{h_\nu} A_{\rho]}\,,
\end{aligned}
\ee
and similar redefinitions of the six-form. 
An analogous redefinition for the barred gauge parameters will be given later. 

As discussed in  Section \ref{sec:fielddec}, the  components of the  metric, warp factor, three and six-form potentials and the dual graviton  $\tilde{g}$ 
with the same number of external legs fit into $E_{6(6)}$ representations 
\begin{align}
\label{app:gmscalars}
& G^{-1} \ \leftrightarrow\ \{\Delta,\,g , \, A , \, \tilde A  \} \, \\[1mm]
\label{app:Avec}
& \mathcal{A}_\mu \,=\, h_\mu  + A_{\mu} + \tilde{A}_{\mu}\, , \\[1mm]
\label{app:Btens}
& \mathcal{B}_{\mu\nu} \,=\, A_{\mu\nu} +   \tilde{A}_{\mu\nu} + \tilde g_{\mu\nu } \,, \\[1mm]
\label{app:Ctens}
& \mathcal{C}_{\mu\nu\rho\,} \,=\,  A_{\mu\nu \rho} + \tilde{A}_{\mu\nu \rho} + \tilde g_{\mu\nu\rho}\,,
\end{align}
where $G^{MN}$ is the inverse generalised metric,   $\mathcal{A}_\mu \in E$ is a generalised vector,  $\mathcal{B}_{\mu\nu} \in N$ is a    weighted  dual vector and
$\mathcal{C}_{\mu\nu\rho\,} $ is a section of the weighted $E_{6(6)}$ adjoint  bundle $ (\det T^*) \otimes  {\rm ad} F$.   The same holds for the gauge parameter, which we arrange into a generalised vector $\Lambda$, a weighted  dual vector $\bar{\Xi}_\mu$ and a section of a sub-bundle of the ${\bf 78}$, $\bar{\Phi}_{\mu \nu}$,
\be
\begin{aligned}
& \Lambda = v  +   \lambda +  \tilde{\lambda}  \, , \\
& \bar{\Xi}_\mu =  \bar{\lambda}_\mu + \bar{\tilde{\lambda}}_{\mu} + \ldots  \, , \\
& \bar{\Phi}_{\mu \nu} =  \bar{\lambda}_{\mu \nu} + \bar{\tilde{\lambda}}_{\mu\nu} + \ldots   \, . \\
\end{aligned}
\ee
In \eqref{app:gmscalars}--\eqref{app:Ctens} we introduced the  dual graviton  to give the full $E_{6(6)}$ representation.  However in this paper 
we will not discuss the dual graviton since it is not relevant for the truncation we are interested in.

\vspace{0.2cm}

We can now decompose the gauge transformations given above. 
We find that the fields with no or purely internal legs transform as 
\begin{align}
\label{app:scalartr}
\delta\,{\rme^{2\Delta}} \,&=\, \mathcal{L}_v \,\rme^{2\Delta}\,,\nn\\[1mm]
\delta {g} \,&=\, \mathcal{L}_{v}\, {g} \,,\nn\\[1mm]
\delta {A} \,&=\, \mathcal{L}_{v} A -  \dd \lambda \,, \nn\\[1mm]
\delta {\tilde A} \,&=\, \mathcal{L}_{v}  \tilde A - \dd \tilde{\lambda} + \frac{1}{2} \dd \lambda \wedge  A \,,
\end{align}
where  the Lie derivative $\mathcal{L}$ and the exterior derivative $\dd$ are taken with respect to the {\it internal} coordinates only, although the fields and gauge parameters depend on both the internal and external coordinates.  When repackaging all the fields with no external legs into the inverse generalised metric,  the transformations \eqref{app:scalartr} become the action of the generalised Lie derivative along the generalised vector $\Lambda$,
\be
\delta_\Lambda G^{-1}  = L_\Lambda G^{-1} \, . 
\ee

Consider now the fields with one external leg. Their gauge  transformations are
\begin{align}\label{var_one_ext_index}
\delta h_\mu \,&=\, -\partial_\mu v + \mathcal{L}_v h_\mu \,,\nn\\[1mm]
\delta A_\mu \,&=\, -   \partial_\mu \lambda +   \dd \overline{\lambda}_\mu -   \iota_{h_\mu} \dd\lambda  + \mathcal{L}_v A_\mu \, , \nn\\[1mm]
\delta \tilde{A}_\mu \,&=  -  \partial_\mu  \tilde{\lambda} +   \dd \overline{\tilde{\lambda}}_\mu -   \iota_{h_\mu} \dd \tilde{\lambda}  - \dd \lambda \wedge A_\mu   + \mathcal{L}_v \tilde{A}_\mu \,,
\end{align}
and it is straightforward to verify that they can be  recast into 
\be
\label{app:varA1}
\delta \mathcal{A}_\mu = - \partial_\mu \Lambda + L_\Lambda \mathcal{A}_\mu + \dd \overline{\Xi}_\mu \, , 
\ee
where
\be
\begin{aligned}
 L_\Lambda \mathcal{A}_\mu & = ( \mathcal{L}_v  h_\mu) +( \mathcal{L}_v A_\mu - \iota_{h_\mu} \dd \lambda ) +  ( \mathcal{L}_v   \tilde{A}_\mu -
  \iota_{h_\mu} \dd \tilde{\lambda}   - A_\mu \wedge \dd \lambda )\,, \\
 L_{\mathcal{A}_\mu} \Lambda  & = ( \mathcal{L}_{ h_\mu} v ) + ( \mathcal{L}_{h_\mu} \lambda  - \iota_{v} \dd A_\mu ) +  ( \mathcal{L}_{h_\mu} \tilde{\lambda} -  
   \iota_v  \dd \tilde{ A}_\mu - \lambda \wedge  \dd A_\mu )\,.
\end{aligned}
\ee
By redefining   the gauge parameters\footnote{In components the redefinition \eqref{app:Xired} reads
\be
\begin{aligned}
\overline{\lambda}_\mu = &    \lambda_\mu  -  \iota_{h_{\mu}} \lambda -  \iota_v A_\mu \,, \\
\overline{\tilde \lambda}_\mu =&  \tilde{\lambda}_\mu -  \iota_{h_{\mu}} \tilde{ \lambda}  - \iota_v  \tilde{A}_\mu + \lambda \wedge A_\mu  \, . 
\end{aligned} 
\ee}
\be
\label{app:Xired}
\overline{\Xi}_\mu = \Xi_\mu  - \mathcal{A}_\mu \otimes_N \Lambda  \, ,
\ee
with
$ \mathcal{A}_\mu \otimes_N \Lambda   =  (\iota_{h_{\mu}} \lambda + \iota_v A_\mu ) +  (\iota_{h_{\mu}} \tilde{ \lambda}  + \iota_v  \tilde{A}_\mu - \lambda \wedge A_\mu )$,  and using 
 \eqref{derid} to compute 
\be
\label{dXi}
\dd \overline{\Xi}_\mu = \dd  \Xi_\mu  - L_{ \mathcal{A}_\mu} \Lambda - L_\Lambda  \mathcal{A}_\mu \, ,
\ee
we bring the variation  \eqref{app:varA1} to an appropriate form  to compare with  five-dimensional gauged supergravity 
\be
\label{deltaAmuf}
\delta \mathcal{A}_\mu = - \partial_\mu \Lambda - L_{ \mathcal{A}_\mu} \Lambda  + \dd \Xi_\mu \, . 
\ee

\vspace{0.2cm}

The variations of the fields with two  external legs are 
\be
\begin{aligned}
\label{app:deltaA2}
\delta A_{\mu\nu} \,=\, & -  2\, \partial_{[\mu} \overline{\lambda}_{\nu]}  -  \dd \overline{\lambda}_{\mu\nu} +  \iota_{h_{[\mu}}\partial_{\nu]} \lambda - \iota_{h_{[\mu}}\dd \overline{\lambda}_{\nu]} + \mathcal{L}_v A_{\mu\nu} - \iota_{\partial_{[\mu}v} A_{\nu]}\,, \\[1mm] 
\delta \tilde A_{\mu\nu} \, = & - 2 \partial_{[\mu} \overline{ \tilde{\lambda}}_{\nu]}  - \dd \overline{\tilde{\lambda}}_{\mu \nu}  +  \iota_{h_{[\mu}}  \partial_{\nu]}  \tilde{\lambda}  - \iota_{h_{[\mu}}   \dd \overline{\tilde{\lambda}}_{\nu]} +  \mathcal{L}_v  \tilde{A}_{\mu \nu}      -   \iota_{  \partial_{[\mu} v}  \,\lrcorner\,  \tilde{A}_{\nu]}   \\
 & + ( \partial_{[\mu} \lambda -  \dd \overline{\lambda}_{[\mu}) \wedge   A_{\nu]}    + \dd \lambda \wedge  A_{\mu \nu}  \,.   
\end{aligned}
\ee
By a lengthy but straightforward computation   \eqref{app:deltaA2} can be written as 
\be 
\begin{aligned}
\label{deltaA22}
\delta  \mathcal{B}_{\mu\nu} \,&=\, -  2\,   \partial_{[\mu}  \Xi_{\nu]}    -  2 L_{\mathcal{A}_{[\mu }}  \Xi_{\nu ]}  + \mathcal{H}_{\mu \nu}  \otimes_N \Lambda  - \delta \mathcal{A}_{[\mu} \otimes_N \mathcal{A}_{\nu]}  \\ 
&  \quad +    L_{\mathcal{A}_{[\mu }} \mathcal{A}_{\nu ]}  \otimes_N \Lambda + 2    L_{\mathcal{A}_{[\mu }} \Lambda  \otimes_N \mathcal{A}_{\nu ]} +
 L_{ \Lambda} \mathcal{A}_{[\mu }    \otimes_N \mathcal{A}_{\nu ]}  \\ 
& \quad -  \dd [  \overline{ \Phi}_{\mu \nu}  -  2  \mathcal{A}_{[\mu} \times_{\rm ad} \bar{\Xi}_{\nu]}   -  \mathcal{B}_{\mu \nu} \times_{\rm ad}  \Lambda ]  \,,
\end{aligned}
\ee
where we defined the field strength 
\be
\label{Hfstr}
\mathcal{H}_{\mu \nu} =  \dd \mathcal{B}_{\mu \nu}  +  L_{\mathcal{A}_{[\mu }} \mathcal{A}_{\nu ]} + 2    \partial_{[\mu} \mathcal{A}_{\nu ]}  \, . 
\ee
Applying  the Leibniz rule  for the generalised Lie derivative  and  \eqref{derid} one can show that
\be
\begin{aligned}
 L_{\mathcal{A}_{[\mu }} \mathcal{A}_{\nu ]}  \otimes_N \Lambda + 2    L_{\mathcal{A}_{[\mu }} \Lambda  \otimes_N \mathcal{A}_{\nu ]} +
 L_{ \Lambda} \mathcal{A}_{[\mu }    \otimes_N \mathcal{A}_{\nu ]}   =  \dd [\mathcal{A}_{[\mu} \times_{\rm ad} (\mathcal{A}_{\nu]} \otimes_N \Lambda)] 
\end{aligned}
\ee
and  the variation of $\mathcal{B}_{\mu\nu}$
can be written in a form compatible with five-dimensional gauged supergravity
\be 
\begin{aligned}
\label{deltaBf}
\delta \mathcal{B}_{\mu\nu} \,&=\, -  2\,   \partial_{[\mu}  \Xi_{\nu]}    -  2 L_{\mathcal{A}_{[\mu }}  \Xi_{\nu ]}  + \mathcal{H}_{\mu \nu}  \otimes_N \Lambda  - \delta \mathcal{A}_{[\mu} \otimes_N \mathcal{A}_{\nu]}    - \dd  \Phi_{\mu \nu}   
\end{aligned}
\ee 
where we have made the following redefinition of the gauge parameters\footnote{In components
\be
 \begin{aligned}
\lambda_{\mu \nu} = &    ( \overline{ \lambda}_{\mu \nu}   - \iota_v A_{\mu \nu}  -  2   \iota_{h_{[\mu}}  \lambda_{\nu]}  + \iota_{h_{[\mu}} \iota_{ h_{\nu]}} \lambda  +  \iota_{h_{[\mu}} \iota_v A_{\nu]} ) \,, \\
\tilde \lambda_{\mu \nu} = &  ( \overline{\tilde {\lambda}}_{\mu \nu} -  \iota_v \tilde{A}_{\mu \nu} - 2   \iota_{h_{[\mu}}  \tilde {\lambda}_\nu   +  \iota_{h_{[\mu}}  \iota_{h_{\nu}} \tilde{ \lambda}  +  \iota_{h_{[\mu}}   \iota_v  \tilde{A}_{\nu]}     -2    \lambda_{[\mu}  \wedge A_{\nu]} 
-  \lambda \wedge  A_{\mu \nu} \\
& -  \lambda \wedge    \iota_{h_{[\mu}}  A_{\nu]} + ( \iota_v  A_{[\mu})   \wedge A_{\nu]}   )  \, . 
\end{aligned}
\ee}
 \be
 \begin{aligned}
\Phi_{\mu \nu}   = 
  \overline{\Phi}_{\mu \nu}  +    2 \mathcal{A}_{[\mu} \times_{\rm ad} \Xi_{\nu]}   +   \mathcal{B}_{\mu \nu} \times_{\rm ad}  \Lambda - \mathcal{A}_{[\mu} \times_{\rm ad} (\mathcal{A}_{\nu]} \otimes_N \Lambda) \,.
\end{aligned}
\ee
 
\vspace{0.2cm}
 
Finally we should consider the  variations of the fields with three external legs. To our purposes it is enough to study the three-form 
\be
\begin{aligned}
\label{deltaA3}
 \delta  A_{\mu\nu\rho} &=    \mathcal{L}_v  A_{\mu\nu\rho}   - 3 \partial_{[\mu} \overline{\lambda}_{\nu \rho]}   -3    \iota_{h_{[\mu}} ( 2\, \partial_{[\nu} \overline{\lambda}_{\rho]} +  \dd \overline{\lambda}_{\nu\rho]}  )   \\
 &\ \quad +  2  \iota_{h_{[\mu}}   \iota_{h_{\nu}} (  \partial_{\rho]} \lambda - \dd \overline{\lambda}_{\rho]}  )  -  2   \,  \iota_{\partial_{[\mu} v}  \iota_{h_\nu} A_{\rho]}    \, . 
\end{aligned}
\ee

In generalised geometry  \eqref{deltaA3} embeds in the lowest component of the variation of the tensor $\mathcal{C}_{\mu \nu \rho}$ in  \eqref{app:Ctens}. 
We introduce the modified field strength for the three-form field $\mathcal{C}_{\mu \nu \rho}$,
\be
\mathcal{H}_{\mu \nu \rho} = -  \dd \mathcal{C}_{\mu \nu \rho} +3 \partial_{[\mu}  \mathcal{B}_{ \nu \rho]} + 3 L_{ \mathcal{A}_{[\mu}} \mathcal{B}_{ \nu \rho]} + 
  \mathcal{A}_{[\mu} \otimes_N (  3  \partial_{\nu}  \mathcal{A}_{ \rho]} +  L_{ \mathcal{A}_{\nu}}   \mathcal{A}_{ \rho]} )\,, 
\ee
and by manipulations similar to what we did previously we can recast the gauge variations as 
\be
\begin{aligned}
\label{app:deltaC}
\delta \mathcal{C}_{\mu \nu \rho} = & -  3  \partial_{[\mu}  \Phi_{ \nu \rho]}  -  3 L_{ \mathcal{A}_{[\mu}} \Phi_{ \nu \rho]}  +  3 \mathcal{H}_{[\mu \nu } \times_{\rm ad} \Xi_{\rho]}  +  \mathcal{H}_{\mu \nu \rho}  \times_{\rm ad} \Lambda  \\
& -3   \mathcal{B}_{ [\mu \nu}    \times_{\rm ad} \delta \mathcal{A}_{\rho]}  
 - \mathcal{A}_{[\mu}  \times_{\rm ad} ( \mathcal{A}_{\nu}     \times_{\rm ad} \delta \mathcal{A}_{\rho]}  )\,,
\end{aligned}
\ee
up to terms involving a four-form gauge parameter, which would continue the tensor hierarchy.

\vspace{0.2cm}

The five-dimensional gauge transformations are obtained by plugging the reduction ansatz  in the variations \eqref{deltaAmuf},  \eqref{deltaBf} and \eqref{app:deltaC}.
The fields  $\mathcal{A}_\mu{}^I(x)$ are expanded as in \eqref{ansatz_Avec} 
\be
\label{app:1fans}
\mathcal{A}_\mu \,=  \, \mathcal{A}_\mu{}^I(x)\, K_I \,, 
\ee
where $K_I$ are the generalised vectors that are singlets of the  $G_S$ structure.  In \eqref{ansatz-Bform},  the two-from fields are expanded on the weighted duals $K^I_\flat $ of the generalised vectors $K_I$.  These are elements of the bundle $N$ and can also written as
\be
K^J_\flat \,=\, D^{IJK}  K_J \otimes_N K_K
\ee
where the tensor  $D^{IJK}$ satisfies $D^{IKL}  C_{JKL} = 1/2 \delta^I_J$ where  $C_{IJK}$ is defined in \eqref{eq:CKKK_cond}. So the two-forms are expanded as 
\be
\mathcal{B}_{\mu\nu} \,=  \,  \mathcal{B}_{\mu\nu \,I}(x)\,  D^{IJK} K_J \otimes_N K_K  \, . 
\ee
The gauge parameters have a similar expansion
\be
\label{gpans}
\Lambda  = -  \Lambda^I(x) K_I\, , \qquad   \Xi_{\mu} = - \tfrac{1}{2} \Xi_{\mu, I }(x) D^{IJK}  K_J \otimes_N K_K\, . 
\ee

With  the  ansatze   \eqref{app:1fans}  and  \eqref{gpans} for $\mathcal{A}_{\mu}$ and the gauge parameters,  the variations \eqref{deltaAmuf} of the one-forms become
\be
 \delta  \mathcal{A}_\mu{}^I(x)  =   \partial_\mu   \Lambda^I(x)  +  f_{JK}^I \mathcal{A}^J_\mu(x)  \Lambda^K(x)  - f_{(JK)}^I D^{JKL} \Xi_{\mu, L}   \,  ,
\ee
where we used  the algebra of the vectors $K_I$  \eqref{eq:K-alg}  and  \eqref{derid}. 

The variations of  two-forms are reduced in a similar way.  We expand the field strength  $\mathcal{H}^{J}_{\mu \nu} $  as in \eqref{app:1fans} and use 
again  the generalised Lie derivative of vectors $K_I$ given in   \eqref{eq:K-alg}. In this way we obtain for the gauge variations of the five-dimensional two-forms $\mathcal{B}_{\mu\nu, I }(x)$ 
\be
\begin{aligned}
\delta   \mathcal{B}_{\mu\nu, I }  \,   = &  \,    \mathcal{D}_{[\mu} \Xi_{\nu],I}   - 2 C_{IJK}  \mathcal{H}^{J}_{\mu \nu}   \Lambda^{K}  - 2  C_{IJK}  \delta \mathcal{A}^{J}_{[\mu}   \mathcal{A}^{K}_{\nu]}     - \Theta_I{}^A \Phi_{A\, \mu \nu} \,,   \\
\end{aligned}  
\ee
where $\Phi_{\mu\nu}=\Phi_{A\,\mu\nu}J^{\flat\,A}$ and 
\be
  \mathcal{D}_{[\mu} \Xi_{\nu],I}(x)  =    \partial_{[\mu}  \Xi_{\nu],I}(x)    + 2   X_{IJ}^K  \mathcal{A}^J_{[\mu } (x)   \Xi_{\nu],K} (x)\, ,
\ee
with
\be
X_{IJ}^K  =C_{ILM} D^{K  M N } f^L_{J N}\, .
\ee
This is in agreemement with five-dimensional supergravity. The variation $\delta\mathcal{C}_{\mu\nu\rho}$ reduces analogously.

\section{Parameterisation of $S^4$ and generalised frames}\label{app:param_and_frames_S4}

The six-dimensional geometry of interest in this paper is given by a four-sphere $S^4$ fibered over a Riemann surface $\mathbb{\Sigma}$.
In this appendix we describe $S^4$ as a foliation of $S^3$ over an interval and review the generalised frames on $S^4$.

\subsection{Parameterisation of $S^3$}

In terms of the standard Euler angles $0\leq\theta\leq \pi$, $0\leq\phi\leq 2\pi$, $0\leq\psi\leq 4\pi$, the unit metric on the round $S^3$ reads
\be\label{roundmetricS3}
g_{S^3}\, =\, 
\tfrac{1}{4}\left(\dd\theta^2 +  \dd\phi^2 + \dd \psi^2 + 2 \cos\theta\,\dd\phi\,\dd\psi  \right)\,.
\ee
The Killing vectors generating its $\SO(4)\simeq \SU(2)_{\rm left}\times\SU(2)_{\rm right}$ isometries can be split into $\SU(2)_{\rm left}$-invariant Killing vectors $\xi_\alpha$, $\alpha=1,2,3$, generating the $\SU(2)_{\rm right}$ isometries, and $\SU(2)_{\rm right}$-invariant Killing vectors $\tilde \xi_\alpha$, generating the $\SU(2)_{\rm left}$ isometries.
The left-invariant vectors read
\begin{align}\label{left_inv_Killing_vectors}
\xi_1 &= -\cot\theta\sin\psi \,\partial_\psi +\cos\psi \,\partial_\theta + \frac{\sin\psi}{\sin\theta}\,\partial_\phi\,,\nn\\
 \xi_2 &= -\cot\theta\cos\psi \,\partial_\psi -\sin\psi \,\partial_\theta + \frac{\cos\psi}{\sin\theta}\,\partial_\phi\,,\nn\\
\xi_3 &= \partial_\psi\,,
\end{align}
while the right-invariant ones are
\begin{align}\label{right_inv_Killing_vectors}
\tilde\xi_1 &= \frac{\sin\phi}{\sin\theta}\,\partial_\psi +\cos\phi\,\partial_\theta - \cot\theta\sin\phi\,\partial_\phi\,,\nn\\
\tilde\xi_2 &= -\frac{\cos\phi}{\sin\theta}\,\partial_\psi +\sin\phi\,\partial_\theta + \cot\theta\cos\phi\,\partial_\phi\,,\nn\\
\tilde\xi_3 &= \partial_\phi\,.
\end{align}
These satisfy
\be
\mathcal{L}_{\xi_\alpha}\xi_\beta = \epsilon_{\alpha\beta\gamma}\,\xi_\gamma\,,\qquad \mathcal{L}_{\tilde\xi_\alpha}\tilde\xi_\beta = -\epsilon_{\alpha\beta\gamma}\,\tilde\xi_\gamma\,,\qquad \mathcal{L}_{\xi_\alpha}\tilde\xi_\beta = 0\,,
\ee
where $\mathcal{L}$ is the ordinary Lie derivative.
We also introduce the one-form counterparts of these Killing vectors, namely left-invariant one-forms $\sigma_\alpha$ and right-invariant one-forms $\tilde\sigma_\alpha$. These satisfy
\begin{align}
&\iota_{\xi_\alpha}\sigma_\beta = \delta_{\alpha\beta}\,,\quad\qquad \iota_{\tilde\xi_\alpha}\tilde\sigma_\beta = \delta_{\alpha\beta}\,,\nn\\[1mm]
&\dd\sigma_\alpha = -\tfrac{1}{2} \,\epsilon_{\alpha\beta\gamma} \sigma_\beta\wedge \sigma_\gamma\,,\qquad
\dd\tilde\sigma_\alpha = \tfrac12 \,\epsilon_{\alpha\beta\gamma} \tilde\sigma_\beta\wedge\tilde\sigma_\gamma\,,
\end{align}
and their coordinate expression is
\begin{align}\label{leftinv_oneforms}
\sigma_1 &= \cos\psi \,\dd\theta +\sin\psi \sin\theta\,\dd\phi\,,\nn\\
\sigma_2 &= -\sin\psi\,\dd\theta + \cos\psi\sin\theta\,\dd\phi\,,\nn\\
\sigma_3 &= \dd\psi + \cos\theta\,\dd\phi\,,
\end{align}
\begin{align}\label{rightinv_oneforms}
\tilde\sigma_1 &= \cos\phi \,\dd\theta +\sin\phi \sin\theta\,\dd\psi  \,,\nn\\
\tilde\sigma_2 &=  \sin\phi\,\dd\theta - \cos\phi\sin\theta\,\dd\psi \,, \nn\\
\tilde\sigma_3 &=  \dd\phi + \cos\theta\,\dd\psi \,.
\end{align}
The metric \eqref{roundmetricS3} may also be expressed as
\be
g_{S^3}\, =\, \tfrac{1}{4}\left( \sigma_1^2 + \sigma_2^2 + \sigma_3^2\right)\, =\, \tfrac{1}{4}\left( \tilde\sigma_1^2 + \tilde\sigma_2^2 + \tilde\sigma_3^2\right)\,.
\ee
We fix the orientation on $S^3$ by defining the volume form as
\be
{\rm vol}_{S^3}   \,=\,  \tfrac{1}{8}\,\sigma_1\wedge\sigma_2\wedge\sigma_3 \,=\, \tfrac{1}{8}\,\tilde\sigma_1\wedge\tilde\sigma_2\wedge\tilde\sigma_3 \,=\, \tfrac{1}{8}\sin\theta\,\dd\theta\wedge\dd\phi\wedge\dd\psi \,.
\ee

\subsection{Parameterisation of $S^4$}

The round four-sphere of radius $R$ can be described via constrained $\mathbb{R}^5$ coordinates $R\, y^i$, $i=1,\ldots,5$, satisfying $\delta_{ij}y^iy^j=1$. In these coordinates, the metric and the volume form read
\begin{equation}
\label{metricvolS4_ycoords}
g_{4} = R^2 \,\delta_{ij} \dd y^i \dd y^j\,,\qquad
\vol_{4} = \tfrac{1}{4!}R^4\,\epsilon_{i_1i_2i_3i_4 i_5} \,y^{i_1} \dd y^{i_2}\wedge \dd y^{i_3}\wedge \dd y^{i_4} \wedge \dd y^{i_5}\,.
\end{equation}
The constrained coordinates can be mapped into angular coordinates $\{\zeta,\theta,\phi,\psi\}$, where $0\leq\zeta\leq \pi$, and $\{\theta,\phi,\psi\}$ are the Euler angles on $S^3$ introduced above, as
\begin{align}\label{emb_coords_S4}
y^1 + \ii\, y^2 &= \sin\zeta\, \cos\tfrac{\theta}{2}\,\rme^{\frac{\ii}{2} \left(\phi+\psi\right)} \,,\nn \\[1mm]
y^3 + \ii\, y^4 &= \sin\zeta\, \sin\tfrac{\theta}{2}\,\rme^{\frac{\ii}{2} \left(\phi-\psi\right)} \,,\nn\\[1mm]
y^5 &=  \cos\zeta \,.
\end{align}
Then the metric and volume form in \eqref{metricvolS4_ycoords} become
\begin{align}
g_4 &= R^2\left(\dd \zeta^2 + \sin^2\zeta\, \dd s^2_{S^3}\right)\,,\nn\\[1mm]
&= R^2\left[\dd \zeta^2 + \tfrac{1}{4}\sin^2\zeta\, \left( \dd\theta^2 + \dd\phi^2+\dd\psi^2 + 2\cos\theta\,\dd\phi\,\dd\psi  \right)\right]\,,\nn\\[1mm]
{\rm vol_4} &=  R^4\sin^3\zeta \,\dd\zeta\wedge {\rm vol}_{S^3} =  \frac{1}{8}R^4\sin^3\zeta\sin\theta \,\dd\zeta\wedge\dd\theta\wedge\dd\phi\wedge\dd\psi\,.
\end{align}

We denote by $v_{ij}=v_{[ij]}$ the Killing vector fields generating the isometries of $S^4$. These satisfy the $\so_5$ algebra,
\be\label{algebra_Killing_v}
\mathcal{L}_{ v_{ij}} v_{kl} \, =\, R^{-1}\left(\delta_{ik}v_{lj} - \delta_{il}v_{kj} - \delta_{jk}v_{li} + \delta_{jl} v_{ki} \right)\ .
\ee
Demanding that the constrained coordinates transform in the fundamental representation,
\be
\mathcal{L}_{v_{ij}}y_k \,\equiv\, \iota_{v_{ij}}\dd y_k \,=\, R^{-1}\left( y_i \delta_{jk} -  y_j\delta_{ik}\right)\, ,
\ee
and using the map \eqref{emb_coords_S4},
we can work out the expression for the Killing vectors in the basis defined by the angular coordinates $\{\zeta,\theta,\phi,\psi\}$. In particular, we obtain the following embedding of the $\SU(2)_{\rm right}$ and $\SU(2)_{\rm left}$ generators given in \eqref{left_inv_Killing_vectors}, \eqref{right_inv_Killing_vectors} into the $\SO(5)$ generators:
 \begin{align}
\tfrac{2}{R}\,\xi_1 = v_{13}+ v_{24}\,,\qquad \tfrac{2}{R}\,\xi_2 = v_{14}-v_{23}\,,\qquad \tfrac{2}{R}\,\xi_3 = v_{12}-v_{34}\,,\nn\\[1mm]
\tfrac{2}{R}\,\tilde\xi_1 = v_{13}- v_{24}\,,\qquad \tfrac{2}{R}\,\tilde\xi_2 = v_{23}+v_{14}\,,\qquad \tfrac{2}{R}\,\tilde\xi_3 = v_{12}+v_{34} \,.
\end{align}

\subsection{Generalised frames on $S^4$}
\label{app:S4par}
In generalised geometry all spheres are generalised parallelisable as they admit globally defined frames on their exceptional tangent bundle  \cite{Lee:2014mla}. 
In particular  the generalised tangent bundle on $S^4$ is 
\be
\begin{aligned}
\label{app:S4gentanb}
 E_4  & \,\simeq\, T S^4 \oplus \Lambda^2  T^\ast S^4 \, , 
 \end{aligned}
\ee 
and its fibres transform in the ${\bf 10}$ of  the structure group  $\SL(5 ,\mathbb{R})$. We will also need the bundles 
\be
\begin{aligned}
\label{app:S4bundles}
 N_4 &  \,\simeq\, T^\ast S^4 \oplus \Lambda^4  T^\ast S^4 \,, \\
N^\prime_4 &   \,\simeq\,  \bbR  \oplus  \Lambda^3  T^\ast S^4\,,  \\
\end{aligned}
\ee 
whose fibres transform in the ${\bf 5}$  and ${\bf 5^\prime}$ representations, respectively.    
These bundles admit globally defined frames, which in constrained coordinates read
\begin{equation}
\label{frames_S4}
\begin{aligned}
   E_{ij} &= v_{ij} + R^2*_4\! (\dd y_i\wedge\dd y_j) + \iota_{v_{ij}} A_{S^4}  
      \qquad\in\, \Gamma(E_4) \,, \\
   E_i &= R\,\dd y_i - y_i \vol_{4} + R\,\dd y_i\wedge A_{S^4} 
      \qquad\quad\ \; \in\, \Gamma(N_4) \,, \\
   E'_i &= y_i + R\, *_4\! \dd y_i + y_i A_{S^4} 
     \qquad\qquad\qquad\; \  \in\, \Gamma(N_4')\, , 
\end{aligned}
\end{equation}
where the Hodge star $*_4$ is computed using \eqref{metricvolS4_ycoords}, and the three-form potential $A_{S^4}$ must satisfy 
\be\label{flux_MN_background}
\dd A_{S^4} = 3R^{-1} \vol_{4}\,.
\ee
This is the flux relevant for the AdS$_7\times S^4$ supersymmetric Freund-Rubin solution to eleven-dimensional supergravity; the twist over the Riemann surface discussed in the main text will modify it. The $E_{ij}$ are generalised Killing vectors generating the $\so_5$ algebra via the action of the generalised Lie derivative,
\begin{equation}
   L_{E_{ij}} E_{kl} = - R^{-1} \left(
      \delta_{ik}E_{jl} -\delta_{il}E_{jk} + \delta_{jl} E_{ik}
      - \delta_{jk} E_{il} \right) \, . 
\end{equation}
In the main text we will need the following linear combinations,
\begin{align}\label{Xi_tildeXi}
&\Xi_1 = E_{13}+E_{24}\,,\qquad \Xi_2 = E_{14}-E_{23} \,,\qquad \Xi_3 = E_{12}-E_{34}\,,\nn\\
&\ti\Xi_1 = E_{13}-E_{24}\,,\qquad \ti\Xi_2 = E_{14}+E_{23} \,,\qquad \ti\Xi_3 = E_{12}+E_{34}\,.
\end{align}

Using the map \eqref{emb_coords_S4}, the frame elements \eqref{frames_S4} can equivalently be expressed in terms of angular coordinates on $S^4$. In particular, choosing a gauge such that the potential $A_{S^4}$ satisfying \eqref{flux_MN_background} is $\SU(2)_{\rm left}\times \SU(2)_{\rm right}$ invariant,
\be\label{3form_on_S4}
A_{S^4} = \tfrac{1}{32}R^3\left(\cos(3\zeta) - 9 \cos\zeta \right)\,\sigma_1\wedge\sigma_2\wedge\sigma_3\,,
\ee
we find that the combinations \eqref{Xi_tildeXi} are expressed in terms of the $\SU(2)_{\rm left}$ and $\SU(2)_{\rm right}$ invariant tensors as
\begin{align}
\Xi_\alpha 
&=\, \tfrac{2}{R} \,\xi_\alpha + \tfrac{R^2}{2}  \dd \left( \cos\zeta\;\sigma_\alpha \right)\,,\nn\\[1mm]
\ti\Xi_\alpha 
&=  \tfrac{2}{R} \,\tilde\xi_\alpha - \tfrac{R^2}{2}\,\dd \left( \cos\zeta\;\tilde\sigma_\alpha \right)\,.
\end{align}
The $\Xi_\alpha$ can be seen as left-invariant generalised Killing vectors generating the $\SU(2)_{\rm right}\subset\SO(4)\subset\SO(5)$ generalised isometries, 
while $\tilde \Xi_\alpha$ are right-invariant generalised Killing vectors generating the $\SU(2)_{\rm left}$ generalised isometries. 
We will also need the expressions for $E_i$ and $E_i'$ in terms of angular coordinates, in fact just for $i=5$. These read
\be
E_5 = -R \,\sin\zeta\,\dd \zeta + \tfrac{R^4}{8}\,\sin(2\zeta)\, \dd \zeta\wedge \sigma_1\wedge\sigma_2\wedge\sigma_3\,,
\ee
\be
E_5'= 
\cos\zeta - \tfrac{R^3}{16} \left(\cos(2\zeta) +3\right) \sigma_1\wedge\sigma_2\wedge\sigma_3 \,.
\ee
Notice that $\dd E_5' = \tfrac{1}{R} \,E_5$.

\section{Details on the generalised $U(1)$ structure of MN1 solution}
\label{app:U1MN}

In this appendix we give the details of the construction  of the $U(1)$ structure discussed in Section \ref{sec:U1str}.  In order to identify the correct $U(1)$ subgroup of $E_{6(6)} \times \bbR^+$
and its commutant it is convenient  to decompose $E_{6(6)}$ under its maximal compact subgroup $\USp(8)$
and then express the $\USp(8)$ representations in terms of $\Cliff(6)$ gamma matrices.  For the latter step we also need the decomposition of $E_{6(6)}$ under   $SL(6) \times SL(2)$.  
We first give a brief summary of the decomposition of 
 $E_{6(6)}$ under $\USp(8)$ and $SL(6) \times SL(2)$ and then we apply this to the construction of the $U(1)$ structure, which reduces to simple gamma matrix algebra.

\subsection{$\USp(8)$ and $SL(6) \times SL(2)$ decompositions}
\label{Usp8SL6}

In this section we mostly use  the conventions of \cite{Grana:2016dyl}. 
Consider first the decomposition of $E_{6(6)}$ under  $\USp(8)$. 
We denote by $M,N , \dots = 1, \dots, 27$ the $E_{6(6)}$ indices and by $\alpha, \beta, \ldots = 1, \dots,  8$ the  $\USp(8)$ ones.

The fundamental of  $E_{6(6)}$  is irreducible under $\USp(8)$  and is defined by an anti-symmetric traceless tensor 
\be
V^{\alpha \beta} = V^{[\alpha \beta]}  \qquad  V^\alpha_\alpha =0  \, . 
\ee
The $\USp(8)$  indices  are raised and lowered by the  $\USp(8)$ symplectic form $\Omega^{\alpha \beta}$ and its inverse.
The dual vectors in the ${\bf \overline{27}}$ are denoted by $Z_{\alpha \beta}$. 
 The adjoint of  $E_{6(6)}$ decomposes as 
\be
 {\bf 78}  = {\bf 36}  + {\bf 42} \,,
\ee
where  the  ${\bf 36}$  is the adjoint of $\USp(8)$ and the ${\bf 42}$ contains the non-compact generators. 
The elements of the  ${\bf 36}$ are $8 \times 8$  matrices $\mu^\alpha{}_\beta$  satisfying 
\be
\mu_{\alpha \beta} =  \mu_{\beta \alpha}
\ee
with  $\mu_{\alpha \beta} = (\Omega^{-1})_{\alpha \gamma} \mu^\gamma{}_\beta$.
The non compact generators  $\mu_{\alpha \beta \gamma \delta} \in {\bf 42} $ are anti-symmetric  tensors  satisfying
\be
\mu^{\alpha \beta \gamma \delta} (\Omega^{-1})_{\alpha \beta} = 0  \, . 
\ee

The adjoint action on the ${\bf 27}$ is 
\begin{equation}
\label{delta27}
(\mu V)^{\alpha \beta} = \mu^\alpha{}_\gamma V^{\gamma \beta} -  \mu^\beta{}_\gamma V^{\gamma \alpha} - \mu^{\alpha \beta \gamma \delta} V_{\gamma \delta}
\end{equation}
and the $E_{6(6)}$ commutators are
\begin{align}
\label{comm36}
& [ \mu, \nu]^{\alpha \beta} = \mu^{\alpha}{}_\gamma  \nu^{\gamma \beta} - \frac{1}{3} \mu^{\alpha \gamma \delta \epsilon} \nu_{ \gamma \delta \epsilon}{}^{ \beta} - (\mu \leftrightarrow \nu) \\
\label{comm42}
& [ \mu, \nu]^{\alpha \beta \gamma \delta} = - 4 \mu^{[\alpha}{}_\epsilon  \nu^{\beta \gamma \delta] \epsilon }
 - (\mu \leftrightarrow \nu) \,.
\end{align}

Given the generalised vectors $V$, $V^\prime$, $V^{\prime \prime}$ and the duals  $Z$, $Z^\prime$, $Z^{\prime \prime}$,  the  $E_{6(6)}$  quadratic form becomes
\be
 \langle V,Z \rangle  = V^{\alpha \beta} Z_{\alpha \beta} \, ,   
\ee
and the cubic invariants are 
\be
\begin{aligned}
\label{app:cubicSU8}
& c(V, V^\prime, V^{\prime \prime}) =   V^\alpha{}_\beta   V^{\prime \beta}{}_\gamma  V^{\prime \prime \gamma}{}_\alpha \, ,  \\ 
& c^*(Z, Z^\prime, Z^{\prime \prime}) =   Z_\alpha{}^\beta   Z_\beta^{\prime \, \,  \gamma}{}  Z_\gamma^{\prime \prime \alpha} \, . 
\end{aligned} 
\ee

We will also need the projection into the adjoint of the product of a generalised vector $V$ and a dual generalised vector $Z$
\be
\label{adjprUsp8}
\begin{aligned}
(V \times Z)^{\alpha \beta} & =  2 V^{(\alpha}{}_\gamma Z^{|\gamma| \beta)} \\
(V \times Z)^{\alpha \beta \gamma \delta}  &= 6 \left( V^{[\alpha \beta} Z^{\gamma \delta]} + V^{[\alpha}{}_\epsilon Z^{|\epsilon| \beta} \Omega^{\gamma \delta} 
+ \tfrac{1}{3} \tr (VZ) \Omega^{[\alpha \beta} \Omega^{\gamma \delta]} \right) \, .
\end{aligned}
\ee

\vspace{0.2cm}

Consider now  the decomposition of $E_{6(6)}$ under 
{$SL(6) \times SL(2)$. 
 We denote the  $SL(6)$ indices with $m,n, \ldots = 1, \dots,  6$  and the 
$SL(2)$ indices $ \hat{\imath}, \hat{\jmath}  \ldots  = 1, 2$.  Under $SL(6) \times SL(2)$  the ${\bf 27}$ and $\rep{ \overline{27}}$
 decompose as 
\be
\begin{aligned}
\label{27Sl6split}
{\bf 27}  & = ({\bf \bar 6}, {\bf 2}) + ({\bf 15} ,{\bf 1}) \qquad  V^M = (v^{ \hat{\imath}}_m , V^{mn}) \,,\\ 
{\bf \overline{27}}  & = ({\bf 6}, {\bf  \bar 2}) + ({\bf \overline{15}} ,{\bf 1})  \qquad  Z_M = (z_{ \hat{\imath}}^m , Z_{mn})\,,
\end{aligned} 
\ee
where $V^{mn}$ and $Z_{mn}$ are anti-symmetric.  The components in \eqref{27Sl6split}  are related to the $GL(6)$ tensors \eqref{app:gentan} and \eqref{app:genvec} as follows
\be
\begin{aligned}
& V = \omega  \qquad v^1 =  v \qquad  v^2\, \lrcorner\, \rm{vol}_6 = \sigma \, ,  \\ 
& Z = \hat{\omega}  \qquad  z_1 =  \hat{\rm \vol}_6  \,\lrcorner\, \sigma     \qquad  z_2  = \hat{v}  \, . 
\end{aligned} 
\ee

The adjoint  of  $E_{6(6)}$ decomposes as 
\begin{align}
\label{adSL6split}
{\bf 78} = ({\bf 35}, {\bf 1}) \oplus  ({\bf 1}, {\bf 3})  \oplus ({\bf 20}, {\bf 2})   \qquad  \mu^M{}_N = ( \mu^m{}_n \,,  \mu^{\hat{\imath}}{}_{\hat{\jmath}} \, , \mu^ {\hat{\imath}}_{mnp})  \, , 
\end{align} 
where $\mu^m{}_n$ are real, traceless,  $6 \times 6$ matrices generating $SL(6)$, $\mu^{ \hat{\imath}}{}_{ \hat{\jmath}}$ are real and traceless and generate $SL(2)$ and $\mu^{ \hat{\imath}}_{mnp}$ are a pair of
real fully antisymmetric tensors in the  $({\bf 20}, {\bf 2})$. 
The matrices $\mu^m{}_n$ are identified with the traceless part of the $GL(6)$ matrix $r$, with the trace given by the diagonal non-compact generator of $SL(2)$, where we have also set $l=\frac{1}{3}\tr(r)$,
\be
\mu^m{}_n = r^m{}_n - \tfrac{1}{6} \tr(r) \delta^m{}_n  \qquad    \mu^1{}_1 = - \mu^2{}_2 =  \tfrac{1}{2} \tr(r) \, . 
\ee
The compact and remaining non-compact generator of $SL(2)$ are identified with the combinations of six-form and six-vector transformation $\tilde{a}\pm \tilde{\alpha}$.

The tensors $\mu^{ \hat{\imath}}_{mnp}$ correspond to the  three-forms and three-vectors
\be
\mu^1_{mnp} = \alpha^{mnp}\,, \qquad \mu^2_{mnp} = a_{mnp} \, .
\ee

\vspace{0.2cm}

Using $\Cliff(6,\bbR)$ gamma matrices one can relate $\USp(8)$ and $SL(6) \times SL(2)$  representations.  We  introduce the doublet of matrices 
\be
\hat{\Gamma}^m_{ \hat{\imath}} = \big(\hat{\Gamma}^m,\, \ii\,  \hat{\Gamma}^m \hat{\Gamma}^7\big)
\,,   \qquad \hat{\imath}= 1,2 \, . 
\ee
Then the ${\bf 27}$  and ${\bf \overline{27}}$ of $\USp(8)$ are  given in terms of $SL(6) \times SL(2)$ representation by 
\be
\begin{aligned}
\label{27US}
V^{\alpha \beta}  & = \frac{1}{2 \sqrt{2}} \big[   v^{ \hat{\imath}}_m  (\hat{\Gamma}^m_{\hat{\imath}})^{\alpha \beta} + \frac{\ii}{2} V^{mn}  (\hat{\Gamma}_{mn7})^{\alpha \beta}   \big] \,,\\
Z_{\alpha \beta}  & = \frac{1}{2 \sqrt{2}} \big[    z^m_{ \hat{\imath}} (\hat{\Gamma}_m^{\hat{\imath}})^{\alpha \beta} + \frac{\ii}{2} Z_{mn}  (\hat{\Gamma}^{mn7})^{\alpha \beta}   \big]\,,
\end{aligned}
\ee
where  $\Gamma_{mn7}$ denotes the anti-symmetric product of two gamma's and $\Gamma_7$. 
The  ${\bf 36}$ and the ${\bf 42}$ of $\USp(8)$ are given 
   \begin{align}
 \label{Mmu78}
 & \mu_{\alpha \beta} =  \frac{1}{4}  \big[   \mu^m{}_n  (\hat{\Gamma}_m{}^n)  +  \ii\, \epsilon_{ \hat{\imath}}{}^{ \hat{\jmath}} \mu^{ \hat{\imath}}{}_{ \hat{\imath}} \hat{\Gamma}_7  
 + \tfrac{1}{6} \, \epsilon_{ \hat{\imath}}{}^{ \hat{\jmath}} \mu^{ \hat{\imath}}_{mnp} \hat{\Gamma}^{mn} \hat{\Gamma}^p_{ \hat{\jmath}}  \big]_{\alpha \beta} \,, \\
 & \mu^{\alpha \beta \gamma \delta} = \frac{1}{8}  \left[ - \mu^m{}_n ( \hat{\Gamma}^{ \hat{\imath}}_m \otimes  \hat{\Gamma}^n_{ \hat{\imath}}  -  \hat{\Gamma}_{mp7} \otimes  \hat{\Gamma}^{p n 7} ) + \mu^{ \hat{\imath}}{}_{ \hat{\jmath}}\hat{\Gamma}_{ \hat{\imath}}^m \otimes  \hat{\Gamma}_m^{ \hat{\jmath}}
 + \ii\, \mu^{ \hat{\imath}}_{mnp} \hat{\Gamma}^m_{ \hat{\imath}} \otimes   \hat{\Gamma}^{np7}    \right]^{[\alpha \beta \gamma \delta]} \nonumber 
 \end{align}
where   $\otimes$ denotes   the  tensor product of two gamma's, $\mu^m{}_n$ is traceless and $\mu^{\hat{\imath}}_{mnp}$ are  anti-symmetric in the three lower indices. 
 

\medskip

We take the $\Cliff(6,\bbR)$  gammas  $\hat{\Gamma}_m$ with $m=1, \ldots, 6$ such that
\be  
\hat{\Gamma}_m^T = \hat{C}^{-1}  \hat{\Gamma}_m \hat{C}
\ee
where $\hat{C}$ is the charge conjugation matrix satisfying  $\hat{C}^T=-\hat{C}$,  which we identify  with the $\USp(8)$ symplectic invariant $\Omega$.  The chiral gamma is given by
\be
\hat{\Gamma}_7 =  \ii\,   \hat{\Gamma}^1 \cdots \hat{\Gamma}^6  \, . 
\ee

Since the six-dimensional manifolds we are interested in are $S^4$ fibrations over  a Riemann surface, we further decompose  the $\Cliff(6)$ gamma matrices 
according to  $SO(4) \times SO(2)$. We take $m=5,6$ to be directions along the Riemann surface
\be
\label{app:2+4}
\begin{aligned}
\hat{\Gamma}_{m}  & =  \id  \otimes \Gamma_m  \qquad    m =1,2,3,4\,, \\ 
\hat{\Gamma}_{5 }&  =  \gamma_1 \otimes \Gamma_5\,, \\
\hat{\Gamma}_{6}&  =  \gamma_2  \otimes \Gamma_5   \,,
\end{aligned}
\ee
where $\Gamma_m$ are the  $SO(4)$ gamma matrices  with $\Gamma_5=\Gamma_{1234}$ and  $\gamma_1,\gamma
_2$  are the $SO(2)$ ones.  Then the six-dimensional chirality matrix 
becomes
\begin{equation}
\hat{\Gamma}_7 =  \ii\, \gamma_{12} \otimes \Gamma_5  \, .
\end{equation}

\subsection{The $\U(1)$ structure}
\label{app:U1str}

We can now give the details of the construction  of the $U(1)_S$ structure discussed in Section~\ref{sec:U1str}.  The $\mathcal{N}=2$ solution of \cite{Maldacena:2000mw}  has 
an $U(1)_S$ structure corresponding to the diagonal of the $SO(2)$ holonomy on the Riemann surface $\mathbb{\Sigma}$,  and the $U(1)_{\rm right}$ subgroup of the $SO(5) $ isometry 
of the four-sphere, according to the embedding
\begin{equation}
\label{app:decompose_SO5}
 SO(5) \supset SO(4)\, \simeq\, SU(2)_{\rm left} \times SU(2)_{\rm right} \, \supset  \,SU(2)_{\rm left} \times U(1)_{\rm right}  \, .
\end{equation}
Seen as an element of  $E_{6(6)}$, the $U(1)_S$ corresponds to a compact generator and therefore belongs to $\USp(8)$. 
Using the expression \eqref{Mmu78} for the generators of  $\USp(8)$, and now taking the indices $m=5,6$ for the direction along the Riemann surface, the $U(1)_S$  generator can be written
as 
\be
\label{app:u1sgen}
\mathfrak{u}(1)_S \,=\, \ii\,  \hat{\Gamma}_{56} -\tfrac{\ii}{2}\, (\hat{\Gamma}_{12} - \hat{\Gamma}_{34})  \, , 
\ee
where $ \hat{\Gamma}_{56}$ is the generator of the $SO(2)$ holonomy of $\mathbb{\Sigma}$ and $\frac{\ii}{2} (\hat{\Gamma}_{12} - \hat{\Gamma}_{34}) $ generates $U(1)_{\rm right} \subset  SO(5) $.

To embed this generator in  $E_{6(6)}$  and determine the invariant generalised tensors it is convenient to decompose all   $E_{6(6)}$  representations into  $\USp(8)$ one's and then use the parameterisation of $\USp(8)$ in terms of gamma matrices of Section \ref{Usp8SL6}. In this way the computation of the commutant, $\Comm_{E_{6(6)}}(U(1)_S)$,  of $U(1)_S$ in  $E_{6(6)}$  and the determination of the  $U(1)_S$ singlets reduce to simple  gamma matrix algebra. 

We first compute the commutators  of $U(1)_S$ 
with the generic elements of the $ {\bf 36}$ and ${\bf 42}$ in \eqref{Mmu78}.  This will allow to determine the number of  $U(1)_S$ singlets in the 
${\bf 78}$ and the commutant $C_{E_{6(6)}}(U(1)_S)$. 
Using \eqref{comm36}  we find that there  are eight singlets  in  the ${\bf 36}$. Five correspond to elements of $SO(6) \subset SL(6)$,
\be
\begin{aligned}
& S^{(36)}_1 = \hat{\Gamma}_{56}  \, ,  \\
& S^{(36)}_2 = \hat{\Gamma}_{12} \, ,  \\
& S^{(36)}_3 = \hat{\Gamma}_{34}  \, ,  
\end{aligned}
\qquad 
\begin{aligned}
& S^{(36)}_4 = \frac{1}{2} (\hat{\Gamma}_{24} - \hat{\Gamma}_{13}) \, , \\
& S^{(36)}_5 = \frac{1}{2} (\hat{\Gamma}_{14} + \hat{\Gamma}_{23})  \, , 
\end{aligned}
\ee
two are compact elements of $({\bf \overline{20}}, {\bf 2})$ associated to 
\begin{equation}
\begin{aligned}
& S^{(36)}_6 = \frac{1}{2} (\hat{\Gamma}_{135} +  \hat{\Gamma}_{146}- \hat{\Gamma}_{236}+ \hat{\Gamma}_{245})  \, ,\\
& S^{(36)}_7 = \frac{1}{2} (   \hat{\Gamma}_{136} - \hat{\Gamma}_{145} + \hat{\Gamma}_{235} +  \hat{\Gamma}_{246})  \, , 
\end{aligned}
\end{equation}
and the last one  is the  generator of $SO(2) \subset SL(2)$ corresponding to the anti-symmetric part of $\mu^1{}_2$,
\be
 S^{(36)}_8 =  \ii\, \hat{\Gamma}_7  \, .
 \ee
 
A similar computation gives the singlets in the  ${\bf 42}$:  four  are non  compact elements of $SL(6)$
\be
\begin{aligned}
S^{(42)}_1  & = - \frac{1}{4} ( \hat{\Gamma}_1 \otimes  \hat{\Gamma}^1 +  \hat{\Gamma}_2  \otimes   \hat{\Gamma}^2
 - \hat{\Gamma}_5  \otimes  \hat{\Gamma}^5 -  \hat{\Gamma}_6   \otimes  \hat{\Gamma}^6  )   \\ 
S^{(42)}_2  & = -   \frac{1}{4} (  \hat{\Gamma}_3 \otimes  \hat{\Gamma}^3  - \hat{\Gamma}_5  \otimes  \hat{\Gamma}^5 -  \hat{\Gamma}_6   \otimes  \hat{\Gamma}^6  )  \\
S^{(42)}_3  & = - \frac{1}{2} ( \hat{\Gamma}_1 \otimes  \hat{\Gamma}^4 + \hat{\Gamma}_4  \otimes  \hat{\Gamma}^1 +\hat{\Gamma}_2  \otimes \hat{\Gamma}^3   + \hat{\Gamma}_3 \otimes  \hat{\Gamma}^2  )  \\
S^{(42)}_4  & =  - \frac{1}{2} ( \hat{\Gamma}_2 \otimes  \hat{\Gamma}^4  + \hat{\Gamma}_4  \otimes \hat{\Gamma}^2 -  \hat{\Gamma}_1 \otimes  \hat{\Gamma}^3  - \hat{\Gamma}_3  \otimes \hat{\Gamma}^1 ) \,,\\
\end{aligned}
\ee
two  are the non-compact generators of $SL(2)$
\be
\begin{aligned}
S^{(42)}_5  &=   \frac{1}{4}  ( \hat{\Gamma}^m  \otimes \hat{\Gamma}_m
+   \hat{\Gamma}^m  \hat{\Gamma}_7  \otimes \hat{\Gamma}_m \hat{\Gamma}_7 ) \\ 
S^{(42)}_6  & =   \frac{i}{4}  (  \hat{\Gamma}^m  \otimes \hat{\Gamma}_m   \hat{\Gamma}_7 
   + \hat{\Gamma}^m \hat{\Gamma}_7   \otimes  \hat{\Gamma}_m )    \, , 
\end{aligned}
\ee
and  the remaining ones are  in the $({\bf \overline{20}}, {\bf 2}) $ 
\be
\begin{aligned}
S^{(42)}_7 &=  -  \frac{1}{4}   ( \hat{\Gamma}^6  \otimes   \hat{\Gamma}^{237}  -  \hat{\Gamma}^5 \otimes     \hat{\Gamma}^{137} -   \hat{\Gamma}^5\otimes    \hat{\Gamma}^{247}  -  \Gamma^6 \otimes   \Gamma^{147}  ) \\
S^{(42)}_8 & =  - \frac{1}{4}  (  \hat{\Gamma}^6 \otimes   \hat{\Gamma}^{247}  - \hat{\Gamma}^5 \otimes   \hat{\Gamma}^{147} +  \hat{\Gamma}^5 \otimes   \hat{\Gamma}^{237}  +   \hat{\Gamma}^6 \otimes   \hat{\Gamma}^{137}  )\,.
\end{aligned}
\ee

These singlets generate the  commutant of $U(1)_S$ in  $E_{6(6)}$.  Given the number of singlets this must be
\be
\label{app:U1comm}
C_{E_{6(6)}} (U(1)_S) = \mathbb{R}^+ \times \Spin(3,1) \times \SU(2,1) \times U(1)_S  \, . 
\ee
 From the commutators  \eqref{comm36} and \eqref{comm42} it  is easy to see that the  factor $\bbR^+$ is generated by  the combination
\be
\label{genRp}
J_\bbR = S^{(42)}_1 + S^{(42)}_2    \, . 
\ee
Similarly it is straightforward to identify the generators of  the group $SO(3,1)$ as 
\be
\begin{aligned}
J^{SO(3,1)}_1 & = \frac{\ii}{2}   (S^{(36)}_2 +  S^{(36)}_3 )\, ,  \\ 
J^{SO(3,1)}_2  & = \frac{\ii}{2}   S^{(36)}_4   \,  , \\
J^{SO(3,1)}_3  & = \frac{\ii}{2}   S^{(36)}_5    \, ,
\end{aligned}
\qquad 
\begin{aligned}
K^{SO(3.1)}_1 & =  \frac{\ii}{4} (S^{(42)}_1-S^{(42)}_2)   \, , \\
K^{SO(3,1)}_2 & = -  \frac{\ii}{4} S^{(42)}_3  \, ,  \\
K^{SO(3,1)}_3 & =  \frac{\ii}{4} S^{(42)}_4  \, .  \\
\end{aligned}
\ee
The remaining singlets give $SU(2,1)$. The compact generators are defined as
\be
 \begin{aligned}
J^{SU(2,1)}_1  & = -  \frac{\ii}{2}   S^{(36)}_7 \\
J^{SU(2,1)}_2  & =  \frac{\ii}{2}   S^{(36)}_8    \\
J^{SU(2,1)}_3 & = -  \frac{\ii}{4}   (S^{(36)}_1 +  S^{(36)}_2  -S^{(36)}_3 - S^{(36)}_8  ) \\
 J^{SU(2,1)}_8   & =   -  \frac{\ii}{4 \sqrt{3} }  ( S^{(36)}_1 +  S^{(36)}_2  - S^{(36)}_3  + 3 S^{(36)}_8  )  \, , 
\end{aligned}
\ee
while the non-compact ones are 
\be
\begin{aligned}
J^{SU(2,1)}_4  & = -  \frac{\ii}{2} S^{(42)}_7  \\ 
J^{SU(2,1)}_5  & =   \frac{\ii}{2} S^{(42)}_8 
\end{aligned} 
\qquad \quad 
\begin{aligned}
J^{SU(2,1)}_6  & = -  \frac{\ii}{2 }  S^{(42)}_6   \\ 
J^{SU(2,1)}_7  & =   \frac{\ii}{2} S^{(42)}_5  \, . \\ 
\end{aligned} 
\ee
The compact  singlets give the commutant of $U(1)_S$ into $\USp(6)$,
\be
\Comm_{\USp(8)}(U(1)_S)  = \SU(2) \times \SUH \times U(1) \times U(1)_S  \, . 
\ee 

We also need the $U(1)_S$ singlets in the ${\bf 27}$. Computing the action \eqref{delta27} of $U(1)_S$  on a generic element of the ${\bf 27}$, given in \eqref{27US}, we find five 
singlets 
\be
\label{app:27dec}
\begin{aligned}
{\bf 27} \,&=\,   ({\bf 1},{\bf 1} )_{(0,8)} \oplus   ({\bf 4},{\bf 1} )_{(0,-4)} \oplus  ({\bf 2},{\bf 1} )_{(3,-2)}  \oplus ({\bf  \bar{2}}, {\bf 1} )_{(-3,-2)}  \\ 
\,&\,   \oplus   ({\bf 1},{\bf 3} )_{(2,-4)} \oplus   ({\bf 1},{\bf \bar 3} )_{(-2,-4)}    \oplus   ({\bf \bar 2},{\bf 3} )_{(1,2)} \oplus   ({\bf 2},{\bf \bar 3} )_{(-1, 2)}\,,
 \end{aligned} 
\ee
One is  a singlet of both  $SO(3,1)$ and $SU(2,1)$ and has charge 8 under $\mathbb{R}^+$, 
\be
\label{app:K0}
K_0 \sim   \ii\,  \hat{\Gamma}_{56} \hat{\Gamma}_7  = \id \otimes \Gamma_5 \, , 
\ee
where in the second equality we used \eqref{app:2+4} for the gamma matrices. The other singlets are invariant under 
$SU(2,1)$ and  form a quadruplet of $SO(3,1)$ of charge $-4$ under $\mathbb{R}^+$ 
\be
\label{app:Ka}
\begin{aligned} 
& K_1 \,\sim\,  \ii\,  (  \hat{\Gamma}_{13} -   \hat{\Gamma}_{24}) \hat{\Gamma}_7  \,=\, \gamma_{(2)} \otimes  ( \Gamma_{13} -  \Gamma_{24} )  \, , \\ 
&K_2 \, \sim \,   \ii\, (  \hat{\Gamma}_{14 } +  \hat{\Gamma}_{23}) \hat{\Gamma}_7 \,=\,  \gamma_{(2)} \otimes  ( \Gamma_{14} +  \Gamma_{23} )   \, , \\ 
& K_3 \,\sim\,  \ii\,  ( \hat{\Gamma}_{12 } +  \hat{\Gamma}_{34}) \hat{\Gamma}_7  \,=\, \gamma_{(2)} \otimes  ( \Gamma_{12 } +  \Gamma_{34} )   \, , \\ 
&K_4 \,\sim\,  \ii\,  (  \hat{\Gamma}_{12 } -   \hat{\Gamma}_{34})  \hat{\Gamma}_7  \,=\, \gamma_{(2)} \otimes  ( \Gamma_{12 } -  \Gamma_{34} )  \, . 
\end{aligned} 
\ee

The singlets in the ${\bf 27}$ and ${\bf 78}$ are all we need to specify the generalised $U(1)_S$ structure. However,  the generators of $SO(3,1)$ and $\mathbb{R}^+$  in \eqref{app:U1comm} do not leave the singlets generalised vectors invariant and hence do not  belong to the $U(1)_S$ structure.
Using \eqref{adjprUsp8}, one can show that  they  are obtained as  products of the singlets in the $\rep{27}$ and their duals
\be
\label{app:so31genprod}
J_\alpha^{SO(3,1)} = 2  \ii\, \epsilon_{\alpha\beta\gamma}  (K_\beta \times_{\adj} K^*_\gamma)\,,  \qquad   K_\alpha^{SO(3,1)} = - \ii\, (K_\alpha \times_{\adj} K^*_4) \,,\qquad \alpha =1,2,3\,,
\ee
and
\be
\label{app:Rpgenprod}
J_{\mathbb{R}} = 4  (K_0 \times_{\adj} K^*_0)  - 4  (K_4 \times_{\adj} K^*_4)  \, .
\ee

\vspace{0.2cm}

In summary the generalised  $U(1)_S$ structure is defined by the five generalised vectors and the eight generators of $SU(2,1)$
\be
\label{app:genstr}
\{ K_I , J_A \} \qquad I=0, \ldots, 4 , \quad A= 1, \dots, 8 \, .
\ee

The last step is to derive explicit expressions for these generalised tensors in terms of geometrical objects on the six-dimensional internal manifold $M$.  We use the fact that, in our case,  
 $M$ is a fibration of the four sphere over a  Riemann surface and  that the four-sphere is generalised parallelisable as reviewed in Appendix~\ref{app:S4par}.

We decompose the  six-dimensional bundles 
in representation of $GL(2,\mathbb{R})$, the ordinary structure group on the Riemann surface, and $SL(5,\mathbb{R})$, the exceptional structure group of $S^4$. 
Under 
\be
\Ex{6} \supset \GL(2,\bbR)\times\SL(5,\bbR) \, ,
\ee
the generalised tangent bundle decomposes as 
\be
\begin{aligned}
\label{app:SL5-vecdecomp}
 E &\simeq T\mathbb{\Sigma} \oplus (T^*\mathbb{\Sigma} \otimes N_4)
      \oplus (\Lambda^2 T^*\mathbb{\Sigma} \otimes N_4') \oplus E_4\,, \\
       \rep{27} &= \repp{2}{1} \oplus \repp{2}{5'} \oplus \repp{1}{5}   \oplus \repp{1}{10} \,, 
\end{aligned}
\ee
where $E_4$, $ N_4$ and  $N_4'$ are defined  in  Appendix~\ref{app:S4par}.  Using \eqref{27US} and defining $\Cliff(5,\bbR)$ gamma matrices as 
\be
\label{5dg}
\Gamma_i = \{ \Gamma_1,  \ldots,  \Gamma_5 \}  \, ,
\ee
we can identify the components of the ${\bf 27}$   in  \eqref{app:SL5-vecdecomp} as 
%
\be
\begin{aligned}
& \{  \gamma_1 \otimes \id  , \gamma_2  \otimes \id  \}  \in {\bf (2,1)}   \\ 
& \{ \gamma_1  \otimes \Gamma_I ,  \gamma_2  \otimes \Gamma_I \}    \in  {\bf (2,5)}      \\
& \id  \otimes \Gamma_i    \in   {\bf (1,5)}   \\ 
& \gamma_{(2)}  \otimes \Gamma_{ij}    \in    {\bf (1,10)}  \, . 
\end{aligned} 
\ee
In terms of generalised vectors,  the elements of the  $ {\bf (2,1)}$ embed  as 
 \be 
 \label{app:2,1}
 R^{-1}\, { \hat{e}_1  \choose   \hat{e}_2}   \, , 
 \ee
 while those in  the ${\bf (2,5)}$ and {\bf (1,5)} can be written as
 \be
 \Psi_{i} =  R\,  {e_1  \\ \wedge E_i ,  \choose  e_2 \wedge E_i}   \qquad \mbox{and} \qquad 
   R^2\, {\rm vol}_{\mathbb{\Sigma}} \wedge E^\prime_i\,,\qquad i=1, \ldots, 5\,,
  \ee
where  $ \text{vol}_{\mathbb\Sigma}=e_1\wedge e_2$ is the volume form on the Riemann surface, $R$ is the $S^4$ radius, and $E_i$ and $E_i'$ are the sections of $N_4$ and $N_4'$ defined in Appendix~\ref{app:S4par}. 
The elements of the $ {\bf (1,10)}$  are the $\Xi_\alpha$, $\ti{\Xi}_\alpha$, with $\alpha = 1,2,3$, defined in  \eqref{Xi_tildeXi}, and
$E_{i5}$ with $i=1,2,3,4$.
 
Comparing with \eqref{app:K0} and \eqref{app:Ka},  we see that
\be
K_0 \in  {\bf (1,5)} \sim  \Lambda^2  T^\ast \mathbb\Sigma \otimes  N^\prime_4\,, \qquad K_I \in  {\bf (1,10)} \sim E_4 \, , \quad \text{for}\ I=1,\ldots,4\,,
\ee
and can then be written as generalised vectors on $M$ as
\be
\label{app:Ksing}
K_0 \sim  R^2 \text{vol}_{\mathbb \Sigma}\wedge \, E'_5  \,, \qquad  K_\alpha  \sim   \ti{\Xi}_\alpha   \,, \qquad 
K_4 \sim   \Xi_3   \, ,
\ee
where $\alpha=1,2,3$.
To have the final expressions for these five generalised vectors we still have to implement the twist of $S^4$ as described in Section \ref{sec:U1str}. 
The $E_{6(6)}$ element implementing the twist is 
\begin{align}
\Upsilon \,&=\,  -\frac{R}{2} \, \upsilon \times_{\adj}\Xi_3\,,\nn\\
\,&=\, -\xi_3 \otimes \upsilon - \tfrac{1}{4}\,R^3\, \upsilon \wedge \dd (\cos\zeta\,\sigma_3)\,,
\end{align}
and acts on the frames  $E_{ij}$, $E_i$ as  
\begin{align}\label{twisting_the_E}
\rme^{\Upsilon}\cdot E_{ij} &=E_{ij}+\tfrac{1}{2}\,\upsilon\wedge E_5(\delta_{1[i}\delta_{j]2}+\delta_{3[i}\delta_{j]4})-\tfrac{1}{2}\,\delta_{5[i}P_{j]}{}^k \upsilon \wedge E_k\,, \nonumber \\
\rme^{\Upsilon}\cdot E_i &=E_i+\tfrac{1}{2}\,\upsilon\wedge \big(E'_{[1}\delta_{2]i}+ E'_{[3}\delta_{4]i}\big)\,,
\end{align}
where $P_i{}^j$ is the matrix
\begin{equation}
P_i{}^j=\begin{pmatrix}
&&&-1\\
&&1&\\
&-1&&\\
1&&& 
\end{pmatrix} \, . 
\end{equation} 
It is then straightforward to check that only $K_4$ is modified by the twist, and the expressions \eqref{sec:Ksing} are obtained.

\medskip

Finally we need the expressions for the singlets in the  $\rep{78}$ generating $\SU(2,1)$. Under  $\Ex{6} \supset \GL(2,\bbR)\times\SL(5,\bbR)$ as 
\begin{align}
{\adj} F &   \simeq  \adj F_4 \oplus  (T \Sigma \otimes T^\ast \Sigma)  \oplus  (  T^\ast \Sigma \otimes E_4) 
  \oplus ( \Lambda^2  T^\ast \Sigma \otimes N_4)    \oplus (  T \Sigma \otimes E^\ast_4) \oplus ( \Lambda^2  T \Sigma \otimes N^\ast_4) \nonumber  \\
\rep{78}  & \sim  (\rep{1}, \rep{24}) \oplus ( \rep{4}, \rep{1} ) \oplus ( \rep{2}, \rep{10} ) \oplus ( \rep{1}, \rep{5} ) \oplus  ( \rep{2}, \rep{10} ) \oplus ( \rep{1}, \rep{5} ) 
\end{align}
where  $\adj F_4$ is  the adjoint bundle on $S^4$
\be
\adj F_4  \simeq \mathbb{R} \oplus  (T S^4 \otimes  T^\ast S^4)  \oplus \Lambda^3  T^\ast S^4 \oplus  \Lambda^3 T S^4   \, . 
\ee
The expressions for the singlets  are easily obtained from \eqref{adjprUsp8} as products of the ${\bf 27}$ and ${\bf \overline{27}}$. 
In this way we obtain precisely the expressions given in Eq.~\eqref{sec:Jsing}, where the twisting by $\Upsilon$ can be evaluated with the aid of \eqref{twisting_the_E}.

\section{Parameterisation of the H structure moduli space}\label{app:param_Hstructure}

We discuss here our parameterisation of the coset space $\mathcal{M}_{\rm H}=\frac{\SU(2,1)}{\SU(2)\times\Uni{1}}$ that describes the hypermultiplet structure moduli space.
We model the generators of $\SU(2,1)$ on the matrices ${\sf j}_A$, $A=1,\ldots,8$, defined as:
\be
{\sf j}_{1,2,3} = -\ii \,\lambda_{1,2,3}\,,\qquad {\sf j}_{4,5,6,7} = \lambda_{4,5,6,7}\,,\qquad {\sf j}_8 = -\ii\,\lambda_{8}\,,
\ee
where $\lambda_A$, $A=1,\ldots,8$, are the standard Gell-Mann matrices generating the $\su_{3}$ algebra in the fundamental representation. 
These generators satisfy 
\be
{\sf j}_A^\dagger \,m + m\, {\sf j}_A = 0\,,\qquad \text{with}\ m = {\rm diag}(-1,-1,1)\,,
\ee
\be
{\rm tr}\big(\,{\sf j}_A \,{\sf j}_B\big) \,=\, 2\,\eta_{AB}\,\qquad \text{with}\ \eta = {\rm diag}(-1,-1,-1,1,1,1,1,-1)\,,
\ee
as well as the commutation relations
\begin{align}\label{comm_su21_generators}
&[\,{\sf j}_1 , {\sf j}_2\,] = 2\, {\sf j}_3  \,,\qquad [\,{\sf j}_3 , {\sf j}_1\,] = 2\, {\sf j}_2  \,,\qquad   [\,{\sf j}_2 , {\sf j}_3\,] = 2\, {\sf j}_1 \,,\nn\\
&[\,{\sf j}_4\,,\, {\sf j}_5\,] = -\,(\,{\sf j}_3 + \sqrt 3 \,{\sf j}_8  ) \qquad [\,\tfrac{1}{2}(\,{\sf j}_3 + \sqrt 3 \,{\sf j}_8  )\,,\, {\sf j}_4\,] = 2\, {\sf j}_5  \,,\qquad [\,{\sf j}_5\,,\, \tfrac{1}{2}(\,{\sf j}_3 + \sqrt 3 \,{\sf j}_8  )\,] = 2\, {\sf j}_4 \,, \nn\\
&[\,{\sf j}_6\,,\, {\sf j}_7\,] = -\,(\,-{\sf j}_3 + \sqrt 3 \,{\sf j}_8  ),\qquad [\,\tfrac{1}{2}(\,-{\sf j}_3 + \sqrt 3 \,{\sf j}_8  )\,,\, {\sf j}_6\,] = 2\, {\sf j}_7  ,\qquad [\,{\sf j}_7\,,\, \tfrac{1}{2}(\,-{\sf j}_3 + \sqrt 3 \,{\sf j}_8  )\,] = 2\, {\sf j}_6 \,, \nn\\
&[\,{\sf j}_1 , {\sf j}_4\,] =  {\sf j}_7   \,,\qquad [\,{\sf j}_7 , {\sf j}_1\,] =  {\sf j}_4  \,, \qquad [\,{\sf j}_4 , {\sf j}_7\,] = - {\sf j}_1\,,\nn\\
&[\,{\sf j}_2 , {\sf j}_4\,] =  {\sf j}_6    \,,\qquad [\,{\sf j}_6 , {\sf j}_2\,] =  {\sf j}_4 \,,\qquad [\,{\sf j}_4 , {\sf j}_6\,] = - {\sf j}_2  \,,\nn\\
&[\,{\sf j}_1 , {\sf j}_5\,] = - {\sf j}_6    \,,\qquad [\,{\sf j}_6 , {\sf j}_1\,] = - {\sf j}_5 \,,\qquad [\,{\sf j}_5 , {\sf j}_6\,] =  {\sf j}_1 \,,\nn\\
&[\,{\sf j}_2 , {\sf j}_5\,] = {\sf j}_7   \,,\qquad [\,{\sf j}_7 , {\sf j}_2\,] =  {\sf j}_5   \,,\qquad [\,{\sf j}_5 , {\sf j}_7\,] = - {\sf j}_2 \,,\nn\\
&[\,{\sf j}_1 , {\sf j}_8\,] =  [\,{\sf j}_2 , {\sf j}_8\,] =  [\,{\sf j}_3 , {\sf j}_8\,] = 0\,,\nn\\
&[\,{\sf j}_4 , \sqrt3\,{\sf j}_3-{\sf j}_8\,] =  [\,{\sf j}_5 , \sqrt3\,{\sf j}_3-{\sf j}_8\,]  =   [\,{\sf j}_6 , \sqrt3\,{\sf j}_3+{\sf j}_8\,]= [\,{\sf j}_7 , \sqrt3\,{\sf j}_3+{\sf j}_8\,] = 0\,,
\end{align}
where the first three lines show the three $\su_2$ subalgebras.
Note that  $\{{\sf j}_1,{\sf j}_2,{\sf j}_3,{\sf j}_8\}$ generate the compact subgroup $\SU(2)\times\U(1)\subset\SU(2,1)$. It is convenient 
to choose a solvable parameterisation for the remaining generators, describing the coset space $\frac{\SU(2,1)}{\SU(2)\times\U(1)}$. Following the Appendix D of \cite{Ceresole:2014vpa}, we define\footnote{We rearrange the indices of their $3\times 3$ matrices as $1_{\rm there}\to 3_{\rm here}$, $2_{\rm there}\to 1_{\rm here}$, $3_{\rm there}\to 2_{\rm here}$.} 
\begin{align}\label{coset_generators_j}
&T_1 = \frac{1}{2\sqrt 2} \left(\,  {\sf j}_1 -  {\sf j}_2 - {\sf j}_4 -{\sf j}_5 \right)\,,\qquad T_2 = \frac{1}{2\sqrt 2} \left( \, {\sf j}_1 +  {\sf j}_2 + {\sf j}_4 -  {\sf j}_5 \right)\,,\nn\\[1mm]
&T_{\bullet} = \frac{1}{4} \left(  2\,{\sf j}_7  + {\sf j}_3-\sqrt3 \,{\sf j}_8 \right)\,,\qquad \ \quad H_0 = \frac{1}{2}\, {\sf j}_6\,.
\end{align}
These span the Borel subalgebra of the $\SU(2,1)$ algebra and satisfy the commutation relations
\be
[H_0, T_\bullet] = T_\bullet\,,\qquad [H_0,T_1]=\frac12\, T_1\,,\qquad [H_0,T_2]=\frac12\, T_2\,,\qquad [T_1,T_2]=T_\bullet\,.
\ee
A parameterisation of the coset is obtained by exponentiating the Borel subalgebra as
\be
\label{g_SU21_coset_solvable}
L = \rme^{-( \theta_1+\theta_2) T_1 + (\theta_1-\theta_2) T_2 + \xi\, T_\bullet}\, \rme^{-2\varphi H_0}\,,
\ee
where $\{\varphi,\xi,\theta_1,\theta_2\}$ are the four real coordinates. Starting from the coset representative \eqref{g_SU21_coset_solvable}, we compute the Maurer-Cartan form $L^{-1}\diff L$ and then identify the coset vielbeine as the coefficients of its expansion in the coset generators,
\be\label{MCform_hyper}
L^{-1}\diff L = -2\,\diff \varphi H_0 -\rme^\varphi(\diff\theta_1+\diff\theta_2)  T_1 +\rme^\varphi (\diff \theta_1-\diff \theta_2) T_2 + \rme^{2\varphi}\left( \diff \xi  -\theta_1 \diff \theta_2 + \theta_2 \diff \theta_1 \right) T_\bullet\,.
\ee
In this way we obtain the following Einstein metric on $\frac{\SU(2,1)}{\SU(2)\times\U(1)}\,$,
\be\label{Quaternionic_metric_App}
\diff s^2 = 2\, \diff \varphi^2 + \rme^{2\varphi} \left( \diff \theta_1^2 + \diff \theta_2^2 \right) + \frac12\,\rme^{4\varphi} \left( \diff \xi -\theta_1 \diff \theta_2 + \theta_2 \diff \theta_1\right)^2\,.
\ee
The normalisation is chosen so that the Ricci scalar is $\mathcal{R}=-12$, in agreement with our five-dimensional supergravity conventions. 

In the main text, we need the ``dressed'' $ \su_{2}$ algebra constructed via the adjoint action of the coset representative on the $ \su_{2}$ algebra generated by  $\{{\sf j}_1, {\sf j}_2,{\sf j}_3\}$, that is
\be
\hat {\sf j}_1 = L\, {\sf j}_1\, L^{-1} \,,\qquad \hat {\sf j}_2 = L\, {\sf j}_2\, L^{-1}\,,\qquad \hat {\sf j}_3 = L\, {\sf j}_3\, L^{-1}\,.
\ee
An explicit evaluation using \eqref{g_SU21_coset_solvable} gives
\begin{align}
\hat{\sf j}_1 & =  \tfrac{1}{2}\,\rme^\varphi({\sf j}_1+{\sf j}_5) + \tfrac{1}{4}\,\rme^\varphi\big( \theta_2^2 -3\theta_1^2 +2 \rme^{-2\varphi}\big)({\sf j}_1-{\sf j}_5) + \tfrac{1}{2}\,\rme^\varphi\left( \xi-2\theta_1\theta_2\right)({\sf j}_2+{\sf j}_4) + \tfrac{1}{\sqrt2}\,\rme^\varphi\theta_2{\sf j}_6\nn\\[1mm]
&  -\tfrac{1}{2\sqrt2}\,\rme^\varphi\theta_1\big(3 {\sf j}_3+\sqrt3\, {\sf j}_8 \big) + \tfrac{1}{4\sqrt2}\left[ \rme^\varphi(\theta_1^3+\theta_1\theta_2^2-2\theta_2\xi \big) -2\,\rme^{-\varphi}\theta_1 \right]\big({\sf j}_3+2\,{\sf j}_7-\sqrt3\, {\sf j}_8\big)\,, \nn\\[4mm]
\hat {\sf j}_2 &= \tfrac{1}{2}\,\rme^\varphi ({\sf j}_2-{\sf j}_4) + \tfrac{1}{4}\,\rme^\varphi\big(\theta_1^2-3\theta_2^2+2\rme^{-2\varphi}\big)({\sf j}_2+{\sf j}_4) - \tfrac{1}{2}\,\rme^\varphi\left( \xi+2\theta_1\theta_2\right)({\sf j}_1-{\sf j}_5) - \tfrac{1}{\sqrt2}\rme^\varphi\theta_1{\sf j}_6 \nn\\[1mm]
& -\tfrac{1}{2\sqrt2}\,\rme^\varphi\theta_2 \big(3{\sf j}_3+\sqrt3\,{\sf j}_8\big) + \tfrac{1}{4\sqrt2} \left[ \rme^\varphi \big( \theta_2^3+\theta_1^2\theta_2 + 2\theta_1\xi \big) -2\,\rme^{-\varphi}\theta_2 \right]\big({\sf j}_3+2\,{\sf j}_7-\sqrt3\, {\sf j}_8\big)\,, \nn\\[4mm]
\hat {\sf j}_3 &=  -\tfrac{1}{4\sqrt2}\left[ \rme^{2\varphi}\big(\theta_1^3 + \theta_1\theta_2^2 + 2\theta_2\xi\big) - 6\theta_1 \right]({\sf j}_1-{\sf j}_5) -\tfrac{1}{4\sqrt2}\left[ \rme^{2\varphi}\big(\theta_2^3 + \theta_1^2\theta_2 - 2\theta_1\xi \big) - 6\theta_2 \right]({\sf j}_2+{\sf j}_4) \nn\\[1mm]
&+\tfrac{1}{2\sqrt2}\,\rme^{2\varphi}\left[\theta_1 ({\sf j}_1+{\sf j}_5) +\theta_2({\sf j}_2-{\sf j}_4) \right] -\tfrac{1}{2}\,\rme^{2\varphi}(\xi\, {\sf j}_6 + {\sf j}_7) + \tfrac{1}{8} \left[ 2 - \rme^{2\varphi}\big(\theta_1^2+\theta_2^2\big) \big(3{\sf j}_3+\sqrt3\,{\sf j}_8\big) \right] \nn\\[1mm]
& +\tfrac{1}{32} \left[ \rme^{2\varphi}\big( \theta_1^2+\theta_2^2 \big)^2 + 4 \,\rme^{2\varphi} \big(1+\xi^2\big)  -12 \big( \theta_1^2+\theta_2^2 \big) + 4\,\rme^{-2\varphi} \right]  \big({\sf j}_3+2\,{\sf j}_7-\sqrt3\, {\sf j}_8\big) \,.
\end{align}
Now we can replace the matrices ${\sf j}_A$ with the generalised tensors $J_A$ invariant under the $\U(1)$ generalised structure. This provides our four-parameter family of H structures.

\bibliographystyle{JHEP}
\bibliography{Bibliography}

\end{document}